\documentclass[twocolumn,aps,prl,showpacs,superscriptaddress]{revtex4}
\usepackage[dvips]{graphicx}
\usepackage{bm}
\begin{document}
\title{Single-ion and exchange anisotropy in high-symmetry tetramer single molecule magnets}
\author{Richard A. Klemm}
\email{klemm@phys.ksu.edu} \affiliation{Department of Physics,
Kansas State University, Manhattan, KS 66506 USA}
\author{Dmitri V. Efremov}
\email{efremov@theory.phy.tu-dresden.de} \affiliation{Institut
f{\"u}r Theoretische Physik, Technische Universit{\"a}t Dresden,
01062 Dresden, Germany}
\date{\today}
\begin{abstract}
 For  equal-spin $s_1$ tetramer single molecule magnets with ionic point
groups $g=T_d$, $D_{4h}$, $D_{2d}$,  $C_{4h}$, $C_{4v}$, and
$S_4$, we write the group-invariant single-ion,
Dzaloshinskii-Moriya, and anisotropic near-neighbor and
next-nearest-neighbor exchange Hamiltonians using the respective
local axial and azimuthal vector groups.  In the molecular
representation, the symmetric anisotropic exchange interactions
renormalize the near-neighbor and next-nearest-neighbor isotropic
exchange interactions to $\tilde{J}_g$ and $\tilde{J}_g'$.    The
local single-ion and symmetric anisotropic exchange interactions
generate the site-independent
 anisotropy interactions in the molecular
representation, $J^g_z(\mu_1^g)$, $J_{1,z}^g(\mu_{12}^g)$, and
$J_{2,z}^g$, respectively, where $\mu_1^g$ and $\mu_{12}^g$
represent the group-consistent sets of parameters required to
diagonalize the single-ion and symmetric near-neighbor and
next-nearest-neighbor anisotropic exchange Hamiltonians,
respectively.    The site-independent group-invariant
near-neighbor and next-nearest-neighbor Dzaloshinskii-Moriya
interactions generally may be written as vectors of rank three and
two, ${\bm d}^g$ and ${\bm d}_{\perp}'^{,g}$, respectively. The
local anisotropy interactions also generate site-dependent
single-ion, near-neighbor, and next-nearest-neighbor anisotropic
exchange interactions in the molecular representation.   Using our
exact, compact forms for the single-ion matrix elements, we
evaluate the eigenstate energies to first order in
$J^g_z(\mu_1^g)$, $J_{1,z}^g(\mu_{12}^g)$, and $J_{2,z}$. There
are two types of ferromagnetic (FM) $(\tilde{J}_g>0$ and
antiferromagnetic (AFM) $(\tilde{J}_g<0$ tetramers.  In Type I,
$\tilde{J}_g'-\tilde{J}_g>0$, the tetramer acts as two dimers with
the maximal  pair quantum numbers $s_{13}=s_{24}=2s_1$ at low
temperatures $T$. Type II tetramers with
$\tilde{J}_g'-\tilde{J}_g<0$ are frustrated, with minimal values
of the  pair quantum numbers $s_{13}$ and $s_{24}$ at low $T$. For
both Type I and II AFM tetramers, we evaluate the first-order
level-crossing inductions analytically for arbitrary $s_1$, and
illustrate the results for $s_1=1/2,1,3/2$.
 Accurate Hartree expressions
  for the thermodynamics, electron
paramagnetic resonance  (EPR) absorption and inelastic neutron
scattering cross-section are given.  A procedure to extract the
effective microscopic parameters for Types I and II FM tetramers
using EPR is given.
\end{abstract}
\vskip0pt\vskip0pt \maketitle
\section{I. Introduction}
Single molecule magnets (SMM's) have been a topic of great
interest for more than a decade,\cite{general} because of their
potential uses in quantum computing and/or magnetic
storage,\cite{moregeneral} which are possible due to magnetic
quantum tunneling  (MQT) and entangled states.  In fits to a
wealth of data,  the Hamiltonian within an SMM was assumed to be
the  Heisenberg exchange interaction plus weaker total (global, or
giant) spin anisotropy interactions, with a fixed overall total
spin quantum number $s$.\cite{general} MQT and entanglement were
only studied in this simple model.

The simplest SMM's  are dimers.\cite{ek,ek2} Surprisingly, two
antiferromagnetic dimers, an Fe$_2$ and a Ni$_2$, appear to have
substantial single-ion anisotropy without any appreciable total
spin anisotropy.\cite{Shapira,Mennerich,ek2} Although the most
common SMM's have ferromagnetic (FM) intramolecular interactions
and contain $n\ge8$ magnetic ions,\cite{Dalal} a number of
intermediate-sized FM SMM's with $n=4$ and rather simple molecular
structures were recently studied.  The Cu$_4$ tetramer
 Cu$_4$OCl$_6$(TPPO)$_4$, where TPPO is triphenylphosphine oxide, has four spin 1/2 ions
on the corners of a regular tetrahedron, with an $s=2$ ground
state and approximate $T_d$ symmetry.\cite{Black1,Black2,Black3}
In this case, there are no single-ion anisotropy effects, but
anisotropic symmetric exchange interactions were thought to be
responsible for the zero-field energy
splittings.\cite{Black1,Buluggiu} The Co$_4$,
Co$_4$(hmp)$_4$(MeOH)$_4$Cl$_4$, where hmp is
hydroxymethylpyridyl, and Cr$_4$,
[Cr$_4$S(O$_2$CCH$_3$)$_8$(H$_2$O)$_4$](NO$_3$)$_2\cdot$H$_2$O,
compounds have $s=6$ ground states with spin 3/2 ions on the
corners of tetrahedrons.\cite{Co4,Cr4} Those  compounds have $S_4$
and approximate $D_{2d}$ symmetry, respectively.\cite{Co4,Cr4} A
number of high symmetry $s=4$ ground state Ni$_4$ structures with
spin 1 ions were
reported.\cite{Ni4,Ni4Maria,Edwards,Ni4S4,Hendrickson} Two of
these, [Ni(hmp)(ROH)Cl]$_4$, where R is an alkyl group, such as
methyl, ethyl, or 3,3-dimethyl-1-butyl and hmp is
2-hydroxymethylpyridyl, form tetramers with precise $S_4$ group
symmetry.\cite{Hendrickson,Ni4S4} Two others, Ni$_4$(ROH)L$_4$,
where R is methyl or ethyl and H$_2$L is
salicylidene-2-ethanolamine, had approximate $S_4$ symmetry,
although the precise symmetry was only $C_1$.\cite{Ni4}  Several
planar Mn$_4$ compounds with the Mn$^{+3}$ spin 2 ions on the
corners of squares were made, with overall $s=8$ tetramer ground
states.\cite{Boskovic} Although two of these complexes had only
approximate $S_4$ symmetry, one of these complexes,
Mn$_4$Cl$_4$(L')$_4$, where H$_2$L' is
4-$t$-butyl-salicylidene-2-ethanolamine, had perfect $S_4$
symmetry.\cite{Boskovic} Inelastic neutron scattering (INS)
experiments provided strong evidence for single-ion anisotropy in
Co$_4$ and a Ni$_4$ with approximate $S_4$ symmetry.\cite{Co4,Ni4}
The presence of single-ion or exchange anisotropy actually
precludes the total spin $s$ from begin a good quantum
number.\cite{ek2} Fits to electron paramagnetic resonance (EPR)
Ni$_4$ data assuming a fixed $s$ were also problematic, suggesting
single-ion or exchange anisotropy in that tetramer, as
well.\cite{Hillprivate}

Recently there have been microscopic treatments of
dimers,\cite{ek2} trimers, and tetramers, including Zeeman
$g$-tensor anisotropy, single-ion anisotropy, and anisotropic
exchange interactions.\cite{Bocabook} Most of those treatments and
their recent extensions to more general systems expressed the
single-spin matrix elements  only  in terms of  Wigner $3j$, $6j$,
and $9j$ symbols.\cite{Bocabook,WG} While such treatments are very
helpful in fitting experimental data, more compact analytic forms
are desirable to study microscopic models of FM SMM's in which the
MQT and entanglement issues crucial for quantum computing can be
understood. We constructed the single-ion and anisotropic
near-neighbor (NN) and next-nearest-neighbor (NNN) exchange SMM
 Hamiltonians from the respective local
axial and azimuthal vector groups for equal-spin tetramer SMM's
with point group symmetries $g=T_d$, $D_{4h}$, $D_{2d}$, $C_{4h}$,
$C_{4v}$, and $S_4$, and found  compact analytic expressions for
the single-spin matrix elements of four general spins.
Surprisingly, each local vector group generates site-dependent
molecular single-ion and exchange anisotropy. We evaluate
 the magnetization, specific heat,
EPR and INS transitions  in the Hartree approximation, and provide a
procedure for extracting the effective site-independent microscopic
parameters using EPR.
\begin{figure}
\includegraphics[width=0.23\textwidth]{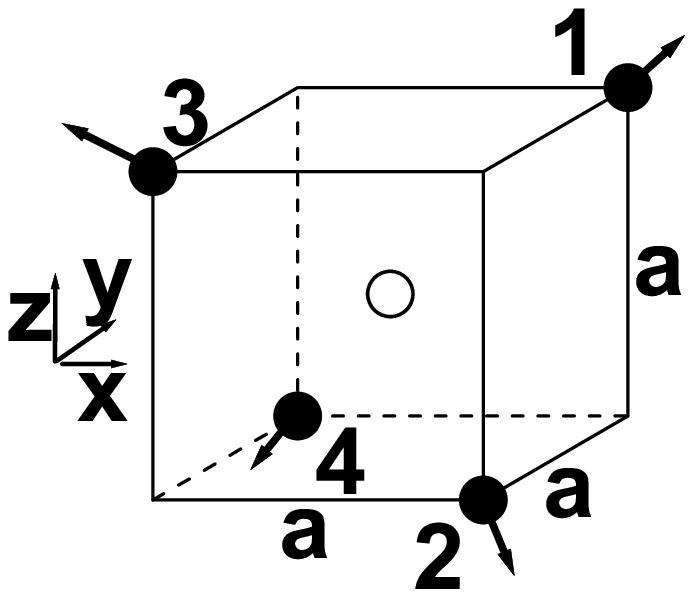}\hskip5pt\includegraphics[width=0.23\textwidth]{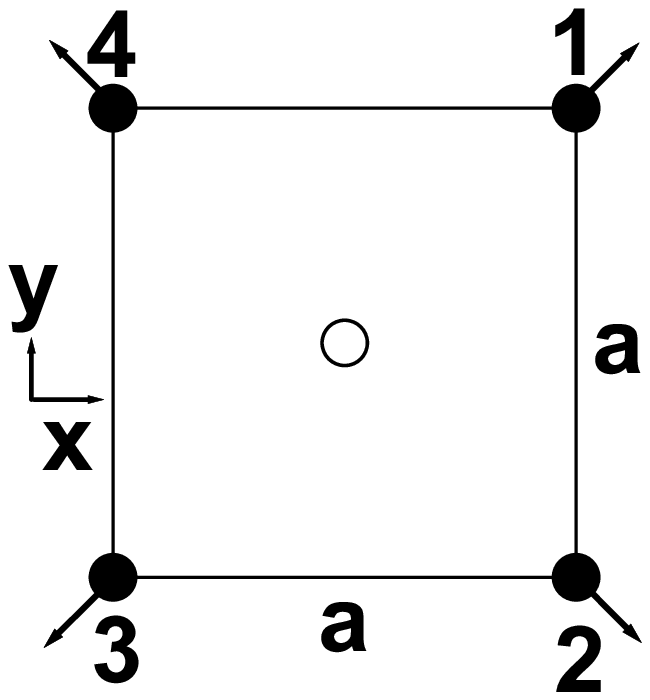}
\caption{$T_{d}$ (left) and $D_{4h}$ (right)  ion sites (filled).
Circle: origin. Arrows: local axial $\hat{\bm z}_n^{T_d}$ (left),
azimuthal $\hat{\bm x}_n^{D_{4h}}$ (right) single-ion vectors. The
axial vectors $\hat{\bm z}_n^{D_{4h}}=\hat{\bm z}$, normal to the
ionic plane.}\label{fig1}
\end{figure}

\begin{figure}
\includegraphics[width=0.23\textwidth]{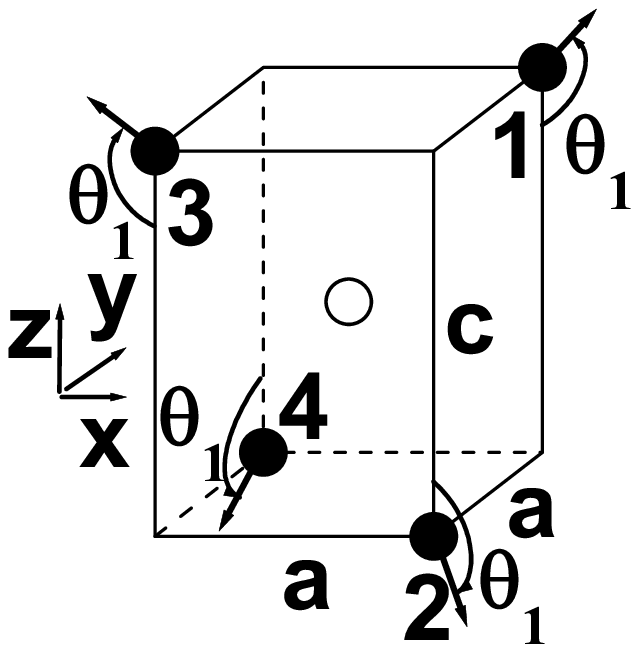}\hskip2pt\includegraphics[width=0.235\textwidth]{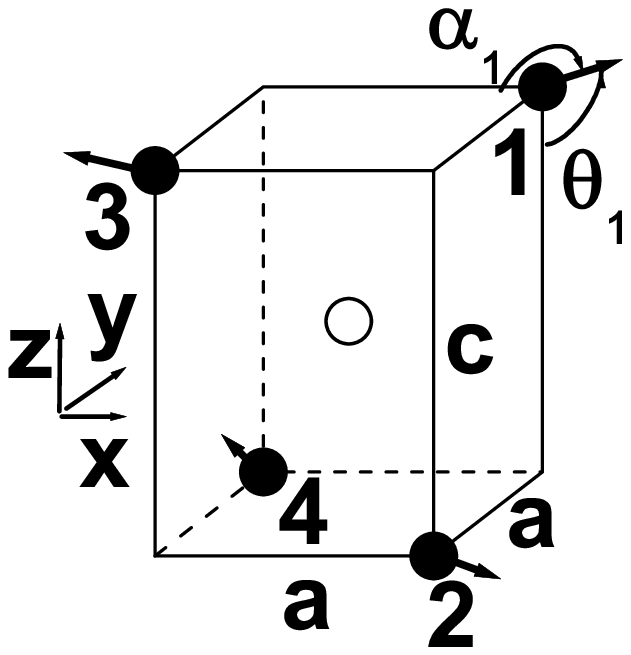}
\caption{$D_{2d}$ (left) and $S_4$ (right) ion sites (filled).
Circle: origin.  Arrows:  local axial single-ion vectors.  The
$g=D_{2d},S_4$ axial vectors $\hat{\bm z}^g_1$ make the angles
$\theta_1^g$ with the $z$ axis, and the $S_4$ axial vector
$\hat{\bm z}^{S_4}_1$ also makes the angle $\alpha_1$ with the $x$
axis, where
$\cos\alpha_1=\sin\theta^{S_4}_1\cos\phi^{S_4}_1$.}\label{fig2}
\end{figure}

\begin{figure}
\includegraphics[width=0.225\textwidth]{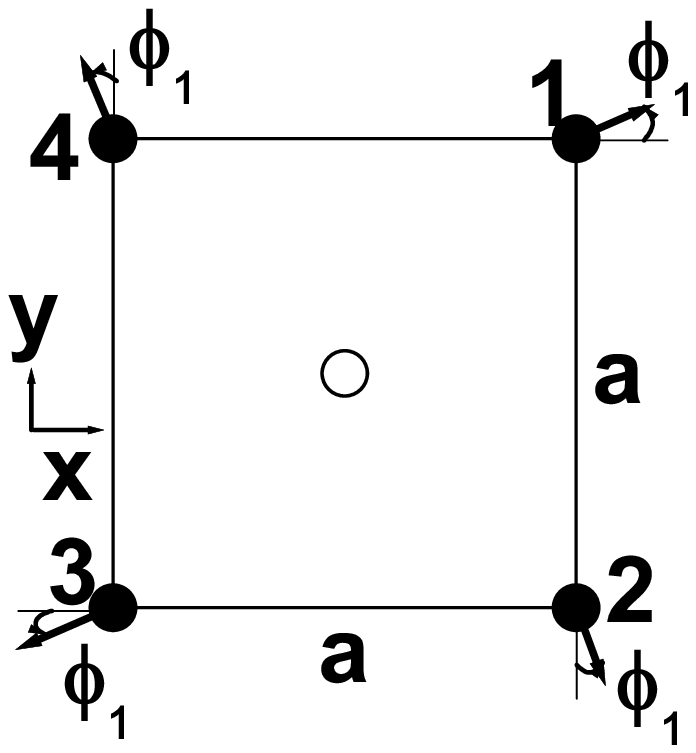}\hskip5pt\includegraphics[width=0.22\textwidth]{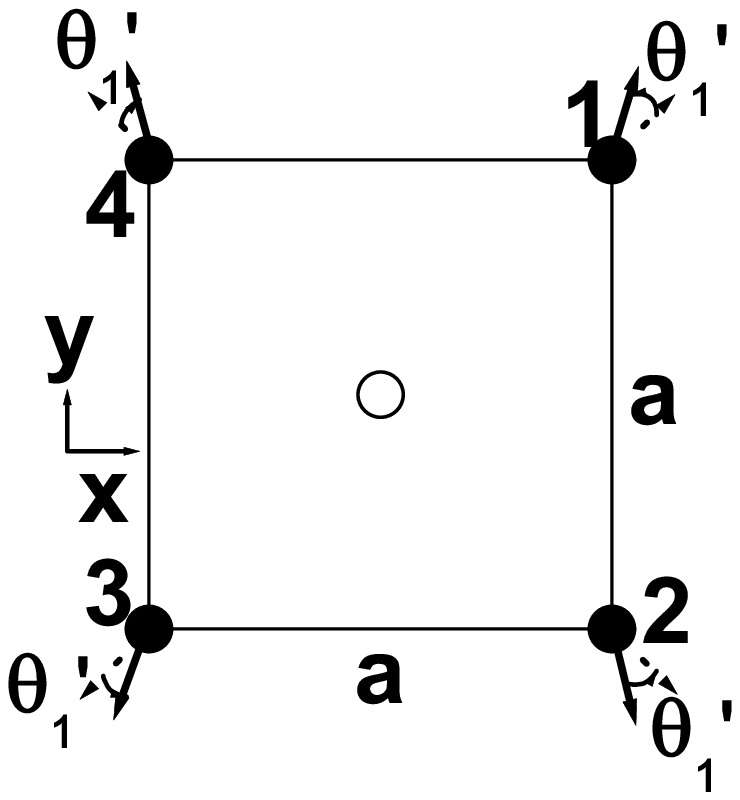}
\caption{$C_{4h}$ (left) and $C_{4v}$ (right) ion sites (filled).
Circle: origin.  Arrows:  local azimuthal $\hat{\bm x}_n^{C_{4h}}$
(left) axial $\hat{\bm z}_n^{C_{4v}}$ (right) single-ion vectors.
The axial vectors $\hat{\bm z}_n^{C_{4h}}=\hat{\bm z}$. The
$C_{4v}$ axial vectors each make the angle
$\theta_1=\pi/2-\theta_1'$ with with $z$ axis. The dotted arrows
(equivalent to the $D_{4h}$ azimuthal single-ion vectors $\hat{\bm
x}_n^{D_{4h}}$) are their projections in the $xy$
plane.}\label{fig3}
\end{figure}
\section{II. Structures and Bare Hamiltonian}

For SMM's with ionic
 site point groups $g=T_d, D_{4h}$, we assume the four equal-spin $s_1$ ions sit  on opposite
corners of a cube or square of side $a$ centered at the origin, as
pictured in Fig. 1. For SMM's with $g=D_{2d},S_4$, we take the
ions to sit on opposite corners of a tetragonal prism with sides
$(a,a,c)$ centered at the origin, as in Fig. 2.   The ions for
$g=C_{4h}, C_{4v}$ also sit on the corners of a square of side $a$
centered at the origin, as pictured in Fig. 3, but the ligand
groups have different symmetries than for the simpler $D_{4h}$
case pictured in Fig. 1. In each case, we take the origin to be at
the geometric center, so that $\sum_{n=1}^4{\bm r}_n=0$, where the
relative ion site vectors
\begin{eqnarray}
{\bm r}_{n}&=&\frac{a}{2}[-\gamma_n^{+}\hat{\bm
x}+\gamma_n^{-}\hat{\bm
y}]-\frac{c}{2}(-1)^n\hat{\bm z},\label{rn}\\
\gamma_n^{\pm}&=&\epsilon_n^{+}(-1)^{n/2}\pm\epsilon_n^{-}(-1)^{(n+1)/2},\label{gamman}
\end{eqnarray}
and $\epsilon^{\pm}_n=[1\pm(-1)^n]/2$.\cite{Tinkham} For clarity,
the values of $\epsilon_n^{\pm}$ and $\gamma_n^{\pm}$ are given in
Table 1. In tetrahedrons with $g=T_d$, $c/a=1$, approximately as
in Cu$_4$.\cite{Black1} In squares with $g=D_{4h}$, $C_{4h}$, or
$C_{4v}$, $c=0$.  The high $D_{4h}$ symmetry is approximately
exhibited by the square  Nd$_4$ compound, Nd$_4$(OR)$_{12}$, where
R is 2,2-dimethyl-1-propyl, in which the Nd$^{+3}$ ions have equal
total angular momentum $j=9/2$.\cite{Nd4,KlemmLuban} We note that
the Mn$_4$ SMM's with approximate or exact $S_4$ symmetry also
have $c=0$.\cite{Boskovic}  In tetragonal prisms with $g=D_{2d}$
or $S_4$, $c/a>1$, approximately as in Co$_4$,\cite{Co4} or
$c/a<1$, as in Mn$_4$ and a Ni$_4$.\cite{Boskovic,Ni4Maria}
$\hat{\bm x}, \hat{\bm y}, \hat{\bm z}$ are the molecular (or
laboratory) axes of each SMM.

\begin{table}
\centerline{\begin{tabular}{ccccc}
\hline $n$&$\epsilon_n^{+}$&$\epsilon_n^{-}$&$\gamma_n^{+}$&$\gamma_n^{-}$\\
\hline 1&0&1&-1&1\\
2&1&0&-1&-1\\
3&0&1&1&-1\\
4&1&0&1&1 \end{tabular}} \caption{Values of $\epsilon_n^{\pm}$ and
$\gamma_n^{\pm}$.}
\end{table}

The most general Hamiltonian quadratic in the four spin operators
${\bm S}_n$ may be written for group $g$ as
\begin{eqnarray}
{\cal H}^g=-\mu_B \sum_{n=1}^4{\bm B}\cdot\tensor{\bm
g}^g_n\cdot{\bm S}_n+\sum_{n,n'=1}^4{\bm S}_n\cdot\tensor{\bm
D}_{n,n'}^g\cdot{\bm S}_{n'},\label{generalH}
\end{eqnarray}
where $\mu_B$ is the Bohr magneton and  ${\bm
B}=B(\sin\theta\cos\phi,\sin\theta\sin\phi,\cos\theta)$ is the
magnetic induction at an arbitrary direction $(\theta,\phi)$
relative to the molecular coordinates $(\hat{\bm x},\hat{\bm
y},\hat{\bm z})$.\cite{Bocabook,BenciniGatteschi}

The exchange matrices $\tensor{\bm D}^g_{n,n'}$ for $n'\ne n$ can
have antisymmetric terms, which are the Dzyaloshinshii-Moriya (DM)
interactions. Of the six molecular group symmetries under study,
the DM interactions are non-vanishing for all but the highest
group symmetry, $T_d$.\cite{Moriya,Waldmann}
 For the other five group symmetries, the  DM interactions  lead to weak
 perturbations to the eigenstate energies.
 Inequivalent ions or ligands, or local structural distortions
from the ideal point group symmetry, as commonly occur in real
systems,\cite{NaV2O5} break these $g$ symmetries, introducing
stronger DM interactions. Such lower symmetry groups will be
discussed elsewhere.\cite{ekfuture}

For simplicity, we take $\tensor{\bm g}^g_n$ to be diagonal,
isotropic, and site-independent, so that the Zeeman interaction
may be written in terms of a single gyromagnetic ratio
$\gamma\approx2\mu_B$. Thus in the following, $g$ only refers to
the molecular group.    We separate $\tensor{\bm D}^g_{n,n'}$ into
its symmetric and antisymmetric parts, $\tensor{\bm
D}^g_{n,n'}=\tensor{\bm D}^{g,s}_{n,n'}+\tensor{\bm
D}^{g,a}_{n,n'}$, respectively.  For $n'=n$, the single-ion
$\tensor{\bm D}^g_{n,n}$ is necessarily symmetric, so $\tensor{\bm
D}^{g,a}_{n,n}=0$.  For each $g$, the four  $\tensor{\bm
D}^{g,s}_{n,n}$ contain the local single-ion structural
information, and the six distinct symmetric $\tensor{\bm
D}^{g,s}_{n,n'}$ contain the local symmetric exchange structural
information, which lead to the isotropic, or Heisenberg, exchange
interactions, and the remaining symmetric anisotropic exchange
interactions. The six distinct antisymmetric
 $\tensor{\bm
D}^{g,a}_{n,n'}$  contain additional local structural information
which lead to the DM interactions. Physically, the symmetric
anisotropic exchange interactions also contain the dipole-dipole
interactions, which can be even larger in magnitude than the terms
originating from actual anisotropic
exchange.\cite{BenciniGatteschi}

As is well known, each of the symmetric rank-three tensors (or
matrices) can be diagonalized by three rotations:  a rotation by
the angle $\phi^g_{n,n'}$ about the molecular $z$ axis, then a
rotation by the angle $\theta^g_{n,n'}$ about the rotated
$\tilde{x}$ axis, followed by a rotation by the angle
$\psi^g_{n,n'}$ about the rotated $\tilde{z}$
axis.\cite{Goldstein} This necessarily leads to the three
principal axes $\hat{\tilde{\bm x}}^g_{n,n'}$, $\hat{\tilde{\bm
y}}^g_{n,n'}$, and $\hat{\tilde{\bm z}}^g_{n,n'}$.   For the
single-ion axes with $n'=n$, we denote these principal axes to be
$\hat{\tilde{\bm x}}^g_n$, $\hat{\tilde{\bm y}}^g_n$, and
$\hat{\tilde{\bm z}}^g_n$, respectively,  which are written
explicitly in Sec. III.  The non-vanishing matrix elements in
these locally-diagonalized symmetric matrix coordinates are
$\tilde{D}^{g,s}_{n,n',xx}$, $\tilde{D}^{g,s}_{n,n',yy}$ and
$\tilde{D}^{g,s}_{n,n',zz}$. Since the structural information in
each of the $\tensor{\tilde{\bm D}}^{g,s}_{n,n'}$ depends upon the
local environment, in the absence of molecular group $g$ symmetry,
each of these angles would in principle be different from one
another. We characterize the single-ion and symmetric anisotropic
exchange vector sets by those for the spin at site 1, the NN pair
at sites 1 and 2, and the NNN pair at sites 1 and 3, respectively,
\begin{eqnarray}
\{\mu_1^g\}&=&(\theta_1^g,\phi_1^g,\psi_1^g),\\
\{\mu_{12}^g\}&=&(\theta_{12}^g,\phi_{12}^g,\psi_{12}^g),\\
\{\mu_{13}^g\}&=&(\theta_{13}^g,\phi_{13}^g,\psi_{13}^g),
\end{eqnarray}
respectively, where $\theta_1^g=\theta_{1,1}^g$,
$\theta_{12}^g=\theta^g_{1,2}$, and
$\theta_{13}^g=\theta^g_{1,3}$.

Since the antisymmetric exchange matrix cannot generally be
diagonalized, instead of writing a different general set of
exchange vectors, we first employ the Moriya rules for the
interactions between each pair of spins in the molecular
representation,\cite{Moriya,BenciniGatteschi} and then impose the
group operations in that representation.  This leads to an
additional set of as many as five parameters for the NN and NNN DM
interactions, which we write as the three- and two-vectors, ${\bm
d}^g$ and ${\bm d}_{\perp}'^{,g}$, respectively.

For the six high-symmetry groups under study, we analyze the
effects of molecular group symmetry upon these coordinates. As
shown in the following, the molecular group symmetry can greatly
simplify the theoretical analysis of the measurable quantities, so
that only the various interaction magnitudes and these five
angular sets are required. Some group symmetries impose severe
restrictions upon the elements of the angular sets.

In the absence of any anisotropy interactions, the bare Hamiltonian
${\cal H}^g_0$ is given by the Zeeman and Heisenberg interactions,
\begin{eqnarray}
{\cal H}_0^g&=&-\gamma {\bm B}\cdot{\bm S}-J_g'({\bm S}_1\cdot{\bm
S}_3+{\bm S}_2\cdot{\bm S}_4)\nonumber\\
& &-J_g({\bm S}_1\cdot{\bm S}_2+{\bm S}_2\cdot{\bm S}_3+{\bm
S}_3\cdot{\bm S}_4+{\bm S}_4\cdot{\bm S}_1),\label{H0bare}
\end{eqnarray}
which can be rewritten as
\begin{eqnarray}
 {\cal
H}_0^g&=&-\frac{J_g}{2}{\bm S}^2-\gamma{\bm B}\cdot{\bm
S}-\frac{(J_g'-J_g)}{2}({\bm S}_{13}^2+{\bm S}_{24}^2),\label{H0}
\end{eqnarray}
where we have dropped the irrelevant constants, and \begin{eqnarray} J_{T_d}'&=&J_{T_d},\label{JTdp}\\
J_{g}'&\ne&J_{g}'\label{Jgp} \end{eqnarray} for
$g=D_{2d},S_4,D_{4h}, C_{4h}$, and $C_{4v}$.  In terms of the
diagonalized matrix elements,
$-2J_g=\tilde{D}^{g,s}_{1,2,xx}+\tilde{D}^{g,s}_{1,2,yy}$ and
$-2J_g'=\tilde{D}^{g,s}_{1,3,xx}+\tilde{D}^{g,s}_{1,3,yy}$, for
instance.

We note that for $c/a<1$, $-J_g$ and $-J_g'$ are the NN and NNN
Heisenberg interactions for $D_{2d},S_4$, as for $C_{4v}, C_{4h}$
and $D_{4h}$ with $c=0$, but for $c/a>1$, $-J_g$ and $-J_g'$ are
the NNN and NN Heisenberg interactions, respectively, for $D_{2d}$
and $S_4$. In Eq. (\ref{H0}), ${\bm S}_{13}={\bm S}_1+{\bm S}_3$,
${\bm S}_{24}={\bm S}_2+{\bm S}_4$, and ${\bm S}={\bm S}_{13}+{\bm
S}_{24}$ is the total spin operator.\cite{KlemmLuban}

\section{III. The single-ion and anisotropic exchange Hamiltonians}

In order to properly take account of the molecular group $g$
symmetries, it is useful to describe the single-ion and symmetric
anisotropic exchange interactions in terms of the {\it local}
coordinates.  We write the local Hamiltonian for these
interactions, and require that it must be  invariant under all
$\lambda$ operations ${\cal O}_{\lambda}^g$ of $g$.  As noted
above, we write the antisymmetric exchange interactions in the
molecular representation, employing the group symmetry
interactions relevant to the particular pair involved in each
exchange, and then impose the overall group symmetry.  In Secs.
IV-IX, we then impose the group symmetries on these interactions
for $C_{4h}, D_{4h}, C_{4v}, S_4, D_{2d}$, and $T_d$ molecular
group symmetries, respectively.

\subsection{A. Local single-ion Hamiltonian}

To do so for the single-ion anisotropy, we first define the vector
bases for each group $g$. We define the local axial vector basis
to be $\{\hat{\tilde{\bm z}}_n^g\}$ for each $g$, and the
orthogonal azimuthal vector bases to be $\{\hat{\tilde{\bm
x}}_n^g\}$ and $\{\hat{\tilde{\bm y}}_n^g\}$.  The elements of
these bases are the vectors that diagonalize the single ion matrix
from $\tensor{\bm D}_{n,n}^g$ to $\tensor{\tilde{\bm
D}}_{n,n}^g$.\cite{Goldstein} Since we employ these vectors
repeatedly, we write them here for simplicity of presentation.
 The diagonalized vectors $\hat{\tilde{\bm x}}_{n}^g$, etc., may be written in the molecular
 $(\hat{\bm x},\hat{\bm y},\hat{\bm z})$ representation as
\begin{eqnarray}
\hat{\tilde{\bm x}}_{n}^g&=&\left(\begin{array}{c}
\cos\phi_n^g\cos\psi_n^g-\cos\theta_n^g\sin\phi_n^g\sin\psi_n^g\\
\cos\psi_n^g\sin\phi_n^g+\cos\theta_n^g\cos\phi_n^g\sin\psi_n^g\\
\sin\theta_n^g\sin\psi_n^g\end{array}\right),\label{xng}\\
\hat{\tilde{\bm y}}_{n}^g&=&\left(\begin{array}{c}
-\cos\phi_n^g\sin\psi_n^g-\cos\theta_n^g\sin\phi_n^g\cos\psi_n^g\\
-\sin\psi_n^g\sin\phi_n^g+\cos\theta_n^g\cos\phi_n^g\cos\psi_n^g\\
\sin\theta_n^g\cos\psi_n^g\end{array}\right),\label{yng}\\
\hat{\tilde{\bm z}}_{n}^g&=&\left(\begin{array}{c}
\sin\theta_n^g\sin\phi_n^g\\
-\sin\theta_n^g\cos\phi_n^g\\
\cos\theta_n^g\end{array}\right),\label{zng}
\end{eqnarray}
which satisfy $\hat{\tilde{\bm x}}^g_n\times\hat{\tilde{\bm
y}}_n^g=\hat{\tilde{\bm z}}_n^g$. We then write the most general
quadratic single-ion anisotropy interaction as
\begin{eqnarray}
{\cal H}^{g,\ell}_{si}&=&-\sum_{n=1}^4\Bigl(J_{a,n}^g({\bm
S}_n\cdot\hat{\tilde{\bm
z}}^g_n)^2\nonumber\\
& &+J_{e,n}^g[({\bm S}_n\cdot\hat{\tilde{\bm x}}^g_n)^2-({\bm
S}_n\cdot\hat{\tilde{\bm y}}^g_n)^2]\Bigr),\label{Hsi}
\end{eqnarray}
in terms of the site-dependent axial and azimuthal interactions
$J_{a,n}^g, J_{e,n}^g$, analogous in notation  to that for
homoionic dimers.\cite{ek2} In terms of the diagonalized matrix
elements,
$-J_{a,n}^g=\tilde{D}^{g,s}_{n,n,zz}-(\tilde{D}^{g,s}_{n,n,xx}+\tilde{D}^{g,s}_{n,n,yy})/2$
and
$-J_{e,n}^g=(\tilde{D}^{g,s}_{n,n,xx}-\tilde{D}^{g,s}_{n,n,yy})/2$.

 We require ${\cal H}^{g,\ell}_{si}$ to be invariant under all
allowed $g$ symmetries.  As we shall see for the six high-symmetry
cases under study, this forces $J_{a,n}^g$ and $J_{e,n}^g$ to be
independent of $n$ for each $g$ in these local coordinates, and
places constraints upon the single-ion parameter set
$\{\theta_n^g,\phi_n^g,\psi_n^g\}$.

\subsection{B. Local symmetric anisotropic exchange Hamiltonian}

In addition to the single-ion interactions, the other microscopic
anisotropic interactions are the anisotropic exchange
interactions, which include the dipole-dipole interactions. We
neglect intermolecular dipole-dipole interactions, which can be
important at very low temperatures.\cite{Marisol} As for the
single-ion interactions, we first construct the symmetric
anisotropic exchange Hamiltonian ${\cal H}_{ae}^g$ in the local
group coordinates.   In this case, there are distinct local vector
sets for the NN and NNN exchange interactions. Diagonalization of
the symmetric anisotropic exchange matrix $\tensor{\bm
D}_{n,n'}^{g,s}$ leads to $\tensor{\tilde{\bm D}}_{n,n'}^{g,s}$
and the vector basis $\{\hat{\tilde{\bm x}}_{n,n'}^g,
\hat{\tilde{\bm y}}_{n,n'}^g, \hat{\tilde{\bm z}}_{n,n'}^g\}$,
given by Eqs. (\ref{xng})-(\ref{zng}) with the subscript $n$
replaced by $n,n'$.

The local symmetric anisotropic exchange Hamiltonian ${\cal
H}_{ae}^{g,\ell}$ is then generally given by
\begin{eqnarray}
 {\cal
H}_{ae}^{g,\ell}&=&-\sum_{m=1}^2\sum_{n=1}^{6-2m}\Bigl[J^{f,g}_{n,n+m}({\bm
S}_n\cdot\hat{\tilde{\bm z}}_{n,n+m}^g)({\bm
S}_{n+m}\cdot\hat{\tilde{\bm
z}}_{n,n+m}^g)\nonumber\\
& &+J^{c,g}_{n,n+m}\Bigl(({\bm S}_n\cdot\hat{\tilde{\bm
x}}_{n,n+m}^g)({\bm
S}_{n+m}\cdot\hat{\tilde{\bm x}}_{n,n+m}^g)\nonumber\\
& &\qquad-({\bm S}_n\cdot\hat{\tilde{\bm y}}_{n,n+m}^g)({\bm
S}_{n+m}\cdot\hat{\tilde{\bm
y}}_{n,n+m}^g)\Bigr)\Bigr],\label{Hae}
\end{eqnarray}
where we define ${\bm S}_5\equiv{\bm S}_1$, as if the four NN
spins were on a ring.  In Eq. (\ref{Hae}), the axial and azimuthal
interaction strengths
$-J_{n,n'}^{f,g}=\tilde{D}^{g,s}_{n,n',zz}-(\tilde{D}^{g,s}_{n,n',xx}+\tilde{D}^{g,s}_{n,n',yy})/2$
and
$-J_{n,n'}^{c,g}=(\tilde{D}^{g,s}_{n,n',xx}-\tilde{D}^{g,s}_{n,n',yy})/2$,
as for the single-ion interaction strengths, but in this case
$n'=n+m$, where $m=1,2$.

Since the structures allow for two distinct bond lengths, with
four NN's and 2 NNN's for $c/a<1$, there are two sets of local
axial and azimuthal coordinates.  The vector sets
$\{\hat{\tilde{\bm z}}_{n,n+1}^g\}$, $\{\hat{\tilde{\bm
x}}_{n,n+1}^g\},$ and $\{\hat{\tilde{\bm y}}_{n,n+1}^g\}$ each
have four elements, and are the bases for the NN axial and
azimuthal symmetric anisotropic exchange vectors. In addition, the
vector sets $\{\hat{\tilde{\bm z}}_{n,n+2}^g\}$,
$\{\hat{\tilde{\bm x}}_{n,n+2}^g\},$ and $\{\hat{\tilde{\bm
y}}_{n,n+2}^g\}$ each have two elements, and are the NNN symmetric
anisotropic exchange vector bases. These are both different from
the single-ion vector bases.

For the six high-symmetry groups under study,  the groups
symmetries force the  interaction strengths $J_{n,n+m}^{f,g}$ and
$J_{n,n+m}^{c,g}$ to be independent of $n$, but are generally
different for NN $(m=1$) and NNN ($m=2$) interactions.  The group
symmetries also place restrictions upon the  NN and NNN symmetric
anisotropic exchange parameter sets
$\{\theta_{n,n+m}^g,\phi_{n+m}^g,\psi_{n,n+m}^g\}$ for $m=1,2$.
These restrictions are much stronger for the NNN parameter set
than for the NN parameter set.

\subsection{C. Antisymmetric anisotropic exchange Hamiltonian}

As noted above, since the antisymmetric anisotropic exchange
matrix with three real distinct matrix elements cannot generally
be diagonalized by a unitary transformation, there is no
particular vector basis resulting from the diagonalization.  We
therefore write the antisymmetric anisotropic exchange, or DM,
Hamiltonian ${\cal H}_{DM}^g$ in the molecular
representation,\cite{BenciniGatteschi}
\begin{eqnarray}
{\cal H}_{DM}^g&=&\sum_{m=1}^2\sum_{n=1}^{6-2m}{\bm
d}^{g}_{n,n+m}\cdot\Bigl({\bm S}_n\times{\bm
S}_{n+m}\Bigr).\label{HDM}
\end{eqnarray}
We note that in these molecular coordinates, the DM interaction
three-vectors ${\bm d}^{g}_{n,n+m}$ depend explicitly upon the
exchange bond indices $n,n+m$ for each group $g$. We then employ
the local group symmetries to relate them to one another.

The rules for the directions of the ${\bm d}_{n,n+m}^g$ were given
by Moriya,\cite{Moriya} and were employed for a dimer example by
Bencini and Gatteschi,\cite{BenciniGatteschi}  The Moriya rules
are: (1) ${\bm d}^g_{n,n'}$ vanishes if a center of inversion
connects ${\bm r}_n$ and ${\bm r}_{n'}$. (2) When a mirror plane
contains ${\bm r}_n$ and ${\bm r}_{n'}$, ${\bm d}^g_{n,n'}$ is
normal to the mirror plane.  (3) When a mirror plane is the
perpendicular bisector of ${\bm r}_n-{\bm r}_{n'}$, ${\bm
d}_{n,n'}^g$ lies in the mirror plane. (4) When a two-fold
rotation axis is the perpendicular bisector of ${\bm r}_n-{\bm
r}_{n'}$, then ${\bm d}_{n,n'}^g$ is orthogonal to ${\bm r}_n-{\bm
r}_{n'}$. (5) When ${\bm r}_n-{\bm r}_{n'}$ is a $p$-fold rotation
axis with $p>2$, then ${\bm d}_{n,n'}^g$ is parallel to ${\bm
r}_n-{\bm r}_{n'}$. As noted above, we shall incorporate these
rules  in the molecular representation.

In the following, we impose the Moriya rules on each anisotropic
exchange pair, and then impose the required group symmetries on
the six pairs. For each $g$, the symmetries relate the NN and NNN
DM interactions to one another, greatly restricting the number of
parameters.  For the six groups under study, the group symmetries
place restrictions upon the ${\bm d}_{n,n+m}^g$, causing them to
be independent of $n$, so that only the two three-vectors ${\bm
d}_{12}^g$ and ${\bm d}_{13}^g$ can describe the full DM
interactions of each tetramer group.  However,   some components
of the DM interaction vectors can have site-dependent signs.

In Sections IV-IX, we evaluate ${\cal H}_{si}^g$, ${\cal
H}_{ae}^g$, and ${\cal H}_{DM}^g$ for $g=C_{4h}, D_{4h}, C_{4v},
S_4, D_{2d}$, and $T_d$, respectively.

\section{IV.  $C_{4h}$ group symmetry}

We first discuss the simplest group symmetries, $C_{4h}$. Besides
the trivial identity operation, the allowed group operations
${\cal O}_{\lambda}^{C_{4h}}$ for $\lambda=1,2,3$ are clockwise
and counterclockwise rotations by $\pi/2$ about the $z$ axis, and
reflections in the $xy$ plane.\cite{Tinkham}  These operations are
represented respectively by the matrices
\begin{eqnarray}
{\cal O}_{1,2}^{C_{4h}}&=&\left(\begin{array}{ccc} 0&\pm1&0\\
\mp1&0&0\\
0&0&1\end{array}\right),\label{O12C4h}\\
{\cal O}_3^{C_{4h}}&=&\left(\begin{array}{ccc} 1&0&0\\
0&1&0\\
0&0&-1\end{array}\right).\label{O3C4h}
\end{eqnarray}

\subsection{$C_{4h}$ single-ion anisotropy}

We require ${\cal O}_{\lambda}^{C_{4h}}{\cal
H}_{si}^{C_{4h}}={\cal H}_{si}^{C_{4h}}$ for $\lambda=1,2,3$.  We
first consider the effects of ${\cal O}_1^{C_{4h}}$.  We note that
${\cal O}_1^{C_{4h}}{\bm r}_n={\bm r}_{n+1}$. Hence, this
operation rotates the crystal by $\pi/4$ about the $z$ axis.
Graphically, this is accomplished by relabelling Fig. 3(a) with
$n\rightarrow n-1$, so that ${\bm S}_1$ is now at ${\bm r}_2$,
etc.  Thus, \begin{eqnarray} {\cal O}_1^{C_{4h}}{\cal
H}_{si}^{C_{4h},\ell}&=&-\sum_{n=1}^4\Bigl(J_{a,n}^{C_{4h}}({\bm
S}_n\cdot\hat{\tilde{\bm
z}}^{C_{4h}}_{n+1})^2\nonumber\\
& &+J_{e,n}^{C_{4h}}[({\bm S}_n\cdot\hat{\tilde{\bm
x}}^{C_{4h}}_{n+1})^2-({\bm S}_n\cdot\hat{\tilde{\bm
y}}^{C_{4h}}_{n+1})^2]\Bigr),\nonumber\\
\end{eqnarray}
where ${\cal O}_1^{C_{4h}}\hat{\tilde{\bm
z}}_n^{C_{4h}}=\hat{\tilde{\bm z}}_{n+1}^{C_{4h}}$, etc. In order
that the axial part of ${\cal H}_{si}^{C_{4h},\ell}$ is invariant
under the transformation, we require
\begin{eqnarray}
J_{a,n}^{C_{4h}}&=&J_{a,n-1}^{C_{4h}}=J_a\label{Jan}
\end{eqnarray}
and \begin{eqnarray}
\sin\theta_n^{C_{4h}}\sin\phi_n^{C_{4h}}&=&-\sin\theta_{n-1}^{C_{4h}}\cos\phi_{n-1}^{C_{4h}},\label{C4hsi1}\\
\sin\theta_n^{C_{4h}}\cos\phi_n^{C_{4h}}&=&\sin\theta_{n-1}^{C_{4h}}\sin\phi_{n-1}^{C_{4h}},\label{C4hsi2}\\
\cos\theta_n^{C_{4h}}&=&\cos\theta_{n-1}^{C_{4h}}.\label{C4hsi3}
\end{eqnarray}
There are two possible solutions for these recursion relations.
One possibility is
\begin{eqnarray}
\theta_n^{C_{4h}}&=&0,\pi,\\
 \phi_n^{C_{4h}}&=&{\rm arbitrary}.
 \end{eqnarray}
 The second possible solution is
 \begin{eqnarray}
 \theta_n^{C_{4h}}&=&\theta_{n-1}^{C_{4h}},\\
 \phi_n^{C_{4h}}&=&\phi_{n-1}^{C_{4h}}-\frac{\pi}{2}.
\end{eqnarray}
However, imposition of the $xy$ mirror plane symmetry, ${\cal
O}_3^{C_{4h}}$, such as setting ${\cal O}_3^{C_{4h}}\hat{\bm
z}^{C_{4h}}_n=\pm\hat{\bm z}^{C_{4h}}_n$ forces either the even
parity $\theta_n^{C_{4h}}=0$ or the odd parity
$\theta_n^{C_{4h}}=1$.
 We
 choose the odd parity solution, setting
\begin{eqnarray}
\theta_n^{C_{4h}}&=&0,\label{thetanC4h}
\end{eqnarray}
 which implies
$\hat{\tilde{z}}_n^{C_{4h}}=\hat{\bm z}$, the axis of high
symmetry.

 Now, we also must also
require the azimuthal part of ${\cal H}_{si}^{C_{4h},\ell}$ to be
invariant under ${\cal O}_1^{C_{4h}}$. Using Eq.
(\ref{thetanC4h}), we see from Eqs. (\ref{xng}) and (\ref{yng})
that $\hat{\tilde{x}}_n^{C_{4h}}=\hat{\bm
y}\sin\chi_n^{C_{4h}}+\hat{\bm x}\cos\chi_n^{C_{4h}}$, etc., where
\begin{eqnarray}
\chi_n^{C_{4h}}&=&\phi_n^{C_{4h}}+\psi_n^{C_{4h}}.
\end{eqnarray}
Then, invariance of ${\cal H}_{si}^{C_{4h},\ell}$ under ${\cal
O}_1^{C_{4h}}$ requires
\begin{eqnarray}
J_{e,n}^{C_{4h}}&=&J_{e,n-1}^{C_{4h}}=J_e\label{Jen}
\end{eqnarray}
and
\begin{eqnarray}
\cos\chi^{C_{4h}}_{n}&=&\sin\chi_{n-1}^{C_{4h}},\\
\sin\chi^{C_{4h}}_{n}&=&-\cos\chi_{n-1}^{C_{4h}},
\end{eqnarray}
or that
\begin{eqnarray}
\chi^{C_{4h}}_{n}&=&\chi^{C_{4h}}_{n-1}-\frac{\pi}{2}.
\end{eqnarray}
We note that for these vector groups, ${\cal
H}_{si}^{C_{4h},\ell}$ is automatically invariant under ${\cal
O}_2^{C_{4}}$ and ${\cal O}_3^{C_{4}}$.  In the latter case,
${\cal O}_3^{C_{4}}\hat{\tilde{\bm z}}_n^{C_{4h}}=-\hat{\tilde{\bm
z}}_n^{C_{4h}}$ has odd parity.

 Since ${\cal O}_1^{C_{4h}}$
interchanges the $x$ and $y$ components of the azimuthal vectors,
 it is useful to define the functions
\begin{eqnarray}
\delta_n^{+}(\phi)&=&\epsilon_n^{+}(-1)^{n/2}\sin\phi+\epsilon_n^{-}(-1)^{(n+1)/2}\cos\phi,\label{deltanp}\\
\delta_n^{-}(\phi)&=&\epsilon_n^{+}(-1)^{n/2}\cos\phi-\epsilon_n^{-}(-1)^{(n+1)/2}\sin\phi.\label{deltanm}
\end{eqnarray}  We note that $\delta_n^{\pm}(\pi/4)=\gamma_n^{\pm}/\sqrt{2}$ and $\delta_n^{\pm}(3\pi/4)=\pm\gamma_n^{\mp}/\sqrt{2}$.
 For clarity, these new functions are listed in Table II.  Some
 useful relations are
 \begin{eqnarray}
\Bigl(\delta_n^{+}(\phi)\Bigr)^2+\Bigl(
\delta_n^{-}(\phi)\Bigr)^2&=&1,\\
\Bigl(\delta_n^{+}(\phi)\Bigr)^2-\Bigl(
\delta_n^{-}(\phi)\Bigr)^2&=&(-1)^n\cos(2\phi),\\
\delta_n^{+}(\phi)\delta_n^{-}(\phi)&=&\frac{(-1)^n}{2}\sin(2\phi).
\end{eqnarray}

 \begin{table}
\centerline{\begin{tabular}{ccc}
\hline $n$&$\delta_n^{+}(\phi)$&$\delta_n^{-}(\phi)$\\
\hline 1&$-\cos\phi$&$\sin\phi$\\
2&$-\sin\phi$&$-\cos\phi$\\
3&$\cos\phi$&$-\sin\phi$\\
4&$\sin\phi$&$\cos\phi$ \end{tabular}} \caption{Values of the
functions $\delta_n^{\pm}(\phi)$.}\label{tableII}
\end{table}

Hence, the single-ion vector groups for $g=C_{4h}$ symmetry are
$\{\hat{\bm x}_n^{C_{4h}}\}$, $\{\hat{\bm y}_n^{C_{4h}}\}$,
$\{\hat{\bm z}_n^{C_{4h}}\}$, where the elements are given by
\begin{eqnarray}
\hat{\tilde{\bm x}}_n^{C_{4h}}&=&
-\delta_n^{+}(\chi_1^{C_{4h}})\hat{\bm
x}+ \delta_n^{-}(\chi_1^{C_{4h}})\hat{\bm y},\label{xnC4h}\\
\hat{\tilde{\bm
y}}_n^{C_{4h}}&=&-\delta_n^{-}(\chi_1^{C_{4h}})\hat{\bm
x}- \delta_n^{+}(\chi_1^{C_{4h}})\hat{\bm y},\label{ynC4h}\\
 \hat{\tilde{\bm
z}}_n^{C_{4h}}&=&\hat{\bm z}.\label{znC4h}
\end{eqnarray}
  The
single-ion parameter set for $g=C_{4h}$ is
\begin{eqnarray}
\{\mu_1^{C_{4h}}\}&=&(0,\chi_1^{C_{4h}}),\\
\chi_1^{C_{4h}}&=&\phi_1^{C_{4h}}+\psi_1^{C_{4h}}.\label{mu1C4h}
\end{eqnarray}

\subsection{$C_{4h}$ symmetric anisotropic exchange}

The construction of the symmetric anisotropic exchange vector
groups is entirely analogous to those of the single-ion
anisotropy.  We find
\begin{eqnarray}
J^{f,C_{4h}}_{n,n+m}&=&J_{f}^{n-1,n+m-1}=J_{f,m},\\
&\equiv&J_{f,m},\\
J^{c,C_{4h}}_{n,n+m}&=&J_{c}^{n-1,n+m-1}=J_{c,m},\\
&\equiv&J_{c,m},
\end{eqnarray}
and the near-neighbor and next-nearest-neighbor vector groups are
\begin{eqnarray}
\hat{\tilde{\bm x}}_{n,n+1}^{C_{4h}}&=&
-\delta_n^{+}(\chi_{12}^{C_{4h}})\hat{\bm
x}+ \delta_n^{-}(\chi_{12}^{C_{4h}})\hat{\bm y},\label{xnp1C4h}\\
\hat{\tilde{\bm
y}}_{n,n+1}^{C_{4h}}&=&-\delta_n^{-}(\chi_{12}^{C_{4h}})\hat{\bm
x}- \delta_n^{+}(\chi_{12}^{C_{4h}})\hat{\bm y},\label{ynp1C4h}\\
 \hat{\tilde{\bm
z}}_{n,n+1}^{C_{4h}}&=&\hat{\bm z},\label{znp1C4h}
\end{eqnarray}
and
\begin{eqnarray}
\hat{\tilde{\bm x}}_{n,n+2}^{C_{4h}}&=&
-\delta_n^{+}(\chi_{13}^{C_{4h}})\hat{\bm
x}+ \delta_n^{-}(\chi_{13}^{C_{4h}})\hat{\bm y},\label{xnnp1C4h}\\
\hat{\tilde{\bm
y}}_{n,n+2}^{C_{4h}}&=&-\delta_n^{-}(\chi_{13}^{C_{4h}})\hat{\bm
x}- \delta_n^{+}(\chi_{13}^{C_{4h}})\hat{\bm y},\label{ynnp2C4h}\\
 \hat{\tilde{\bm
z}}_{n,n+2}^{C_{4h}}&=&\hat{\bm z},\label{znnp2C4h}
\end{eqnarray}
respectively, where
\begin{eqnarray}
\chi_{1p}^{C_{4h}}&\equiv&\chi_{1,p}^{C_{4h}}=\phi_{1p}^{C_{4h}}+\psi_{1p}^{C_{4h}}
\end{eqnarray}
for $p=2,3$.

\subsection{C. $C_{4h}$ antisymmetric exchange interactions}

With antisymmetric exchange, we first impose the Moriya rules for
$C_{4h}$ symmetry.  We first note that ${\cal O}_1^{C_{4h}}{\cal
O}_1^{C_{4h}}=-{\cal O}_3^{C_{4h}}$, so that the origin is a
center of inversion for the next-nearest neighbor bonds with
$g=C_{4h}$.\cite{Tinkham}  This implies
\begin{eqnarray}{\bm
d}_{n,n+2}^{C_{4h}}&=&0.
\end{eqnarray}

Since all four spins lie in the $xy$ mirror plane, ${\cal
O}_3^{C_{4h}}$, we have\begin{eqnarray} {\bm
d}_{n,n+1}^{C_{4h}}&=&d_{n,n+1}^{C_{4h}}\hat{\bm z}.
\end{eqnarray}
Then, imposing the ${\cal O}_1^{C_{4h}}$ rotational symmetry on
${\cal H}_{DM}^{C_{4h}}$, we find
\begin{eqnarray}
d_{n,n+1}^{C_{4h}}&=&d_{n-1,n}^{C_{4h}}\\
&\equiv&d_z
\end{eqnarray}
is a scalar independent of $n$.  Thus, the antisymmetric
anisotropic exchange interaction for $C_{4h}$ symmetry is simply
\begin{eqnarray}
{\cal H}_{DM}^{C_{4h}}&=&d_z\sum_{n=1}^4\hat{\bm z}\cdot({\bm
S}_n\times{\bm S}_{n+1}).\label{HDMC4h}
\end{eqnarray}

\section{V. $D_{4h}$ group symmetry}

We now consider the higher symmetry $g=D_{4h}$.  This group
contains the three symmetries of $C_{4h}$, so that ${\cal
O}_{\lambda}^{D_{4h}}={\cal O}_{\lambda}^{C_{4h}}$ for
$\lambda=1,2,3$.  In addition, there are four two-fold rotations
about axes normal to the principle $z$ axis.  These rotation axes
are the $x$ and $y$ axes and the $y=\pm x$ diagonals.  These
$D_{4h}$ operations are represented by
\begin{eqnarray}
{\cal O}_{4,5}^{D_{4h}}&=&\left(\begin{array}{ccc} \pm1&0&0\\
0&\mp1&0\\
0&0&-1\end{array}\right),\\
{\cal O}_{6,7}^{D_{4h}}&=&\left(\begin{array}{ccc} 0&\pm1&0\\
\pm1&0&0\\
0&0&-1\end{array}\right).
\end{eqnarray}

\subsection{A. $D_{4h}$ single-ion anisotropy}

We begin with the same invariances of ${\cal
H}_{si}^{D_{4h},\ell}$ as for $C_{4h}$ symmetry, leading to the
single-ion vector groups given by Eqs. (\ref{xnC4h})-(\ref{znC4h})
with $C_{4h}\rightarrow D_{4h}$. As for $C_{4h}$ symmetry,
$J_{a,n}^{D_{4h}}=J_{a,n-1}^{D_{4h}}=J_a$ and
$J_{e,n}^{D_{4h}}=J_{e,n-1}^{D_{4h}}=J_e$.  We then impose the
additional symmetries ${\cal O}_{\lambda}^{D_{4h}}$ for
$\lambda=4-7$.  We note that ${\cal O}_6^{D_{4h}}{\bm r}_1={\bm
r}_1$, for example. Hence, we require ${\cal O}_4^{D_{4h}}\hat{\bm
x}_{1,3}=\pm\hat{\bm x}_{2,4}$, ${\cal O}_5^{D_{4h}}\hat{\bm
x}_{1,2}=\pm\hat{\bm x}_{4,3}$, ${\cal O}_6^{D_{4h}}\hat{\bm
x}_{1,3}=\pm\hat{\bm x}_{1,3}$, ${\cal O}_6^{D_{4h}}\hat{\bm
x}_{2,4}=\pm\hat{\bm x}_{4,2}$, and ${\cal O}_7^{D_{4h}}\hat{\bm
x}_{1,3}=\pm\hat{\bm x}_{3,1}$, ${\cal O}_7^{D_{4h}}\hat{\bm
x}_{2,4}=\pm\hat{\bm x}_{2,4}$. The even (odd) parity choices lead
to $\sin\chi_1^{D_{4h}}=\pm\cos\chi_1^{D_{4h}}$, and we make the
even parity choice,
\begin{eqnarray}
\chi_1^{D_{4h}}&=&\pi/4,
\end{eqnarray}
consistent with Fig. 1.  We then obtain the single-ion vector sets
for $D_{4h}$ symmetry,
\begin{eqnarray}
\hat{\bm x}_n^{D_{4h}}&=&\frac{1}{\sqrt{2}}(-\gamma_n^{+}\hat{\bm x}+\gamma_n^{-}\hat{\bm y}),\label{xnD4h}\\
\hat{\bm y}_n^{D_{4h}}&=&-\frac{1}{\sqrt{2}}(\gamma_n^{-}\hat{\bm
x}+\gamma_n^{+}\hat{\bm y}),\label{ynD4h}\\
\hat{\bm z}_n^{D_{4h}}&=&\hat{\bm z},\label{znD4h}\\
\{\mu_1^{D_{4h}}\}&=&(0,\pi/4),\label{mu1D4h}
\end{eqnarray}
where the second entry refers to
$\chi_1^{D_{4h}}=\phi_1^{D_{4h}}+\psi_1^{D_{4h}}=\pi/4$.  We note
that $\hat{\bm x}_n^{D_{4h}}=\hat{\bm r}_n$, as pictured in Fig.
1.

\subsection{B. $D_{4h}$ symmetric anisotropic exchange}

As for the single-ion interactions, we begin with the symmetric
anisotropic exchange vector groups for $C_{4h}$.  As for $C_{4h}$
symmetry, we have
$J^{c,D_{4h}}_{n,n+m}=J^{c,D_{4h}}_{n-1,n+m-1}=J_{c,m}$ and
$J^{f,D_{4h}}_{n,n+m}=J^{f,D_{4h}}_{n-1,n+m-1}=J_{f,m}$.  We then
impose the additional symmetries, setting ${\cal
O}_4\hat{\tilde{\bm x}}_{12}^{D_{4h}}=\hat{\tilde{\bm
x}}_{21}^{D_{4h}}=-\hat{\tilde{\bm x}}_{12}^{D_{4h}}$ and ${\cal
O}_4\hat{\tilde{\bm x}}_{13}^{D_{4h}}=\hat{\tilde{\bm
x}}_{24}^{D_{4h}}$, for examples. Thus, it is elementary to obtain
the restrictions
\begin{eqnarray}
\cos\chi_{12}^{D_{4h}}&=&0,\\
\cos\chi_{13}^{D_{4h}}&=&\sin\chi_{12}^{D_{4h}},
\end{eqnarray}
and we choose the solutions
\begin{eqnarray}
\chi_{12}^{D_{4h}}&=&\frac{\pi}{2},\\
\chi_{13}^{D_{4h}}&=&\frac{\pi}{4}.
\end{eqnarray}
These solutions are consistent with the remaining symmetries
${\cal O}_{\lambda}^{D_{4}}$ for $\lambda=5,6,7$.  Thus, we obtain
the near-neighbor and next-nearest-neighbor $D_{4h}$ symmetric
anisotropic exchange vector groups,
\begin{eqnarray}
\hat{\tilde{\bm
x}}_{n,n+1}^{D_{4h}}&=&-\frac{1}{2}(\gamma_n^{+}+\gamma_n^{-})\hat{\bm
x}+\frac{1}{2}(\gamma_n^{-}-\gamma_n^{+})\hat{\bm y},\label{xnnpaD4h}\\
\hat{\tilde{\bm
y}}_{n,n+1}^{D_{4h}}&=&\frac{1}{2}(\gamma_n^{+}-\gamma_n^{-})\hat{\bm
x}-\frac{1}{2}(\gamma_n^{+}+\gamma_n^{-})\hat{\bm y},\label{ynnp1D4h}\\
\hat{\tilde{\bm y}}_{n,n+1}^{D_{4h}}&=&\hat{\bm
z},\label{znnp1D4h}
\end{eqnarray}
and
\begin{eqnarray}
\hat{\tilde{\bm
x}}_{n,n+2}^{D_{4h}}&=&\frac{1}{\sqrt{2}}(-\gamma_n^{+}\hat{\bm
x}+\gamma_n^{-}\hat{\bm y}),\label{xnnp2D4h}\\
\hat{\tilde{\bm
y}}_{n,n+2}^{D_{4h}}&=&-\frac{1}{\sqrt{2}}(\gamma_n^{-}\hat{\bm
x}+\gamma_n^{+}\hat{\bm y}),\label{ynnp2D4h}\\
\hat{\tilde{\bm z}}_{n,n+2}^{D_{4h}}&=&\hat{\bm
z},\label{znnp2D4h}
\end{eqnarray}
respectively.

\subsection{C. $D_{4h}$ antisymmetric anisotropic exchange}

For $D_{4h}$ symmetry, the origin is a center of inversion for the
next-nearest-neighbor  DM interactions, forcing them to vanish, as
for $C_{4h}$.   Moreover, it is easy to show that tetramers with
$D_{4h}$ symmetry, the near-neighbor exchange given by Eq.
(\ref{HDMC4h}) also in invariant under ${\cal
O}_{\lambda}^{D_{4h}}$.  Hence, the most general antisymmetric
anisotropic exchange Hamiltonian for $D_{4h}$ symmetry is
\begin{eqnarray}
{\cal H}_{DM}^{D_{4h}}&=&d_z\sum_{n=1}^4\hat{\bm z}\cdot({\bm
S}_n\times{\bm S}_{n+1}),\label{HDMD4h}
\end{eqnarray}
precisely the same as for $C_{4h}$ symmetry.

\section{VI.  $C_{4v}$   group symmetry}

For $C_{4v}$, the group operations ${\cal O}_{\lambda}^{C_{4v}}$
for $\lambda=1,\ldots,6$ are clockwise and counterclockwise
rotations by $\pi/2$ about the $z$ axis (equivalent to ${\cal
O}_{1,2}^{C_{4h}}$ in Eq. (\ref{O12C4h})), and reflections in the
$xz$, $yz$, and diagonal mirror planes containing the $z$ axis and
the lines $y=\pm x$.\cite{Tinkham} The additional operations are
represented respectively by the matrices
\begin{eqnarray}
{\cal O}_{3,4}^{C_{4v}}&=&\left(\begin{array}{ccc} \pm1&0&0\\
0&\mp1&0\\
0&0&1\end{array}\right),\label{O34C4v}\\
{\cal O}_{5,6}^{C_{4v}}&=&\left(\begin{array}{ccc} 0&\pm1&0\\
\pm1&0&0\\
0&0&1\end{array}\right),\label{O56C4v}
\end{eqnarray}
where the upper signs refer to the lower subscript numbers.

\subsection{A. $C_{4v}$ single-ion anisotropy}

As for $C_{4h}$ and $D_{4h}$ symmetries, we first impose the
${\cal O}_1^{C_{4v}}$ invariance, the rotation by $\pi/2$ about
the $z$ axis.  This again leads to
$J_{a,n}^{C_{4v}}=J_{a,n-1}^{C_{4v}}=J_a$ and
$J_{e,n}^{C_{4v}}=J_{e,n-1}^{C_{4v}}=J_e$, independent of $n$. As
for the axial single-ion anisotropy with $C_{4h}$ symmetry, this
leads to Eqs. (\ref{C4hsi1})-(\ref{C4hsi3}), with the two choices
for the recursion relations for $\theta_n^{C_{4v}}$ and
$\phi_n^{C_{4v}}$.  However, with $C_{4v}$ symmetry, the $xy$
plane is not a mirror plane. Instead, we impose the $xz$ mirror
plane operation ${\cal O}_3^{C_{4v}}$, and require ${\cal
O}_3^{C_{4v}}\hat{\tilde{\bm z}}_2^{C_{4v}}=\pm\hat{\tilde{\bm
z}}_1^{C_{4v}}$. Choosing the even parity case  leads to the
equation
\begin{eqnarray}
\sin\theta_1^{C_{4v}}\sin\phi_1^{C_{4v}}&=&-\sin\theta_1^{C_{4v}}\cos\phi_1^{C_{4v}},
\end{eqnarray}
which is satisfied by either choosing $\sin\theta_1^{C_{4v}}=0$ or
$\sin\phi_1^{C_{4v}}=-\cos\phi_1^{C_{4v}}$.  The former solution
is appropriate for the higher-symmetry $C_{4h}, D_{4h}$ cases in
which the $xy$ plane is a molecular mirror plane, pictured in the
right panel of Fig. 1, but this is not required for the most
general tetramers exhibiting $C_{4v}$ symmetry, for which the
ligands on opposite sides of the ring may be different. We
therefore choose the more general case, setting
$\phi_1^{C_{4v}}=3\pi/4$, and leaving $\theta_1^{C_{4v}}$
arbitrary.  We then construct $\hat{\tilde{\bm z}}_3^{C_{4v}}$ and
$\hat{\tilde{\bm z}}_4^{C_{4v}}$ from ${\cal
O}_1^{C_{4v}}\hat{\tilde{\bm z}}_2^{C_{4v}}=\hat{\tilde{\bm
z}}_3^{C_{4v}}$ and ${\cal O}_1^{C_{4v}}\hat{\tilde{\bm
z}}_3^{C_{4v}}=\hat{\tilde{\bm z}}_4^{C_{4v}}$. The resulting
vector group $\{\hat{\tilde{\bm z}}_n^{C_{4v}}\}$ is invariant
under the diagonal mirror plane operations ${\cal
O}_{5,6}^{C_{4v}}$, as ${\cal O}_5^{C_{4v}}\hat{\tilde{\bm
z}}_{2,4}^{C_{4v}}=\hat{\tilde{\bm z}}_{4,2}^{C_{4v}}$, ${\cal
O}_5^{C_{4v}}\hat{\tilde{\bm z}}_{1,3}^{C_{4v}}=\hat{\tilde{\bm
z}}_{1,3}^{C_{4v}}$, and ${\cal O}_6^{C_{4v}}\hat{\tilde{\bm
z}}_{2,4}^{C_{4v}}=\hat{\tilde{\bm z}}_{2,4}^{C_{4v}}$, ${\cal
O}_6^{C_{4v}}\hat{\tilde{\bm z}}_{1,3}^{C_{4v}}=\hat{\tilde{\bm
z}}_{3,1}^{C_{4v}}$. Hence, these diagonal mirror planes introduce
no restrictions upon $\theta_1^{C_{4v}}$. We also note that the
$yz$ mirror plane operations ${\cal O}_4\hat{\tilde{\bm
z}}_{1,4}^{C_{4v}}=\hat{\tilde{\bm z}}_{4,1}^{C_{4v}}$ and ${\cal
O}_4\hat{\tilde{\bm z}}_{2,3}^{C_{4v}}=\hat{\tilde{\bm
z}}_{3,2}^{C_{4v}}$.

Similarly, from $\hat{\tilde{\bm x}}_1^{C_{4v}}$ given by Eq.
(\ref{xng}) with $n=1$, $g=C_{4v}$, we construct $\hat{\tilde{\bm
x}}_2^{C_{4v}}={\cal O}_1^{C_{4v}}\hat{\tilde{\bm
x}}_1^{C_{4v}}={\cal O}_2^{C_{4v}}\hat{\tilde{\bm x}}_1^{C_{4v}}$.
After imposing our choice $\phi_1^{C_{4v}}=3\pi/4$, we find that
these two equations are only satisfied if $\cos\psi_1^{C_{4v}}=0$.
We therefore choose $\psi_1^{C_{4v}}=-\pi/2$, so that
$\chi_1^{C_{4v}}=\phi_1^{C_{4v}}+\psi_1^{C_{4v}}=\pi/4$, as for
$D_{4h}$ symmetry.

By successive applications of ${\cal O}_1^{C_{4v}}$ and ${\cal
O}_3^{C_{4v}}$, we obtain the three vector groups
$\{\hat{\tilde{\bf x}}_n^{C_{4v}}\}$, $\{\hat{\tilde{\bf
y}}_n^{C_{4v}}\}$ and $\{\hat{\tilde{\bf z}}_n^{C_{4v}}\}$, with
the elements
\begin{eqnarray}
\hat{\tilde{\bm
x}}_n^{C_{4v}}&=&\frac{1}{\sqrt{2}}\left(\begin{array}{c}
-\gamma_n^{+}\cos\theta_1^{C_{4v}}\\
\gamma_n^{-}\cos\theta_1^{C_{4v}}\\
-\sqrt{2}\sin\theta_1^{C_{4v}}\end{array}\right),\label{xnC4v}\\
\hat{\tilde{\bm
y}}_n^{C_{4v}}&=&-\frac{1}{\sqrt{2}}\left(\begin{array}{c}
\gamma_n^{-}\\
\gamma_n^{+}\\
0\end{array}\right),\label{ynC4v}\\
\hat{\tilde{\bm
z}}_n^{C_{4v}}&=&\frac{1}{\sqrt{2}}\left(\begin{array}{c}
-\gamma_n^{+}\sin\theta_1^{C_{4v}}\\
\gamma_n^{-}\sin\theta_1^{C_{4v}}\\
\sqrt{2}\cos\theta_1^{C_{4v}}\end{array}\right),\label{znC4v}
\end{eqnarray}
where the $\gamma_m^{\pm}$ are given in Eq. (\ref{gamman}) and
Table I. These forms are also consistent with the $yz$ and
diagonal mirror plane operations (${\cal O}_{\lambda}^{C_{4v}}$
operations for $\lambda=4,5,6$), and leave Eq. (\ref{Hsi})
invariant.\cite{Tinkham} We note that some of the reflections of
some of the $\hat{\tilde{\bm y}}_n^{C_{4v}}$ have odd parity.  We
also note that these vectors reduce to those for $D_{4h}$
symmetry, Eqs. (\ref{xnD4h})-(\ref{znD4h}) in the
$\theta_1^{C_{4v}}\rightarrow0$ limit.  We reiterate that the
$C_{4v}$ group symmetry imposes restrictions upon
$\phi_1^{C_{4v}}$ and $\psi_1^{C_{4v}}$, leading to
\begin{eqnarray}
\{\mu_1^{C_{4v}}\}&=&(\theta_1^{C_{4v}},3\pi/4,-\pi/2).\label{mu1C4v}
\end{eqnarray}

\subsection{B. $C_{4v}$ symmetric anisotropic exchange}

As for $C_{4h}$ and $D_{4h}$ symmetry, we first impose the
invariance of ${\cal H}_{ae}^{C_{4v}}$ under $\pi/2$ rotations
about the $z$ axis, ${\cal O}_1^{C_{4v}}$.  For the axial part of
${\cal H}_{ae}^{C_{4v}}$, this leads to
\begin{eqnarray}
J_{n,n+m}^{f,C_{4v}}&=&J_{n-1,n+m-1}^{f,C_{4v}}=J_{f,m},\\
J_{n,n+m}^{c,C_{4v}}&=&J_{n-1,m+m-1}^{c,C_{4v}}=J_{c,m},
\end{eqnarray}
plus restrictions upon $\theta_{n,n+m}^{C_{4v}}$,
$\phi_{n,n+m}^{C_{4v}}$, and $\psi_{n,n+m}^{C_{4v}}$, precisely as
for $C_{4h}$ and $D_{4h}$ symmetries.

Now, we impose the mirror plane restrictions of $C_{4v}$.  The
diagonal mirror plane containing the $z$ axis and the line $y=-x$,
${\cal O}_6^{C_{4v}}$ requires ${\cal O}_6^{C_{4v}}\hat{\tilde{\bm
z}}_{12}^{C_{4v}}=\pm\hat{\tilde{\bm
z}}_{32}^{C_{4v}}=\mp\hat{\tilde{\bm z}}_{23}^{C_{4v}}$. By
analogy with $C_{4h}$ and $D_{4h}$, we choose
$\theta_{12}^{C_{4v}}=0$, leaving $\phi_{12}^{C_{4v}}$
unrestricted.  This leads to $\hat{\tilde{\bm
x}}_{12}^{C_{4v}}=\cos\chi_{12}^{C_{4v}}\hat{\bm
x}+\sin\chi_{12}^{C_{4v}}\hat{\bm y}$, where
$\chi_{12}^{C_{4v}}=\phi_{12}^{C_{4v}}+\psi_{12}^{C_{4v}}$.
Imposing the $xz$ mirror plane restriction, ${\cal
O}_3^{C_{4v}}\hat{\tilde{\bm x}}_{12}^{C_{4v}}=\pm\hat{\tilde{\bm
x}}_{12}^{C_{4v}}$ leads to either $\cos\chi_{12}^{C_{4v}}=0$ or
$\sin\chi_{12}^{C_{4v}}=0$.  We then choose
$\chi_{12}^{C_{4v}}=0$. A similar procedure is carried out
separately for the next-nearest-neighbor interactions.

After checking all of the mirror plane operations of $C_{4v}$
symmetry, we find that the $C_{4v}$ symmetric anisotropic exchange
vector group elements may be written as
\begin{eqnarray}
\hat{\tilde{\bm
x}}_{n,n+1}^{C_{4v}}&=&-\frac{1}{2}(\gamma_n^{+}-\gamma_n^{-})\hat{\bm
x}+\frac{1}{2}(\gamma_n^{+}+\gamma_n^{-})\hat{\bm y},\label{xnnp1C4v}\\
\hat{\tilde{\bm
y}}_{n,n+1}^{C_{4v}}&=&\frac{1}{2}(\gamma_n^{+}+\gamma_n^{-})\hat{\bm
x}+\frac{1}{2}(\gamma_n^{+}-\gamma_n^{-})\hat{\bm y},\label{ynnp1C4v}\\
\hat{\tilde{\bm z}}_{n,n+1}^{C_{4v}}&=&\hat{\bm
z},\label{znnp1C4v}
\end{eqnarray}
and the matrix diagonalization parameters satisfy
\begin{eqnarray}
\{\mu_{12}^{C_{4v}}\}&=&(\theta_{12}^{C_{4v}},\chi_{12}^{C_{4v}})=(0,0).\label{mu12C4v}
\end{eqnarray}
The next-nearest-neighbor vector group elements can similarly be
shown to satisfy
\begin{eqnarray}
\hat{\tilde{\bm
x}}_{n,n+2}^{C_{4v}}&=&\frac{1}{\sqrt{2}}(-\gamma_n^{+}\hat{\bm
x}+\gamma_n^{-})\hat{\bm y}),\label{xnnp2C4v}\\
\hat{\tilde{\bm
y}}_{n,n+2}^{C_{4v}}&=&-\frac{1}{\sqrt{2}}(\gamma_n^{-}\hat{\bm
x}+\gamma_n^{+}\hat{\bm y}),\label{ynnp2C4v}\\
\hat{\tilde{\bm z}}_{n,n+2}^{C_{4v}}&=&\hat{\bm
z},\label{znnp2C4v}
\end{eqnarray}
and the matrix diagonalization parameters satisfy
\begin{eqnarray}
\{\mu_{13}^{C_{4v}}\}&=&(\theta_{13}^{C_{4v}},\chi_{13}^{C_{4v}})=(0,\frac{\pi}{4}).\label{mu13C4v}
\end{eqnarray}

\subsection{C. $C_{4v}$ antisymmetric exchange}

We first consider the next-nearest-neighbor antisymmetric exchange
interactions, ${\bm d}_{13}^{C_{4v}}\cdot({\bm S}_1\times{\bm
S}_3)+{\bm d}_{24}^{C_{4v}}\cdot({\bm S}_2\times{\bm S}_4)$. The
relevant mirror planes are the diagonal mirror planes containing
the $z$ axis and the lines $y=\pm x$, ${\cal O}_{5,6}^{C_{4v}}$.
The Moriya rules require that ${\bm d}_{13}^{C_{4v}}$ and ${\bm
d}_{24}^{C_{4v}}$ are perpendicular to the respective mirror
planes containing the interacting spin sites.  Thus, we write
${\bm d}_{13}^{C_{4v}}=d_{13}(\hat{\bm y}-\hat{\bm x})/\sqrt{2}$
and ${\bm d}_{24}^{C_{4v}}=d_{24}(\hat{\bm x}+\hat{\bm
y})/\sqrt{2}$.  Then, we impose the invariance under $\pi/2$
rotations about the $z$ axis, ${\cal O}_1^{C_{4v}}$.  This
operation satisfies ${\cal O}_1^{C_{4v}}{\bm S}_n={\bm S}_{n+1}$
and the coordinates transformations ${\cal O}_1^{C_{4v}}\hat{\bm
x}=-\hat{\bm y}$ and ${\cal O}_1^{C_{4v}}\hat{\bm y}=\hat{\bm x}$.
This symmetry then forces $d_{13}=d_{24}$.

For the next-nearest-neighbor antisymmetric exchange interactions,
the relevant mirror planes are the $xz$ and $yz$ mirror planes,
which bisect the respective near-neighbor interaction sites. The
Moriya rules dictate that  ${\bm d}_{12}^{C_{4v}}$ and ${\bm
d}_{34}^{C_{4v}}$ lie in the $xz$ plane, and that ${\bm
d}_{23}^{C_{4v}}$ and ${\bm d}_{41}^{C_{4v}}$ lie in the $yz$
plane. We therefore write ${\bm d}_{12}^{C_{4v}}=d_{12x}\hat{\bm
x}+d_{12z}\hat{\bm z}$, ${\bm d}_{23}^{C_{4v}}=d_{23y}\hat{\bm
y}+d_{23z}\hat{\bm z}$, ${\bm d}_{34}^{C_{4v}}=d_{34x}\hat{\bm
x}+d_{34z}\hat{\bm z}$, and ${\bm d}_{41}^{C_{4v}}=d_{41y}\hat{\bm
y}+d_{41z}\hat{\bm z}$.  We then impose the cyclic rotation
symmetry ${\cal O}_1^{C_{4v}}$, as for the next-nearest-neighbor
interactions.  Invariance of the Hamiltonian under this symmetry
then requires $d_{12x}=-d_{23y}=-d_{34x}=d_{41y}$ and
$d_{12z}=d_{23z}=d_{34z}=d_{41z}$.

Thus, the combined antisymmetric exchange interaction for
tetramers with $C_{4v}$ symmetry may be written as
\begin{eqnarray}
{\cal H}_{DM}^{C_{4v}}&=&\sum_{n=1}^4\Bigl(d_z\hat{\bm
z}+\frac{d_{\perp}}{2}[(\gamma_n^{-}-\gamma_n^{+})\hat{\bm
x}+(\gamma_n^{+}+\gamma_n^{-})\hat{\bm y})]\Bigr)\nonumber\\
& &\cdot({\bm S}_n\times{\bm S}_{n+1})\nonumber\\
& &+d'\sum_{n=1}^2\Bigl(\hat{\bm y}+(-1)^n\hat{\bm
x}\Bigr)\cdot({\bm S}_n\times{\bm S}_{n+2}),\label{HDMC4v}
\end{eqnarray}
where the three scalar parameters are $d'=d_{13}\sqrt{2}$,
$d_z=d_{12z}$, and $d_{\perp}=d_{12x}$.  Thus, for $C_{4v}$
symmetry, there are both site-independent and site-dependent DM
interactions.  The site-independent DM interactions are the same
as for $C_{4h}$ and $D_{4h}$ symmetry.

\section{VII. $S_4$  group symmetry}

   For $S_4$, the group operations are clockwise and counterclockwise
   operation rotations by $\pi/4$ followed by a reflection in the
   $xy$ plane.\cite{Tinkham}  The structure is pictured in Fig. 2(b). These operations
are represented by the matrices
\begin{eqnarray}
{\cal O}_{1,2}^{S_4}&=&\left(\begin{array}{ccc}0&\pm1&0\\
\mp1&0&0\\
0&0&-1\end{array}\right).\label{O1S4}
\end{eqnarray}

\subsection{A. $S_4$ single-ion anisotropy}

We then construct the vector groups from the site 1 vectors by
successive operations of ${\cal O}_1^{S_4}$, setting
$\hat{\tilde{\bm v}}_{n+1}^{S_4}={\cal O}_1^{S_4}\hat{\tilde{\bm
v}}_n^{S_4}$, where $\hat{\tilde{\bm v}}_n^{S_4}=\hat{\tilde{\bm
x}}_n^{S_4}, \hat{\tilde{\bm y}}_n^{S_4}$ and $\hat{\tilde{\bm
z}}_n^{S_4}$. The inverse operations of ${\cal O}_2^{S_4}$ are
automatically satisfied. With $S_4$ symmetry, there are no mirror
planes to satisfy. However, the improper rotation ${\cal
O}_1^{S_4}$ interchanges the $x$ and $y$ vector components,
complicating the most general vector forms. However, we still
obtain $J_{a,n}^{S_4}=J_{a,n-1}^{S_4}=J_a$ and
$J_{e,n}^{S_4}=J_{e,n-1}^{S_4}=J_e$, both independent of $n$, as
for the three planar symmetries discussed previously.

Hence, the group vectors for $S_4$ symmetry contain the elements
\begin{eqnarray}
\hat{\tilde{\bm x}}_n^{S_4}&=&\left(\begin{array}{c}
-\cos\psi_1^{S_4}\delta_n^{+}(\phi_1^{S_4})-\cos\theta_1^{S_4}\sin\psi_1^{S_4}\delta_n^{-}(\phi_1^{S_4})\\
\cos\psi_1^{S_4}\delta_n^{-}(\phi_1^{S_4})-\cos\theta_1^{S_4}\sin\psi_1^{S_4}\delta_n^{+}(\phi_1^{S_4})\\
(-1)^{n+1}\sin\psi_1^{S_4}\sin\theta_1^{S_4}\end{array} \right),\label{xnS4}\nonumber\\
& &\\
 \hat{\tilde{\bm y}}_n^{S_4}&=&\left(\begin{array}{c}
\sin\psi_1^{S_4}\delta_n^{+}(\phi_1^{S_4})-\cos\theta_1^{S_4}\cos\psi_1^{S_4}\delta_n^{-}(\phi_1^{S_4})\\
-\sin\psi_1^{S_4}\delta_n^{-}(\phi_1^{S_4})-\cos\theta_1^{S_4}\cos\psi_1^{S_4}\delta_n^{+}(\phi_1^{S_4})\\
(-1)^{n+1}\cos\psi_1^{S_4}\sin\theta_1^{S_4}\end{array} \right),\label{ynS4}\nonumber\\
& &\\
 \hat{\tilde{\bm z}}_n^{S_4}&=&\left(\begin{array}{c}
\sin\theta_1^{S_4}\delta_n^{-}(\phi_1^{S_4})\\
\sin\theta_1^{S_4}\delta_n^{+}(\phi_1^{S_4})\\
(-1)^{n+1}\cos\theta_1^{S_4}\end{array} \right),\label{znS4}
\end{eqnarray}
where
\begin{eqnarray}\{\mu_1^{S_4}\}&=&(\theta_1^{S_4},\phi_1^{S_4},\psi_1^{S_4})\label{mu1S4}
\end{eqnarray}
 is unrestricted by the group symmetry, and $\delta_n^{\pm}(\phi)$
 are given by Eqs. (\ref{deltanp}) and (\ref{deltanm}).

 \subsection{B. $S_4$ symmetric anisotropic exchange}

We now consider the most general symmetric anisotropic exchange
vector groups for $S_4$ symmetry.  As for $C_{4v}$ we begin with
the near-neighbor axial vector $\hat{\tilde{\bm z}}_{12}^{S_4}$
using the general Eq. (\ref{zng}) with
$\theta_1^{C_{4v}}\rightarrow\theta_{12}^{S_4}$ and
$\phi_1^{C_{4v}}\rightarrow\phi_{12}^{S_4}$.  We construct the
near-neighbor  axial symmetric anisotropic exchange vector group
by successive operations of ${\cal O}_1^{S_4}$, setting
$\hat{\tilde{\bm z}}_{23}^{S_4}={\cal O}_1^{S_4}\hat{\tilde{\bm
z}}_{12}^{S_4}$, etc. The same procedure can be applied to
construct the azimuthal symmetric anisotropic exchange vector
group. The group symmetry imposes no restrictions upon the three
near-neighbor bond angles $\theta_{12}^{S_4}$, $\phi_{12}^{S_4}$,
and $\psi_{12}^{S_4}$, because the fourth operation, ${\cal
O}_1^{S_4}\hat{\tilde{\bm v}}_{41}^{S_4}=\hat{\tilde{\bm
v}}_{12}^{S_4}$ is automatically satisfied for $\hat{\tilde{\bm
v}}^{S_4}=\hat{\tilde{\bm x}}^{S_4},\hat{\tilde{\bm y}}^{S_4}$,
and $\hat{\tilde{\bm z}}^{S_4}$. Thus, the near-neighbor
anisotropic exchange vector groups with $S_4$ symmetry are fully
general, with $\hat{\tilde{\bm x}}_{n,n+1}^{S_4}$,
$\hat{\tilde{\bm y}}_{n,n+1}^{S_4}$, and $\hat{\tilde{\bm
z}}_{n,n+1}^{S_4}$ obtained from the single-ion $\hat{\tilde{\bm
x}}_{n}^{S_4}$, $\hat{\tilde{\bm y}}_{n}^{S_4}$, and
$\hat{\tilde{\bm z}}_{n}^{S_4}$ vectors in Eqs.
(\ref{xnS4})-(\ref{znS4}) by replacing
$\theta_1^{S_4}\rightarrow\theta_{12}^{S_4}$,
$\phi_1^{S_4}\rightarrow\phi_{12}^{S_4}$, and
$\psi_1^{S_4}\rightarrow\psi_{12}^{S_4}$, respectively.  Thus,
\begin{eqnarray}
\{\mu_{12}^{S_4}\}&=&(\theta_{12}^{S_4},\phi_{12}^{S_4},\psi_{12}^{S_4})\label{mu12S4}
\end{eqnarray}
is unrestricted by the group symmetry.

 However, the next-nearest-neighbor symmetric anisotropic exchange vector
groups are different.  We first set $\hat{\tilde{\bm
z}}_{13}^{S_4}$ equal to Eq. (\ref{zng}) with
$\theta_1^{C_{4v}}\rightarrow\theta_{13}^{S_4}$ and
$\phi_1^{C_{4v}}\rightarrow\phi_{13}^{S_4}$, and construct
$\hat{\tilde{\bm z}}_{24}^{S_4}$ from ${\cal
O}_1^{S_4}\hat{\tilde{\bm z}}_{13}^{S_4}=\hat{\tilde{\bm
z}}_{24}^{S_4}$. But we then require ${\cal
O}_1^{S_4}\hat{\tilde{\bm z}}_{24}=\hat{\bm
z}_{31}=-\hat{\tilde{\bm z}}_{13}$. This forces
$\cos\theta_{13}^{S_4}=0$, and we take $\theta_{13}^{S_4}=\pi/2$.
However, $\phi_{13}^{S_4}$ is unrestricted by the group symmetry.

 Performing the same construction of the
azimuthal symmetric next-nearest-neighbor exchange vector groups,
there is a restriction upon $\psi_{13}^{S_4}$, which can be
satisfied by choosing either $\sin\psi_{13}^{S_4}=0$ or
$\cos\psi_{13}^{S_4}=0$. We choose $\sin\psi_{13}^{S_4}=0$. We
thus obtain
\begin{eqnarray}
\hat{\tilde{\bm
x}}_{n,n+2}^{S_4}&=&-\delta_n^{+}(\phi_{13}^{S_4})\hat{\bm
x}+\delta_n^{-}(\phi_{13}^{S_4})\hat{\bm y},\label{xnnp2S4}\\
\hat{\tilde{\bm y}}_{n,n+2}^{S_4}&=&(-1)^{n+1}\hat{\bm z},\label{ynnp2S4}\\
\hat{\tilde{\bm
z}}_{n,n+2}^{S_4}&=&\delta_n^{-}(\phi_{13}^{S_4})\hat{\bm
x}+\delta_n^{+}(\phi_{13}^{S_4})\hat{\bm y},\label{znnp2S4}
\end{eqnarray}
and  \begin{eqnarray}
\{\mu_{13}^{S_4}\}&=&(\pi/2,\phi_{13}^{S_4},0).\label{mu13S4}
\end{eqnarray}

We note that the axial symmetric  anisotropic exchange vectors
are not necessarily collinear with the bond directions. If all of
these anisotropic exchange interactions arose solely from
dipole-dipole interactions, the axial anisotropic exchange vectors
would lie along the bond directions,  and the azimuthal
anisotropic exchange vectors would be irrelevant. This would imply
$\phi_{13}^{S_4}=3\pi/4$ for the next-nearest-neighbor
interactions, and $\theta_{12}^{S_4}=c/a$ and
$\phi_{12}^{S_4}=\pi$ for the near-neighbor interactions. However,
the group symmetry does not require this, and if some amount of
physical anisotropic exchange (not arising from dipolar
interactions) were also present,  more general
$\phi_{13}^{S_4},\theta_{12}^{S_4}$ and $\phi_{12}^{S_4}$ would be
allowed, and the azimuthal anisotropic exchange vectors would be
relevant, with general $\psi_{13}^{S_4}$ and $\psi_{12}^{S_4}$,
respectively.

\subsection{C. $S_4$ antisymmetric exchange}

We begin with the next-nearest-neighbor antisymmetric exchange
Hamiltonian, which may generally be written as ${\bm
d}_{13}^{S_4}\cdot({\bm S}_1\times{\bm S}_3)+{\bm
d}_{24}^{S_4}\cdot({\bm S}_2\times{\bm S}_4)$, where nominally
${\bm d}_{13}^{S_4}$ and ${\bm d}_{24}^{S_4}$ are three-vectors.
Using the relations ${\cal O}_1^{S_4}{\bm S}_n={\bm S}_{n+1}$,
${\cal O}_1^{S_4}\hat{\bm x}=-\hat{\bm y}$, ${\cal
O}_1^{S_4}\hat{\bm y}=\hat{\bm x}$, and ${\cal O}_1^{S_4}\hat{\bm
z}=-\hat{\bm z}$, we find that the components satisfy
$d_{13z}^{S_4}=d_{24z}^{S_4}=0$, $d_{13y}^{S_4}=d_{24x}^{S_4}$,
and $d_{13x}^{S_4}=-d_{24y}^{S_4}$. There are no mirror planes to
satisfy. There are therefore two distinct parameters for the
next-nearest-neighbor antisymmetric exchange.

For the near-neighbor antisymmetric exchange, we nominally set
${\bm d}_{n,n+1}^{S_4}$ to be general threee-vectors. Then,
requiring the near-neighbor antisymmetric exchange Hamiltonian to
be invariant under the improper rotation ${\cal O}_1^{S_4}$, we
then find that the component satisfy
$d_{12x}^{S_4}=-d_{23y}^{S_4}=-d_{34x}^{S_4}=d_{41y}^{S_4}$,
$d_{12y}^{S_4}=d_{23x}^{S_4}=-d_{34y}^{S_4}=-d_{41x}^{S_4}$, and
$d_{12z}^{S_4}=-d_{23z}^{S_4}=d_{34z}^{S_4}=-d_{41z}^{S_4}$. Thus,
three parameters describe the near-neighbor antisymmetric exchange
for $S_4$ symmetry.

Combining the near-neighbor and next-nearest-neighbor
antisymmetric exchange interactions, we find
\begin{eqnarray}
{\cal H}_{DM}^{S_4}&=&d_z\hat{\bm
z}\cdot\Bigl(\sum_{n=1}^4(-1)^{n+1}{\bm S}_n\times{\bm
S}_{n+1}\Bigr)\nonumber\\
& &+{\bm d}_{\perp}\cdot\Bigl({\bm S}_1\times{\bm S}_2-{\bm
S}_3\times{\bm S}_4\Bigr)\nonumber\\
& &+({\bm d}_{\perp}\times\hat{\bm z})\cdot\Bigl({\bm
S}_2\times{\bm S}_3-{\bm
S}_4\times{\bm S}_1\Bigr)\nonumber\\
& &+{\bm d}_{\perp}'\cdot({\bm S}_1\times{\bm S}_3)+({\bm
d}_{\perp}'\times\hat{\bm z})\cdot({\bm S}_2\times{\bm
S}_4),\label{HDMS4}\nonumber\\
\end{eqnarray}
where $d_z$ is a scalar, and ${\bm d}_{\perp}$ and ${\bm
d}_{\perp}'$ are two-vectors in the $xy$ plane.

Thus, there are only site-dependent and site-independent DM
interactions for $S_4$ symmetry.  The NN DM interactions average
to zero, but the NNN DM interactions generally do not.

\section{VIII. $D_{2d}$ group symmetry}
We now consider the interesting case of $D_{2d}$ symmetry pictured
in Fig. 2. In this case, the operations ${\cal
O}_{\lambda}^{D_{2d}}$ $\lambda=1,\ldots,9$ are rotations by $\pi$
about the $x$, $y$, and $z$ axes ($\lambda=1,2,3$, respectively),
and the two  diagonal mirror planes associated with the principal
rotation axis, $z$.\cite{Tinkham} The first of these is
\begin{eqnarray}
{\cal O}_1^{D_{2d}}&=&\left(\begin{array}{ccc} 1&0&0\\
0&-1&0\\
0&0&-1\end{array}\right),\label{O1D2d}
\end{eqnarray}
and the $\lambda=2,3$ cases have the single $+1$ in the $yy$ and
$zz$ positions, respectively. The two mirror planes associated
with the principal $z$ axis rotation are the planes containing the
$z$ axis and the lines $y=\pm x$, respectively, ${\cal
O}_{4,5}^{D_{2d}}={\cal O}_{5,6}^{C_{4v}}$, given by Eq.
(\ref{O56C4v}).  We note, for example, that the plane bisecting
the sites 1,2 and passing through the sites 3,4 is not a mirror
plane unless $c=a$, which we assume not to be true for $D_{3d}$
symmetry.

We first try to generate the most general axial vector set
satisfying $D_{2d}$ symmetry.  Taking $\hat{\tilde{\bm
z}}_1^{D_{2d}}$ to have the general form of Eq. (\ref{zng}), we
then require it to be invariant under ${\cal O}_4^{D_{2d}}$, which
leaves $\hat{\bm r}_1$ invariant.  It is then easy to see that we
must have $\phi_1^{D_{2d}}=3\pi/4$, but $\theta_1^{D_{2d}}$ is
unrestricted. We then generate the remaining $\hat{\tilde{\bm
z}}_n^{D_{2d}}$ from ${\cal O}_1^{D_{2d}}\hat{\tilde{\bm
z}}_1=\hat{\tilde{\bm z}}_2$, ${\cal O}_2^{D_{2d}}\hat{\tilde{\bm
z}}_1=\hat{\tilde{\bm z}}_4$, and ${\cal
O}_3^{D_{2d}}\hat{\tilde{\bm z}}_1=\hat{\tilde{\bm z}}_3$.  This
is consistent with the invariance of the axial part of the
single-ion Hamiltonian, and also leads to the results that
$J_{a,n}^{D_{2d}}=J_a$, independent of $n$.

We then construct the azimuthal single-ion vectors, beginning with
$\hat{\tilde{\bm x}}_1^{D_{2d}}$ from Eq. (\ref{xng}), using our
result that $\theta_1^{D_{2d}}$ is arbitrary and
$\phi_1^{D_{2d}}=3\pi/4$.   Requiring the invariance of the
azimuthal part of the single-ion Hamiltonian under the mirror
plane represented by ${\cal O}_4^{D_{2d}}$, we find that
$\cos\psi_1^{D_{2d}}=0$, and we choose the solution
$\psi_1^{D_{2d}}=-\pi/2$, so that
$\phi_1^{D_{2d}}+\psi_1^{D_{2d}}=\pi/4$.  Then, we generate the
remaining $\hat{\tilde{\bm x}}_n^{D_{2d}}$ from rotations of
$\hat{\tilde{\bm x}}_1^{D_{2d}}$ about the $x$, $y$, and $z$ axes.
This of course leads to $J_{e,n}^{D_{2d}}=J_e$, independent of
$n$.

We then find that the single-ion vector group elements for
$D_{2d}$ symmetry are
\begin{eqnarray}
\hat{\tilde{\bm
x}}_n^{D_{2d}}&=&\frac{1}{\sqrt{2}}\left(\begin{array}{c}-\gamma_n^{+}\cos\theta_1^{D_{2d}}\\
\gamma_n^{-}\cos\theta_1^{D_{2d}}\\
\sqrt{2}(-1)^n\sin\theta_1^{D_{2d}}\end{array}\right),\label{xnD2d}\\
\hat{\tilde{\bm
y}}_n^{D_{2d}}&=&\frac{(-1)^n}{\sqrt{2}}\left(\begin{array}{c}\gamma_n^{-}\\
\gamma_n^{+}\\
0\end{array}\right),\label{ynD2d}\\
\hat{\tilde{\bm
z}}_n^{D_{2d}}&=&\frac{1}{\sqrt{2}}\left(\begin{array}{c}-\gamma_n^{+}\sin\theta_1^{D_{2d}}\\
\gamma_n^{-}\sin\theta_1^{D_{2d}}\\
\sqrt{2}(-1)^{n+1}\cos\theta_1^{D_{2d}}\end{array}\right),\label{znD2d}
\end{eqnarray}
and the associated single-ion parameters are
\begin{eqnarray}
\{\mu_1^{D_{2d}}\}&=&(\theta_1^{D_{2d}},\frac{3\pi}{4},-\frac{\pi}{2}).\label{mu1D2d}
\end{eqnarray}

 \subsection{B. $D_{2d}$ symmetric anisotropic exchange}

 We first consider the near-neighbor axial exchange vector between
 sites 1 and 2, writing $\hat{\tilde{\bm z}}_{12}$ in the form of
 Eq. (\ref{zng}).  Since we require it to be either odd or even
 under rotations about the $x$ axis, ${\cal O}_1^{D_{2d}}\hat{\tilde{\bm
 z}}_{12}=\mp\hat{\tilde{\bm z}}_{12}$, we then find that
 $\sin\phi_{12}^{D_{2d}}=0$, and we choose $\phi_{12}^{D_{2d}}=0$ with $\theta_{12}^{D_{2d}}$ arbitrary.
 We then employ ${\cal O}_3^{D_{2d}}\hat{\tilde{\bm
 z}}_{12}=\hat{\tilde{\bm z}}_{34}$, ${\cal O}_2^{D_{2d}}\hat{\tilde{\bm
 z}}_{12}=-\hat{\tilde{\bm z}}_{34}$, and ${\cal O}_4^{D_{2d}}\hat{\tilde{\bm
 z}}_{12}=-\hat{\tilde{\bm z}}_{41}$ to generate the axial
 near-neighbor symmetric anisotropic exchange vector group.

 We then consider the azimuthal near-neighbor symmetric
 anisotropic exchange vector $\hat{\tilde{\bm x}}_{12}^{D_{2d}}$,
using Eq. (\ref{xng}). Demanding invariance of the exchange
Hamiltonian under ${\cal O}_1^{D_{2d}}$, or that ${\cal
O}_1^{D_{2d}}\hat{\tilde{\bm x}}_{12}^{D_{2d}}=\mp\hat{\tilde{\bm
x}}_{12}^{D_{2d}}$, we then require $\cos\psi_{12}^{D_{2d}}=0$,
and choose $\psi_{12}^{D_{2d}}=\pi/2$.  We then generate the
remainder of the azimuthal near-neighbor symmetric anisotropic
exchange vector group as above.  We thus obtain
\begin{eqnarray}
\hat{\tilde{\bm
x}}_{n,n+1}^{D_{2d}}&=&-\frac{1}{2}\left(\begin{array}{c}
(\gamma_n^{+}+\gamma_n^{-})\cos\theta_{12}^{D_{2d}}\\
(\gamma_n^{+}-\gamma_n^{-})\cos\theta_{12}^{D_{2d}}\\
(-1)^n\sin\theta_{12}^{D_{2d}}\end{array}\right),\label{xnnp1D2d}\\
\hat{\tilde{\bm
y}}_{n,n+1}^{D_{2d}}&=&\frac{(-1)^n}{2}\left(\begin{array}{c}
(\gamma_n^{-}-\gamma_n^{+})\\
(\gamma_n^{+}+\gamma_n^{-})\\
0\end{array}\right),\label{ynnp1D2d}\\
\hat{\tilde{\bm
z}}_{n,n+1}^{D_{2d}}&=&\frac{1}{2}\left(\begin{array}{c}
(\gamma_n^{+}+\gamma_n^{-})\sin\theta_{12}^{D_{2d}}\\
(\gamma_n^{+}-\gamma_n^{-})\sin\theta_{12}^{D_{2d}}\\
(-1)^{n+1}\cos\theta_{12}^{D_{2d}}\end{array}\right),\label{znnp1D2d}
\end{eqnarray}
and the associated parameter set is
\begin{eqnarray}
\{\mu_{12}^{D_{2d}}\}&=&(\theta_{12}^{D_{2d}},0,\frac{\pi}{2}).\label{mu12D2d}
\end{eqnarray}

 Next, we construct the next-nearest-neighbor group-invariant
symmetric anisotropic exchange vector groups.  Beginning with
$\hat{\tilde{\bm z}}_{13}$ of the form of Eq. (\ref{zng}), we
require invariance under the parallel  diagonal mirror plane,
${\cal O}_4^{D_{2d}}\hat{\tilde{\bm z}}_{13}=\pm\hat{\tilde{\bm
z}}_{13}$.  This leads to either $\sin\theta_{13}^{D_{2d}}=0$ or
$\cos\theta_{13}^{D_{2d}}=0$, and we take
$\theta_{13}^{D_{2d}}=0$.  Then, we similarly require ${\cal
O}_4^{D_{2d}}\hat{\tilde{\bm x}}_{13}=\pm\hat{\tilde{\bm
x}}_{13}$, and require
$\sin\chi_{13}^{D_{2d}}=\cos\chi_{13}^{D_{2d}}$, choosing
$\chi_{13}^{D_{2d}}=\pi/4$, where
$\chi_{13}^{D_{2d}}=\phi_{13}^{D_{2d}}+\psi_{13}^{D_{2d}}$.  Thus,
the next-nearest-neighbor symmetric anisotropic exchange vector
group elements are found to be
\begin{eqnarray}
\hat{\tilde{\bm
x}}_{n,n+2}&=&\frac{1}{\sqrt{2}}(-\gamma_n^{+}\hat{\bm
x}+\gamma_n^{-}\hat{\bm y}),\label{xnnp2D2d}\\
\hat{\tilde{\bm
y}}_{n,n+2}&=&\frac{1}{\sqrt{2}}(\gamma_n^{+}\hat{\bm
x}+\gamma_n^{-}\hat{\bm y}),\label{ynnp2D2d}\\
\hat{\tilde{\bm z}}_{n,n+2}&=&(-1)^{n+1}\hat{\bm
z},\label{znnp2D2d}
\end{eqnarray}
and the associated parmaeter set is
\begin{eqnarray}
\{\mu_{13}^{d_{2d}}\}&=&(\theta_{13}^{D_{2d}},\chi_{13}^{D_{2d}})=(0,\frac{\pi}{4}).
\end{eqnarray}
We note that these vector groups are consistent with
$J_{n,n+m}^{f,D_{2d}}=J_{f,m}$ and $J_{n,n+m}^{c,D_{2d}}=J_{c,m}$,
as for the previous tetramer group symmetries studied.

\subsection{C. $D_{2d}$ antisymmetric exchange}

We first consider the next-nearest-neighbor antisymmetric
anisotropic exchange, which couples the spins at sites 1,3 and at
2,4, which lie parallel to  the $xy$ plane, and hence, the roles
of the diagonal mirror planes are identical to that with $C_{4v}$
symmetry.  Thus, ${\bm d}_{13}^{D_{2d}}||(\hat{\bm y}-\hat{\bm
x})$ and ${\bm d}_{24}^{D_{2d}}||(\hat{\bm y}+\hat{\bm x})$. Then,
we require invariance under rotations about the $x$, $y$, and $z$
axes.  This leads to a single scalar parameter describing the
next-nearest-neighbor antisymmetric anisotropic exchange
interactions.

The near-neighbor antisymmetric anisotropic exchange Hamiltonian
is obtained differently.  We note that each position vector
between each of the near-neighbor sites has an axis of rotation
that bisects it.  Thus, ${\bm d}_{n,n+1}^{D_{2d}}$ must be
orthogonal to the respective axis of rotation.  Specifically,
${\bm d}_{12}^{D_{2d}}=d_{12y}^{D_{2d}}\hat{\bm
y}+d_{12z}^{D_{2d}}\hat{\bm z}$, ${\bm
d}_{23}^{D_{2d}}=d_{23x}^{D_{2d}}\hat{\bm
x}+d_{23z}^{D_{2d}}\hat{\bm z}$, ${\bm
d}_{34}^{D_{2d}}=d_{34y}^{D_{2d}}\hat{\bm
y}+d_{34z}^{D_{2d}}\hat{\bm z}$, and ${\bm
d}_{41}^{D_{2d}}=d_{41x}^{D_{2d}}\hat{\bm
x}+d_{41z}^{D_{2d}}\hat{\bm z}$.  Then, we rotate the
near-neighbor antisymmetric exchange Hamiltonian about the $x$,
$y$, and $z$ axes, and impose invariance under the diagonal mirror
planes.  These combined symmetry operations result in the
following restrictions upon the parameters:
$d_{12y}^{D_{2d}}=d_{23x}^{D_{2d}}=-d_{34y}^{D_{2d}}=-d_{41x}^{D_{2d}}$
and
$d_{12z}^{D_{2d}}=-d_{23z}^{D_{2d}}=d_{34yz}^{D_{2d}}=-d_{41z}^{D_{2d}}$.
Thus, there are only two free parameters describing the
near-neighbor antisymmetric anisotropic exchange interactions. The
combined antisymmetric exchange Hamiltonian for $D_{2d}$ symmetry
may then be written as
\begin{eqnarray}
{\cal H}_{DM}^{D_{2d}}&=&\sum_{n=1}^4({\bm S}_n\times{\bm
S}_{n+1})\cdot\Bigl((-1)^{n+1}d_z\hat{\bm z}\nonumber\\
& &-\frac{d_{\perp}}{2}[(\gamma_n^{+}+\gamma_n^{-})\hat{\bm
x}+(\gamma_n^{+}-\gamma_n^{-})\hat{\bm y}]\Bigr)\nonumber\\
& &+d_{\perp}'\sum_{n=1}^2[\hat{\bm x}+(-1)^n\hat{\bm
y}]\cdot({\bm S}_n\times{\bm S}_{n+2}),\label{HDMD2d}
\end{eqnarray}
where $d_{\perp}'$ is a scalar and we have set the other two
scalars $d_z=d_{12z}^{D_{2d}}$ and $d_{\perp}=d_{12y}^{D_{2d}}$.
The DM interactions for $D_{2d}$ symmetry are similar to those for
$S_4$ symmetry, except that the $S_4$ two-vectors, ${\bm
d}_{\perp}$ and ${\bm d}_{\perp}$, are just the one-vectors,
$d_{\perp}\hat{\bm y}$ and $d_{\perp}'\hat{\bm x}$.

\section{IX. $T_d$ group symmetry}

Aside from the restriction $c=a$, $T_d$ is a very high symmetry
group.\cite{Tinkham} in addition to the identity operation, there
are 23 other distinct operations,\cite{Tinkham} all of which are
necessary to include for our purposes. The operations ${\cal
O}_{\lambda}^{T_d}$ for $\lambda=1,5$ are the same as for $D_{2d}$
symmetry, rotations by $\pi$ about the $x$, $y$, and $z$ axis, and
mirror planes containing the $z$ axis and the lines $y=\pm x$.
There are also four additional mirror planes, two of which contain
the $y$ axis and the lines $z=\pm x$ ($\lambda=6,7$), and two of
which contain the $x$ axis and the lines $y=\pm z$
($\lambda=8,9$). These may be written as
\begin{eqnarray}
{\cal O}_{6,7}^{T_d}&=&\left(\begin{array}{ccc}
0&0&\pm1\\
0&1&0\\
\pm1&0&0\end{array}\right),\\
{\cal O}_{8,9}^{T_d}&=&\left(\begin{array}{ccc}
1&0&0\\
0&0&\pm1\\
0&\pm1&0\end{array}\right).
\end{eqnarray}
In addition, there are four clockwise and four counterclockwise
rotations by $2\pi/3$ about the cube diagonals.  These may be
written as
\begin{eqnarray}
{\cal O}_{10,12}^{T_d}&=&\left(\begin{array}{ccc} 0&\pm1&0\\
0&0&1\\
\pm1&0&0\end{array}\right),\\
{\cal O}_{14,16}^{T_d}&=&\left(\begin{array}{ccc} 0&\pm1&0\\
0&0&-1\\
\mp1&0&0\end{array}\right),
\end{eqnarray}
and ${\cal O}_{2n+1}^{T_d}=\Bigl({\cal
O}_{2n}^{T_d}\Bigr)^{T}=\Bigl({\cal O}_{2n}^{T_d}\Bigr)^{-1}$ for
$n=5,6,7,8$. In addition, there are the six $S_4$ symmetries,
which are clockwise and counterclockwise rotations about the $x$,
$y$, and $z$ axes.\cite{Tinkham}  The first two of these are
${\cal O}_{18,19}^{T_d}={\cal O}_{1,2}^{S_4}$.  The $S_4$
operations in which the axis of rotation are the $x$ and $y$ axes,
respectively, are
\begin{eqnarray}
{\cal O}_{20,21}^{T_d}&=&\left(\begin{array}{ccc} -1&0&0\\
0&0&\pm1\\
0&\mp1&0\end{array}\right),\\
{\cal O}_{22,23}^{T_d}&=&\left(\begin{array}{ccc} 0&0&\pm1\\
0&1&0\\
\mp1&0&0\end{array}\right), \end{eqnarray} respectively.

\subsection{A. $T_d$ single-ion anisotropy}

We first examine the axial single-ion vector for site 1, making
use of Eq. (\ref{zng}).  We require that it remain invariant under
the mirror planes ${\cal O}_{4,6,8}^{T_d}$.  This forces the
equations
$\cos\theta-1^{T_d}=\sin\theta_1^{T_d}\sin\phi_1^{T_d}=-\sin\theta_1^{T_d}\cos\phi_1^{T_d}$,
which has the solution $\phi_1^{T_d}=3\pi/4$ and
$\theta_1^{T_d}=\tan^{-1}(\sqrt{2})$.  Thus, $\hat{\tilde{\bm
z}}_1^{T_d}=\hat{\bm r}_1$.  By successive rotations about the
$x$, $y$, and $z$ axes, we generate the $T_d$ axial vector set. We
obtain
\begin{eqnarray}
\hat{\tilde{\bm z}}_n^{T_d}&=&-\gamma_n^{+}\hat{\bm
x}+\gamma_n^{-}\hat{\bm y}+(-1)^{n+1}\hat{\bm z}\\
&=&\hat{\bm r}_n,\label{znTd}
\end{eqnarray}
and the associated two-parameter set is
\begin{eqnarray}
\{\mu_1^{T_d}\}&=&(\theta_1^{T_d},\phi_1^{T_d})=(\tan^{-1}(\sqrt{2}),3\pi/4).\label{mu1Td}
\end{eqnarray}
 Since these axial vectors are
identical to the position vectors, they automatically satisfy all
of the remaining symmetry operations.  Of course,
$J_{a,n}^{T_d}=J_a$ is clearly independent of $n$, as it was for
the lower symmetry $D_{2d}$ group.

We have attempted to construct an azimuthal single-ion vector set,
but there is no set other than the axial single-ion vector set
that can satisfy all of the symmetry operations.  Hence, there are
no azimuthal single-ion vector groups, and we must infer
$J_{e,n}^{T_d}=0$.

\subsection{B. $T_d$ symmetric anisotropic exchange}

For $T_d$ symmetry, all six of the bond lengths are identical, as
$c/a=1$.  Hence, there are at most three local vector groups, each
with six elements. From Eqs. (\ref{znnp1D2d}) and
(\ref{znnp2D2d}), the axial vector group $\{\hat{\bm
z}_{n,n+m}^{T_d}\}$ is found to have the six elements
\begin{eqnarray}
\hat{\bm
z}_{n,n+m}^{T_d}&=&\frac{\delta_{m,1}}{\sqrt{2}}\Bigl[-\frac{1}{2}\Bigl((\gamma_n^{+}+\gamma_n^{-})\hat{\bm
x}+(\gamma_n^{+}-\gamma_n^{-})\hat{\bm y}\Bigr)\nonumber\\
& &+(-1)^{n+1}\hat{\bm
z}\Bigr]+\frac{\delta_{m,2}}{\sqrt{2}}[\hat{\bm
x}+(-1)^{n+1}\hat{\bm y}].\label{znnpmTd}
\end{eqnarray}

It is easily verified that this axial vector set is a group
satisfying all of the $T_d$ symmetries.   The reason this is true
is actually pretty obvious:  the axial exchange vector group
consists of equally normalized differences between position
vectors, and each position vector necessarily satisfies the group
symmetry. However, there are no azimuthal anisotropic exchange
groups for $T_d$, for the same reasons as for the single-ion group
vectors. For $T_d$, there is therefore only one anisotropic
exchange interaction $J_{n,n+m}^{f,T_d}=J_{f,1}=J_{f,2}$, since
 $J_{c,m}=0$ for $m=1,2$.

 \subsection{C. $T_d$ antisymmetric exchanges vanish}

We now consider the possibility that there might be a
non-vanishing antisymmetric exchange Hamiltonian for tetramers
with $T_d$ symmetry.  From the mirror planes, the Moriya rules
immediately restrict ${\bm d}_{12}^{T_d}=d_{12}(\hat{\bm
z}-\hat{\bm y})/\sqrt{2}$, ${\bm d}_{23}^{T_d}=d_{23}(\hat{\bm
z}+\hat{\bm x})/\sqrt{2}$, ${\bm d}_{34}^{T_d}=d_{34}(\hat{\bm
z}+\hat{\bm y})/\sqrt{2}$, ${\bm d}_{41}^{T_d}=d_{41}(\hat{\bm
z}-\hat{\bm x})/\sqrt{2}$, ${\bm d}_{13}^{T_d}=d_{13}(\hat{\bm
x}-\hat{\bm y})/\sqrt{2}$, and ${\bm
d}_{124}^{T_d}=d_{24}(\hat{\bm x}+\hat{\bm y})/\sqrt{2}$. Then,
rotation by $\pi$ about the $x$, $y$, and $z$ axes (${\cal
O}_{1,2,3}^{T_d}$) forces $d_{24}=d_{13}$,  $d_{41}=d_{23}$, and
$d_{12}=d_{34}$. Then, we require the Hamiltonian to be invarian
under the clockwise rotation by $2\pi/3$ about the cube diagonal
passing through site 1, ${\cal O}_{12}^{T_d}$.  This operation
satisfies ${\cal O}_{12}^{T_d}\hat{\bm x}=\hat{\bm z}$, ${\cal
O}_{12}^{T_d}\hat{\bm y}=\hat{\bm x}$, ${\cal
O}_{12}^{T_d}\hat{\bm z}=\hat{\bm y}$, ${\cal O}_{12}^{T_d}{\bm
S}_1={\bm S}_1$, ${\cal O}_{12}^{T_d}{\bm S}_2={\bm S}_3$, ${\cal
O}_{12}^{T_d}{\bm S}_3={\bm S}_4$, and ${\cal O}_{12}^{T_d}{\bm
S}_4={\bm S}_2$. Invariance of the Hamiltonian then requires
$d_{12}=d_{23}=-d_{13}$.  This reduces the number of parameters to
just one, which we take to be $d_{12}$.  We note that the rotation
about the cube diagonal ${\cal O}_{12}$ leaves this form of ${\cal
H}_{DM}^{T_d}$ invariant, without any further restrictions.
However, we must also have ${\cal H}_{DM}^{T_d}$ invariant under
the six $S_4$ operations, and we choose ${\cal O}_{18}^{T_d}={\cal
O}_1^{S_4}$ as the first example.  It is then easy to see that
setting ${\cal O}_{18}^{T_d}{\cal H}_{DM}^{T_d}={\cal
H}_{DM}^{T_d}$ forces $d_{12}=-d_{12}=0$.  Hence, there are no DM
interactions in tetramers exhibiting $T_d$ group symmetry,

\section{X. The Hamiltonian in the molecular
representation}
\subsection{A.  The molecular single-ion Hamiltonian}

To make contact with experiment, we use Eqs.
(\ref{xnC4h})-(\ref{znC4h}), (\ref{xnD4h})-(\ref{znD4h}),
(\ref{xnC4v})-(\ref{znC4v}), (\ref{xnS4})-(\ref{znS4}),
(\ref{xnD2d})-(\ref{znD2d}), and (\ref{znTd}) to rewrite ${\cal
H}^{g,\ell}_{si}$ in the molecular $(\hat{\bm x},\hat{\bm
y},\hat{\bm z})$ representation,
\begin{eqnarray}{\cal
H}^g_{si}&=&-\sum_{n}\Bigl(J_z^g(\mu_1^g)S_{n,z}^2+(-1)^nJ_{xy}^g(\mu_1^g)(S_{n,x}^2-S_{n,y}^2)\nonumber\\
&
&+\frac{1}{2}\sum_{\alpha\ne\beta}K_{\alpha\beta}^g(n,\mu_1^g)\{S_{n,\alpha},S_{n,\beta}\}\Bigr),\label{Hsimolecular}
\end{eqnarray}
where $\alpha,\beta=x,y,z$ and  the $\mu_1^g\equiv\{\mu_1^g\}$ are
given by Eqs. (\ref{mu1C4h}), (\ref{mu1D4h}), (\ref{mu1C4v}),
(\ref{mu1S4}), (\ref{mu1D2d}), and (\ref{mu1Td}). In Eq.
(\ref{Hsimolecular}), $\{A,B\}=AB+BA$ and we subtracted an
irrelevant constant. The easiest groups for which to evaluate the
parameters in $\tilde{\cal H}_{si}^g$ are $D_{2d}$ and $T_d$. The
site-independent interactions in the molecular representation are
\begin{eqnarray}
J_z^{C_{4h}}&=&J_z^{D_{4h}}=J_a,\label{JzC4h}\\
J_z^{C_{4v}}(\mu_1^{C_{4v}})&=&\frac{1}{2}\Bigl(J_a(3\cos^2\theta_1^{C_{4v}}-1)+3J_e\sin^2\theta_1^{C_{4v}}\Bigr),\label{JzC4v}\nonumber\\
& &\\
J_z^{S_4}(\mu_1^{S_4})&=&\frac{J_a}{2}(3\cos^2\theta_1^{S_4}-1)\nonumber\\
& &-\frac{3}{2}J_e\sin^2\theta_1^{S_4}\cos(2\psi_1^{S_4}),\label{JzS4}\\
J_z^{D_{2d}}(\mu_1^{D_{2d}})&=&\frac{1}{2}\Bigl(J_a(3\cos^2\theta_1^{D_{2d}}-1)+3J_e\sin^2\theta_1^{D_{2d}}\Bigr),\label{JzD2d}\nonumber\\
& &\\
 J_z^{T_d}(\mu_1^{T_d})&=&0.\label{JzTd}
\end{eqnarray}
 We
note  that the form of the site-independent interactions for
$D_{2d}$ and $C_{4v}$ symmetry are identical, and can be obtained
from that for the lowest symmetry, $S_4$,  by setting
$\psi_1^{S_4}\rightarrow3\pi/4$.  In addition, that for the higher
symmetry cases $D_{4h}$ and $C_{4h}$ are also obtained from that
for $S_4$ by setting $\theta_1^{g}\rightarrow0$.  The vanishing
site-independent single-ion interaction for the highest $T_d$
symmetry is obtained from that of the lowest $S_4$ symmetry by
setting both $\theta_1^{S_4}\rightarrow\tan^{-1}(\sqrt{2})$ and
$J_e\rightarrow0$.

In addition, ${\cal H}_{si}^g$ contains the site-dependent
interactions $(-1)^nJ_{xy}^g(\mu_1^g)$ for $g=C_{4h}, S_4$ and
$K_{\alpha\beta}^g(n,\mu_1^g)$ for $g=C_{4h}, D_{4h}, C_{4v}, S_4,
D_{2d}$ and $T_d$, which are listed in the Appendix. These
site-dependent interactions were generated from the
site-independent interaction $J_a$ for  $T_d$, $J_e$ for $C_{4h}$
and $D_{4h}$,  and both $J_a$ and $J_e$ for $C_{4v}$, $D_{2d}$,
and $S_4$ in the local coordinates by the respective group
operations.
 All of these site-dependent interactions
average to zero, but they contribute to the eigenstate energies to
second and higher orders in the interactions $J_a$ and $J_e$. For
$C_{4v}$ and $S_4$, the effective axial site-independent
interactions $J_z^g(\mu_1^g)$ arise from a combination of the local
axial and azimuthal interactions $J_a$ and $J_e$, the details of the
combination depending upon the precise effects of the
 ligand groups near to the ions.  We also note that there is a
 substantial
 difference between $T_d$ and the other three symmetries.
 For $T_d$ symmetry, the average effective interaction vanishes.
 However, for $C_{4v}$,  $D_{2d}$ and $S_4$ symmetry, the effective
 interactions can be large, even if the molecular structure is
 nearly $T_d$ (that is, if $c/a\approx1$). Once this symmetry is  broken,
 the site-independent single-ion interactions $J_z^g$ can become large.

\subsection{B. Symmetric anisotropic exchange in the molecular
representation}
 We then use these local exchange coordinates to
construct the group-satisfying anisotropic exchange Hamiltonian in
the molecular coordinates.  For all four symmetries, there are
similar effects. They first lead to renormalizations of the
isotropic exchange interactions, modifying ${\cal H}_0^g$ to
\begin{eqnarray}
{\cal H}_0^{g,r}&=&-\frac{\tilde{J}_g}{2}{\bm S}^2-\gamma{\bm
B}\cdot{\bm S}-\frac{(\tilde{J}_g'-\tilde{J}_g)}{2}({\bm
S}_{13}^2+{\bm S}_{24}^2),\nonumber\\
\end{eqnarray}
where
\begin{eqnarray}
\tilde{J}_g&=&J_g+\delta J_{g},\\
 \tilde{J}_{g}'&\rightarrow&J_g'+\delta J_g',
\end{eqnarray}
where $J_g'$ is given by Eqs. (\ref{JTdp}) and (\ref{Jgp}). For
simplicity of presentation, we write
\begin{eqnarray}
J_{m,\pm}&=&\frac{1}{2}(J_{c,m}\pm J_{f,m})
\end{eqnarray}
for $m=1,2$.  Then, the isotropic exchange renormalizations may be
written as
\begin{eqnarray}
\delta J_{C_{4h}}&=&\delta J_{D_{4h}}=0,\\
\delta J_{C_{4v}}&=&\frac{1}{2}[J_{f,1}+J_{c,1}\cos(2\psi_{12}^{C_{4v}})],\label{deltaJC4v}\\
\delta J_{D_{2d}}&=&-J_{1,-}\sin^2\theta_{12}^{D_{2d}},\label{deltaJD2d}\\
\delta
J_{S_4}&=&\frac{\sin^2\theta_{12}^{S_4}}{2}[J_{f,1}+J_{c,1}\cos(2\psi_{12}^{S_4})]\label{deltaJS4}\\
\delta J_{T_d}&=&\frac{J_{f,1}}{4},\label{deltaJTd}\\
\delta J_{C_{4h}}'&=&\delta J_{D_{4h}}'=\delta J_{D_{2d}}'=0,\\
\delta J_{C_{4v}}'&=&\delta J_{S_4}'=J_{2,+},\label{deltaJgp}\\
 \delta
J_{T_d}'&=&\frac{J_{f,1}}{2}.\label{deltaJTdp}
\end{eqnarray}
They also lead to addition interactions $\delta{\cal H}_{ae}^g$ in
the molecular frame.  For $g=C_{4v}$, $S_4$, $D_{2d}$, and $T_d$,
these may be written as
\begin{eqnarray}
\delta{\cal
H}_{ae}^{g}&=&\sum_{m=1}^2\sum_{n=1}^{6-2m}\Bigl[J_{m,z}^g(\mu_{1,m+1}^g)S_{n,z}S_{n+m,z}\nonumber\\
&&+J_{m,xy}^g(\mu_{1,m+1}^g)(-1)^{n+1}\nonumber\\
& &\times\Bigl(S_{n,x}S_{n+m,x}-S_{n,y}S_{n+m,y}\Bigr)\Bigr]\nonumber\\
&
&+\sum_{n=1}^2K_{2,xy}^g(\mu_{13}^g)(-1)^{n+1}\nonumber\\
& &\times\Bigl(S_{n,x}S_{n+2,y}+S_{n,y}S_{n+2,x}\Bigr)\nonumber\\
& &+\frac{1}{2}\sum_{n=1}^4\sum_{\alpha,\beta}K_{1,\alpha
\beta}^g(n,\mu_{12}^g)\nonumber\\
&
&\times\Bigl(S_{n,\alpha}S_{n+1,\beta}+S_{n,\beta}S_{n+1,\alpha}\Bigr)\biggr],\label{Haemolecular}
\end{eqnarray}
where $\alpha,\beta=x,y,z$,
$\mu_{12}^g=\{\theta_{12}^g,\phi_{12}^g,\psi_{12}^g\}$, and
$\mu_{13}^g=\{\theta_{13}^g,\phi_{13}^g,\psi_{13}^g\}$.
 The site-independent anisotropic exchange interactions are
found to be
\begin{eqnarray}
J_{1,z}^{C_{4h}}&=&J_{1,z}^{D_{4h}}=-J_{f,1},\label{J1zC4hD4h}\\
J_{1,z}^{C_{4v}}(\mu_{12}^{C_{4v}})&=&\frac{1}{2}[J_{f,1}+3J_{c,1}\cos(2\psi_{12}^{C_{4v}}],\label{J1zC4v}\\
J_{1,z}^{S_4}(\mu_{12}^{S_4})&=&\frac{J_{f,1}}{2}(1-3\cos^2\theta_{12}^{S_4})\nonumber\\
& &+\frac{3J_{c,1}}{2}\sin^2\theta_{12}^{S_4}\cos(2\psi_{12}^{S_4}),\label{J1zS4}\\
J_{1,z}^{D_{2d}}(\mu_{12})&=&\frac{1}{4}[J_{f,1}+3J_{1,-}\sin^2\theta_{12}^{D_{2d}}],\label{J1zD2d}\\
J_{1,z}^{T_d}&=&-\frac{1}{4}J_{f,1}=-2J_{2,z}^{T_d},\label{J1z2zTd}\\
J_{2,z}^{C_{4h}}&=&J_{2,z}^{D_{4h}}=J_{2,z}^{D_{2d}}=-J_{f,2},\label{J2zC4hD4hD2d}\\
J_{2,z}^{S_4}&=&J_{2,z}^{C_{4v}}=J_{2,+}+J_{c,2},\label{J2zg}
\end{eqnarray}
where $J_{2,z}^g(\mu_{13}^g)=J_{2,z}^g$ has fully group-restricted
$\mu_{13}^g$ sets for all symmetries.  The site-dependent
anisotropic exchange interactions are listed in the Appendix.

As we shall see in the following, to first order in the single-ion
and anisotropic exchange interactions, only the site-independent
interactions appear.

\subsection{C. Antisymmetric exchange Hamiltonian}

In Secs. IV-IX, we already presented the antisymmetric exchange
Hamiltonians in the molecular representation.  These may be
combined into a single equation as
\begin{eqnarray}
{\cal H}_{DM}^g&=&\sum_{n=1}^4\bigl({\bm S}_n\times{\bm
S}_{n+1}\bigr)\cdot\Bigl(d^g_z(n)\hat{\bm z}\nonumber\\
& &+\frac{1}{2}(\gamma_n^{-}-\gamma_n^{+}){\bm d}^g_{\perp}\nonumber\\
& &+\frac{1}{2}(\gamma_n^{+}+\gamma_n^{-})(\hat{\bm z}\times{\bm
d}^g_{\perp})\Bigr)\nonumber\\
& &+\frac{1}{2}\sum_{n=1}^2\bigl({\bm S}_n\times{\bm S}_{n+2}\bigr)\cdot\Bigl((\gamma_n^{-}-\gamma_n^{+}){\bm d}'^{,g}_{\perp}\nonumber\\
& &+(\gamma_n^{+}+\gamma_n^{-})(\hat{\bm z}\times{\bm
d}'^{,g}_{\perp})\Bigr),\label{HDM}
\end{eqnarray}
where the relevant parameters are given in Eqs. (\ref{HDMC4h}),
(\ref{HDMD4h}), (\ref{HDMC4v}), (\ref{HDMS4}), (\ref{HDMD2d}), and
they all vanish for $g=T_d$.  We find
\begin{eqnarray}
d_z^{C_{4h}}(n)&=&d_z^{D_{4h}}(n)=d_z^{C_{4v}}(n)=d_z,\\
d_z^{S_4}(n)&=&d_z^{D_{2d}}(n)=d_z(-1)^{n+1},\\
{\bm d}_{\perp}^{C_{4h}}&=&{\bm d}_{\perp}^{D_{4h}}=0,\\
{\bm d}_{\perp}^{C_{4v}}&=&d_{\perp}\hat{\bm x},\\
{\bm d}_{\perp}^{D_{2d}}&=&d_{\perp}\hat{\bm y},\\
{\bm d}_{\perp}^{S_4}&=&{\bm d}_{\perp}=d_{\perp,x}\hat{\bm x}+d_{\perp,y}\hat{\bm y},\\
{\bm d}_{\perp}'^{,C_{4h}}&=&{\bm d}_{\perp}'^{,D_{4h}}=0,\\
{\bm d}_{\perp}'^{,C_{4v}}&=&d_{\perp}'\hat{\bm y},\\
{\bm d}_{\perp}'^{,D_{2d}}&=&d_{\perp}'\hat{\bm x},\\
{\bm d}_{\perp}'^{,S_4}&=&{\bm d}_{\perp}'=d_{\perp,x}'\hat{\bm x}+d_{\perp,y}'\hat{\bm y},\\
d_z^{T_d}(n)&=&{\bm d}_{\perp}^{T_d}={\bm d}_{\perp}'^{,T_d}=0,
\end{eqnarray}
 Tetramers with the lowest-symmetry group $S_4$ require
five parameters to describe the full DM interactions, those with
either  $D_{2d}$ or $C_{4v}$ symmetry require three parameters,
those with either $C_{4h}$ or $D_{4h}$ symmetry require just one
parameter, and tetramers with $T_d$ symmetry do not exhibit any DM
interactions.  We note that for $C_{4v}$, $S_4$, and $D_{2d}$, the
NNN DM interactions, while site-dependent, do not average to zero.

\section{XI. Eigenstates of the full Hamiltonian}

We assume a molecular Hamiltonian of ${\cal H}={\cal
H}_0^{g,r}+{\cal H}^g_{si}+\delta{\cal H}_{ae}$. To take proper
account of ${\bm B}$ in ${\cal H}_0$, we construct our SMM
eigenstates in the induction representation by
\begin{eqnarray}
\left(\begin{array}{c}\hat{\bm x}\\
\hat{\bm y}\\
\hat{\bm z}\end{array}\right)=\left(\begin{array}{ccc}\cos\theta\cos\phi &-\sin\phi &\sin\theta\cos\phi\\
\cos\theta\sin\phi &\cos\phi &\sin\theta\sin\phi\\
-\sin\theta
&0&\cos\theta\end{array}\right)\left(\begin{array}{c}\hat{\tilde{\bm
x}}\\
\hat{\tilde{\bm y}}\\
\hat{\tilde{\bm z}}\end{array}\right),\label{rotation}\nonumber\\
\end{eqnarray}
so that ${\bm B}=B\hat{\tilde{\bm z}}$. A subsequent arbitrary
rotation about $\hat{\tilde{\bm z}}$ does not affect the
eigenstates.\cite{ek2} We then set $\hbar=1$ and write
\begin{eqnarray}
{\bm S}^2|\psi_{s,m}^{s_{13},s_{24}}\rangle&=&s(s+1)|\psi_{s,m}^{s_{13},s_{24}}\rangle,\label{S}\\
{\bm S}_{13}^2|\psi_{s,m}^{s_{13},s_{24}}\rangle&=&s_{13}(s_{13}+1)|\psi_{s,m}^{s_{13},s_{24}}\rangle,\\
{\bm S}_{24}^2|\psi_{s,m}^{s_{13},s_{24}}\rangle&=&s_{24}(s_{24}+1)|\psi_{s,m}^{s_{13},s_{24}}\rangle,\\
S_{\tilde{z}}|\psi_{s,m}^{s_{13},s_{24}}\rangle&=&m|\psi_{s,m}^{s_{13},s_{24}}\rangle,\label{Sz}\\
 S_{\tilde{\sigma}}|\psi_{s,m}^{s_{13},s_{24}}\rangle&=&A_s^{\tilde{\sigma}
m}|\psi_{s,m+\tilde{\sigma}}^{s_{13},s_{24}}\rangle,\label{SxSy}\\
A_s^{m}&=&\sqrt{(s-m)(s+m+1)},\label{Asm}
\end{eqnarray}
 where $S_{\tilde{\sigma}}=S_{\tilde{x}}+i\tilde{\sigma} S_{\tilde{y}}$ with $\tilde{\sigma}=\pm$.
For brevity, we denote $\nu=\{s,m,s_{13},s_{24},\{s_n\}\}$, and
write $|\nu\rangle\equiv|\psi_{s,m}^{s_{13},s_{24}}\rangle$. From
Eqs. (\ref{S}) to (\ref{Sz}), $\langle\nu'|\tilde{\cal
H}^{g,r}_0|\nu\rangle=E_{\nu,0}^g\delta_{\nu',\nu}$, where
\begin{eqnarray}
E_{\nu,0}^g&=&-\frac{\tilde{J}_g}{2}s(s+1)-\gamma
Bm\nonumber\\
&
&-\frac{(\tilde{J}_g'-\tilde{J}_g)}{2}[s_{13}(s_{13}+1)+s_{24}(s_{24}+1)],\label{E0}
\end{eqnarray}
where the $\tilde{J}_g$ and $\tilde{J}_g'$ are given by Eqs.
(\ref{JTdp}), (\ref{Jgp}), and (\ref{deltaJC4v})-
(\ref{deltaJTd}).

\subsection{A. Induction representation}

 We then  transform the molecular ${\cal H}_{si}^g+{\cal H}_{ae}^g$
to the induction representation, yielding $\tilde{\cal
H}_{si}^g+\tilde{\cal H}_{ae}^g$, and make a standard perturbation
expansion for the six microscopic anisotropy energies
$\{J_j\}\equiv(J_a,J_{c,m},J_e,J_{f,m})$ for $m=1,2$ small
relative to $|J_g|, |J_g'|$ and/or  $\gamma B$.\cite{ek2} To do
so, the contain more interesting physics. Compact expressions for
these matrix elements are given in the Appendix.

 At arbitrary $(\theta,\phi)$, the first order
corrections $E_{\nu,1}^g=\langle\nu|\tilde{\cal
H}^g_{si}+\tilde{\cal H}_{ae}^g|\nu\rangle$ to the eigenstate
energies for $g=C_{4v},S_4$, $D_{2d}$, and $T_d$ symmetries are
\begin{eqnarray}
E^{g,\tilde{\mu}^g}_{\nu,1}&=&\frac{\tilde{J}^{g,\overline{\nu}}_{z}(\tilde{\mu}^g)}{2}[m^2-s(s+1)]-\delta \tilde{J}_z^{g,\overline{\nu}}(\tilde{\mu}^g)\nonumber\\
&
&-\frac{[3m^2-s(s+1)]}{2}\tilde{J}^{g,\overline{\nu}}_{z}(\tilde{\mu}^g)\cos^2\theta,\label{E1}\\
\tilde{J}_{z}^{g,\overline{\nu}}(\tilde{\mu}^g)&=&J_z^g(\mu_1^g)a^{+}_{\overline{\nu}}-J_{1,z}^g(\mu_{12}^g)c_{\overline{\nu}}^{-}\nonumber\\
& &\qquad
-\frac{1}{2}J_{2,z}^ga_{\overline{\nu}}^{-},\label{Jztildemu}\\
\delta\tilde{J}_z^{g,\overline{\nu}}(\tilde{\mu}^g)&=&J_z^g(\mu_1^g)b_{\overline{\nu}}^{+}
-\frac{1}{4}J_{1,z}^g(\mu_{12}^g)(b_{\overline{\nu}}^{+}+b_{\overline{\nu}}^{-})\nonumber\\
& &\qquad-\frac{1}{4}J_{2,z}^gb_{\overline{\nu}}^{-},
\end{eqnarray}
where the $a_{\overline{\nu}}^{\pm}$, $b_{\overline{\nu}}^{\pm}$,
and $c_{\overline{\nu}}^{\pm}$ are given in the Appendix, and the
interactions are given by Eqs. (\ref{JzC4v})-(\ref{JzS4}) and
(\ref{J1zC4v})-(\ref{J2zg}). We have used the notations
\begin{eqnarray}
\tilde{\mu}^g&=&\{\theta_1^g,\phi_1^g,\psi_1^g,\theta_{12}^g,\phi_{12}^g,\psi_{12}^g\},\label{tildemu}\\
\overline{\nu}&=&\{s,s_{13},s_{24},s_1\},\label{overlinenu}
\end{eqnarray} which excludes $m$. We note that the site-dependent
interactions in Eqs. (\ref{Hsimolecular}) and (\ref{Haemolecular})
and all of the DM interactions in Eq. (\ref{HDM}) do not
contribute to the first-order correction to the eigenstate
energies.
   Second order corrections
to the energies include contributions from both the site-independent
and site-dependent interactions, and will be presented
elsewhere.\cite{ekfuture}

For all four $g$ symmetries, $E_{\nu,1}^{g,\tilde{\mu}^g}$ has a
form analogous to that of the equal-spin dimer in the absence of
azimuthal single-ion and anisotropic exchange
interactions.\cite{ek2} For these high-symmetry tetramers, to first
order in the anisotropy interactions, the azimuthal single-ion and
anisotropic exchange interactions merely renormalize the effective
respective site-independent axial interactions.  They do contribute
directly to the second order eigenstate energies, but in rather
complicated ways, which we will present elsewhere.\cite{ekfuture}
Thus, to first order, we only have two effective isotropic exchange
and three effective anisotropy interactions, $\tilde{J}_g$,
$\tilde{J}_g'$, $J_z^g(\mu_1^g)$, $J_{1,z}^g(\mu_{12}^g)$, and
$J_{2,z}^g$, which are fixed for a particular SMM. Nevertheless, the
first-order eigenstate energies
$E_{\nu}^g=E_{\nu,0}^g+E_{\nu,1}^{g,\tilde{\mu}^g}$, given by Eqs.
(\ref{E0}) and (\ref{E1}), contain these five effective interaction
strengths in ways that depend strongly upon the quantum numbers
$\overline{\nu}$ and $\theta$. These different
$\overline{\nu},\theta$ dependencies can be employed to provide
definitive measures of the five $\overline{\nu}$-independent
effective isotropic exchange and anisotropy interactions.  To
determine the precise values of the microscopic interactions, one
needs to study the second order eigenstate energies in detail.

\subsection{B. AFM level crossing inductions}

For AFM tetramers, $\tilde{J}_g<0$.    There will be $2s_1+1$ level
crossings, provided that the lowest energy state in each $s$
manifold does not exhibit level repulsion. Letting
$E_{s,m}^{g,\tilde{\mu}^g}(s_{13},s_{24},s_1)=E^g_{\nu,0}+E^{g,\tilde{\mu}^g}_{\nu,1}$,
these occur when
\begin{eqnarray}
E_{s,s}^{g,\tilde{\mu}^g}(s_{13},s_{24},s_1)&=&E_{s-1,s-1}^{g,\tilde{\mu}^g}(s_{13},s_{24},s_1).
\end{eqnarray}

There are two types of AFM level-crossings, depending upon the sign
of $\tilde{J}_g'-\tilde{J}_g$, as seen from Eq. (\ref{E0}).  For
Type I, $\tilde{J}_g<0$ and $\tilde{J}_g'-\tilde{J}_g>0$, the lowest
energy state in each $s$ manifold occurs for $s_{13}=s_{24}=2s_1$,
their maximum values. Hence, the tetramer effectively acts as an AFM
dimer with equal spins $2s_1$.  The parameters for Type I are given
for general $s,s_1$ in the Appendix. For the frustrated Type II,
$\tilde{J}_g<0$ and $\tilde{J}_g'-\tilde{J}_g<0$, the level crossing
occurs for the minimum values of $s_{13}$ and $s_{24}$. It is easy
to see that for even $s$ values, these minima occur at
$s_{13}=s_{24}=s/2$. For odd $s$ values, the situation is doubly
degenerate, and the minima occur at $s_{13},s_{24}=(s\pm1)/2,
(s\mp1)/2$.  Hence, this level-crossing behavior differs
substantially from that of equal spin dimers.\cite{ek2} In the
Appendix, we also listed the formulas for the level-crossing
parameters for Type II.

For both signs of $\tilde{J}_g-\tilde{J}_g'$, the $s$th AFM
level-crossing induction in the first-order approximation may be
written as
\begin{eqnarray}
\gamma B_{s_1,s}^{g,\rm lc
(1)}(\theta)&=&-\tilde{J}_gs+(\tilde{J}_g-\tilde{J}_g')\Theta(\tilde{J}_g-\tilde{J}_g')E\Bigl(\frac{s+1}{2}\Bigr)\nonumber\\
&
&-\frac{J_z^g(\mu_1^g)}{2}(a_2^{+}+2b^{+}+a_1^{+}\cos^2\theta)\nonumber\\
&
&+\frac{J_{1,z}^g(\mu_{12}^g)}{2}[c_2^{-}+\frac{1}{2}(b^{+}+b^{-})+c_1^{-}\cos^2\theta)]\nonumber\\
&
&+\frac{J_{2,z}^g}{4}(a_2^{-}+b^{-}+a_1^{-}\cos^2\theta),\label{levelcrossing}
\end{eqnarray}
where  $\Theta(s)$ is the standard Heaviside step function, $E(x)$
is the largest integer in $x$ and
\begin{eqnarray}
a_{1}^{\pm}&=&s(2s-1)a_{s,s_{13},s_{24}}^{s_1,\,\pm}\nonumber\\
& &-(s-1)(2s-3)a_{s-1,s_{13},s_{24}}^{s_1,\,\pm},\label{am}\\
a_{2}^{\pm}&=&sa_{s,s_{13},s_{24}}^{s_1,\,\pm}-(s-1)a^{s_1,\,\pm}_{s-1,s_{13},s_{24}},\\
b^{\pm}&=&b_{s,s_{13},s_{24}}^{s_1,\,\pm}-b_{s-1,s_{13},s_{24}}^{s_1,\,\pm},\\
c_{1}^{-}&=&s(2s-1)c^{s_1,\,-}_{s,s_{13},s_{24}}\nonumber\\
& &-(s-1)(2s-3)c_{s-1,s_{13},s_{24}}^{s_1,\,-},\\
c_{2}^{-}&=&sc_{s,s_{13},s_{24}}^{s_1,\,-}-(s-1)c_{s-1,s_{13},s_{24}}^{s_1,\,-}.\label{cm}
\end{eqnarray}
These  $a_{j}^{\pm}$, $b^{\pm}$, and $c_{j}^{-}$ for $j=1,2$ are
functions of $s,s_1$ and the case.  For Type II, the functions are
different for even and odd $s$. These functions are given for both
cases with arbitrary $s,s_1$ in the Appendix.

Most notable is that for Type I, the symmetric anisotropic
exchange interactions combine into a simple form, yielding the
effective interaction $J_{\rm eff}^g(s_1)$ given by
\begin{eqnarray}
J_{\rm
eff}^g(s_1)&=&\frac{J_{1,z}^g(\mu_{12}^g)}{2}+\frac{s_1J_{2,z}^g}{4s_1-1}.\label{Jeff}
\end{eqnarray}
Moreover, the symmetric anisotropic exchange contributions to Eq.
(\ref{levelcrossing}) combine to yield the single term
$-c_1^{-}J_{\rm eff}^g(s_1)(1-3\cos^2\theta)/3$ for Type I, where
$c_1^{-}$ is a function of $s,s_1$. However, the single ion
interactions are more complicated, even for Type I, as the
$\theta$-independent and $\theta$-dependent contributions have
different $s,s_1$ dependencies, except for $s_1=1/2$, for which
they both vanish.

One simplification that does occur for Type II AFM tetramers is the
level crossing that occurs as a result of near-neighbor anisotropic
exchange interactions $J_{1,z}^g(\mu_{12}^g)$.  As shown in the
Appendix, the contributions $\gamma B_{1,z}^{g,{\rm
lc}(1)}=J_{1,z}^g(\mu_{12}^g)f_{1,z}(s,\theta)$ to $\gamma
B_{s_1,s}^{g,{\rm lc}(1)}(\theta)$ from these terms are independent
of $s_1$, where
\begin{eqnarray}
f_{1,z}(s,\theta)&=&\left\{\begin{array}{ll}
\frac{(s-1)}{4s}\Bigl(1+(2s-1)\cos^2\theta\Bigr),&
s\>\>{\rm odd}\\
\frac{s}{4(s-1)}\Bigl(1+(2s-3)\cos^2\theta\Bigr),&s\>\>{\rm
even}.\end{array}\right.
\end{eqnarray}  Note in particular that for $s=1$, $f_{1,z}(1,\theta)=0$,
 independent of $s_1$.  However, the single-ion
and next-nearest neighbor anisotropic exchange contributions to the
level crossing inductions depend both upon $s,s_1$ in different ways
for both odd and even $s$, as shown in the Appendix.

\subsubsection{$s_1=1/2$ AFM level crossings}

For the simplest case $s_1=1/2$, as in AFM Cu$_4$ tetramers, the
single-ion terms in Eq. (\ref{levelcrossing}) vanish for both cases,
as for the dimer of equal $s_1=1/2$ spins.\cite{ek2} Using the
results given in the Appendix, the expressions for  the  $\gamma
B_{1/2,s}^{g,{\rm lc}(1)}(\theta)$ functions are particularly
simple.  For the effective dimer Type I,
$\tilde{J}_g'-\tilde{J}_g>0$,
\begin{eqnarray}
\gamma B_{1/2,1}^{g,{\rm
lc}(1)}(\theta)&=&-\tilde{J}_g+\frac{1}{6}J^g_{\rm eff}(1/2)(1-3\cos^2\theta),\\
\gamma B_{1/2,2}^{g,{\rm
lc}(1)}(\theta)&=&-2\tilde{J}_g-\frac{1}{2}J^g_{\rm
eff}(1/2)(1-3\cos^2\theta),
\end{eqnarray}
where $J^g_{\rm eff}(1/2)$ is given by Eq. (\ref{Jeff}) with
$s_1=1/2$, and for the frustrated Type II,
$\tilde{J}_g-\tilde{J}_g'>0$,
\begin{eqnarray}
\gamma B_{1/2,1}^{g,{\rm
lc}(1)}(\theta)&=&-\tilde{J}_g'+\frac{1}{4}J^g_{2,z}(1+\cos^2\theta),\\
\gamma B_{1/2,2}^{g,{\rm
lc}(1)}(\theta)&=&-\tilde{J}_g-\tilde{J}_g'\nonumber\\
& &+ \Bigl(J^g_{\rm
eff}(1/2)-\frac{J_{2,z}^g}{4}\Bigr)(1+\cos^2\theta\Bigr).
\end{eqnarray}
Even in this simplest of all tetramer cases, for which single-ion
interactions do not affect the thermodynamics, there is still a
qualitative different between Type I and Type II AFM $s_1=1/2$
tetramers.  For Type I, there is only one effective isotropic
interaction, $\tilde{J}_g$ and one effective anisotropic
interaction, $J_{\rm eff}^g(1/2)$ that affect the level crossing.
However, for Type II $s_1=1/2$ tetramers, the level crossing is
different for the near-neighbor and next-nearest-neighbor
anisotropic exchange interactions, and both effective isotropic
interactions $\tilde{J}_g$ and $\tilde{J}_g'$ affect the spacing of
the level crossings.   The only effect of the group symmetry is to
provide the particular value of $J^g_{\rm eff}$.  For example,
 $J^{T_d}_{\rm eff}(1/2)=J_{f,1}/8$.  These expressions
also show that Type II has a more complex level-crossing induction
variation than does Type I, as the level-crossing behavior of Type
I is fully described by two parameters $\tilde{J}_g$ and $J^g_{\rm
eff}$, whereas the level-crossing behavior of Type II  depends
$\tilde{J}_g$, $\tilde{J}_g'$, $J_{1,z}^g(\mu_{12}^g)$, and
$J_{2,z}^g$ separately. On the other hand, for this special
$s_1=1/2$ example, the $\theta$ dependencies of the first and
second $\gamma B_{1/2,s}^{g,{\rm lc}(1)}$ are opposite in sign for
Type I, but have the same sign for Type II. In Fig. 4, we
illustrated these behaviors for $J^g_{\rm
eff}(1/2)/\tilde{J}_g=0.2$ and for the general Type I and for Type
II also with $J_{2,z}^g/\tilde{J}_g=0.1$ and
$\tilde{J}_g-\tilde{J}_g'=0.5|\tilde{J}_g|$.

\begin{figure}
\includegraphics[width=0.45\textwidth]{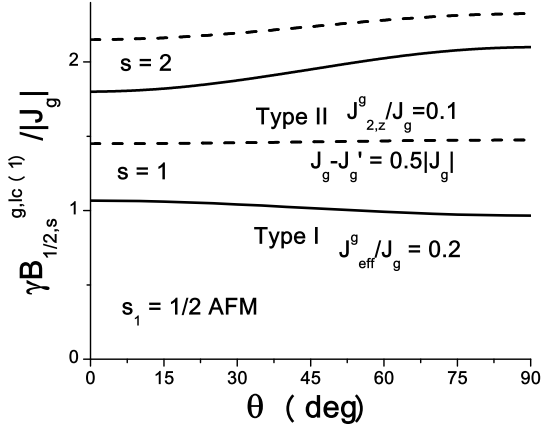}
\caption{Plots of the $s_1=1/2$ first-order level-crossing $\gamma
B_{1/2,s}^{g,\rm lc (1)}(\theta)/|\tilde{J}_g|$ with $J^g_{\rm
eff}/\tilde{J}_g=0.2$, $\tilde{J}_g'-\tilde{J}_g>0$ (solid, Type I)
and $J_{2,z}^g/\tilde{J}_g=0.1$,
 $\tilde{J}_g-\tilde{J}_g'=0.5|\tilde{J}_g|$ (dashed, Type II).}\label{fig4}
\end{figure}

\subsubsection{$s_1\ge1$ AFM level crossings}

 For $s_1>1/2$, single-ion anisotropies are allowed, and the situation
 becomes much more complex than for $s_1=1/2$.  We first consider the simplest
 $s_1>1/2$ case, $s_1=1$, appropriate for AFM Ni$_4$ tetramers.
 Exact expressions for the $s=1,2,3,4$ first-order level-crossing parameters are given  in the Appendix.
 In Figs. 5 and 6, we  plotted the $\theta$-dependence of the
first-order level crossing induction $\gamma B_{1,s}^{g,\rm lc
(1)}(\theta)/|\tilde{J}_g|$ for three special AFM cases of the
general Type I, $\tilde{J}_g'-\tilde{J}_g>0$, and for the
particular Type II example,
$\tilde{J}_g-\tilde{J}_g'=0.5|\tilde{J}_g|$, respectively. In each
curve, we allow only one of the anisotropy interactions (or
effective interactions) to be non-vanishing.  In Fig. 3, the solid
curves are for $J_z^g(\mu_1)/\tilde{J}_g=0.2$ for $g=C_{4h},
D_{4h}, C_{4v}, S_4$, and $D_{2d}$. The dashed curves in Fig. 5
are for $J^g_{\rm eff}(1)/\tilde{J}_g=0.2$ with $g=C_{4h}, D_{4h},
C_{4v}, S_4, D_{2d}$, and $T_d$. For $g=T_d$, these dashed curves
correspond to $J_{f,1}/(24\tilde{J}_g)=0.2$, a huge axial
symmetric exchange anisotropy, but still accessible in our
first-order perturbation due to the small coefficient.  We note
from Fig. 5 that for Type I, the single-ion and symmetric exchange
anisotropies lead to opposite $\theta$-dependencies, both having a
change in sign just before the second level crossing.

In Fig. 6, we illustrate the $s_1=1$ level crossings for Type II,
setting $\tilde{J}_g-\tilde{J}_g'=0.5|\tilde{J}_g|$.  The solid
curves are for $J_z^g(\mu_1^g)/\tilde{J}_g=0.2$ for $g=C_{4h},
D_{4h}, C_{4v}, S_4$, and $D_{2d}$, as in Fig. 5.  Except for
$g=T_d$, Type II does not exhibit a unique effective symmetric
anisotropic exchange interaction, so we plotted the two different
symmetric exchange anisotropy interactions separately. The dashed
and dotted curves are for $J_{1,z}^g(\mu_{12}^g)/\tilde{J}_g=0.4$
and $J_{2,z}^g/\tilde{J}_g=0.4$, respectively, which apply for
$g=C_{4h}, D_{4h}, C_{4v}, S_4,$ and $D_{2d}$.  The dash-dotted
curves correspond to $J_{f,1}/\tilde{J}_g=0.4$ for $g=T_d$.  We
note that for each curve, the Type II isotropic exchange
parameters lead to a larger gap between the $s=2$ and $s=3$ level
crossings.  The sign of the $\theta$-dependencies of the
single-ion (solid) curves changes between $s=2$ and $s=3$.  The NN
symmetric anisotropic exchange interactions vanish for $s=1$, but
increase in magnitude with increasing $s$ for $s=2,3,4$.  The sign
of the $\theta$-dependence of the level crossing due to the NNN
symmetric anisotropic exchange interactions does not change, but
its magnitude increases monotonically.  For $T_d$, there is a sign
change in the $\theta$-dependence of between the $s=3$ and $s=4$
level crossings. Thus, Type II AFM $s_1=1$ tetramers have a richer
set of first-order level-crossing behaviors than do Type I AFM
$s_1=1$ tetramers.

\begin{figure}
\includegraphics[width=0.45\textwidth]{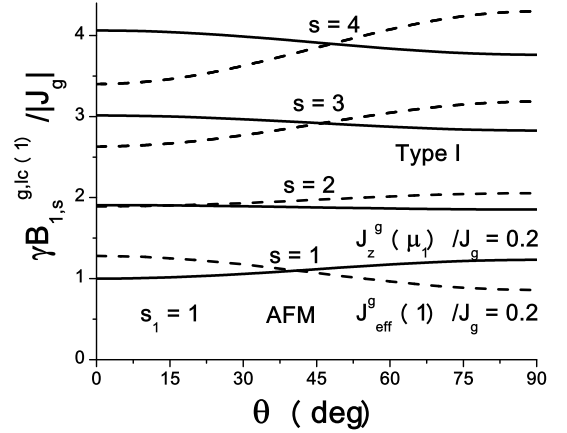}
\caption{Plots of the first-order level-crossing $B_{1,s}^{g,\rm
lc (1)}(\theta)/|\tilde{J}_g|$ for Type I,
$\tilde{J}_g-\tilde{J}_g'<0$ and $s_1=1$. Solid curves:
$J_z^g(\mu_1^g)/\tilde{J}_g=0.2$. Dashed curves: $J^g_{\rm
eff}(1)/\tilde{J}_g=0.2$.}\label{fig5}
\end{figure}

\begin{figure}
\includegraphics[width=0.45\textwidth]{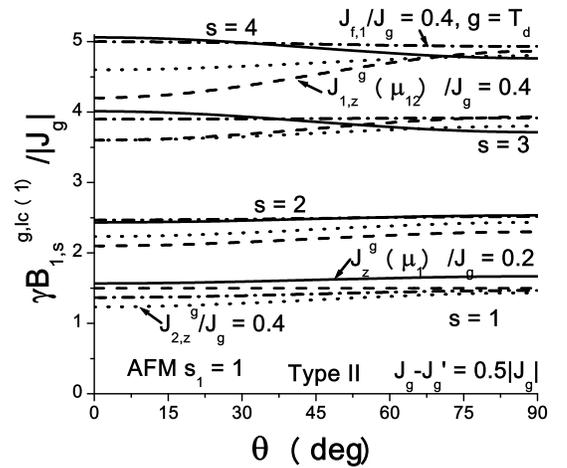}
\caption{Plots of the first-order level-crossing $\gamma
B_{1,s}^{g,\rm lc (1)}(\theta)/|\tilde{J}_g|$ for Type II with
$\tilde{J}_g-\tilde{J}_g'=0.5|\tilde{J}_g|$ and $s_1=1$. Solid
curves: $J_z^g(\mu_1^g)/\tilde{J}_g=0.2$. Dashed curves:
$J_{1,z}^g(\mu_{12}^g)/\tilde{J}_g=0.4$, $g=C_{4h}, D_{4h},
C_{4v}, S_4,D_{2d}$. Dotted curves: $J_{2,z}^g/\tilde{J}_g=0.4$
for $g=C_{4h}, D_{4h}, C_{4v}, S_4,D_{2d}$.
 Dash-dotted curves:  $J_{f,1}/\tilde{J}_g=0.4$ for $g=T_d$.}\label{fig6}
\end{figure}

In Figs. 7 and 8, the analogous Type I and Type II AFM level
crossing inductions are plotted versus $\theta$ for $s_1=3/2$
equal-spin tetramers, such as Cr$_4$.  The notation is the same as
in Figs. 5 and 6.  For Type I $s_1=3/2$ AFM tetramers, the
single-ion and symmetric anisotropic exchange interactions lead to
different $\theta$-dependencies of the level-crossing inductions,
each with a change in sign  in the $\theta$ dependence at about
the second level crossings, as seen in Fig. 7.  For Type II
$s_1=3/2$ AFM tetramers, the sign changes appear between the
second and third level crossings, as shown in Fig. 8.

\begin{figure}
\includegraphics[width=0.45\textwidth]{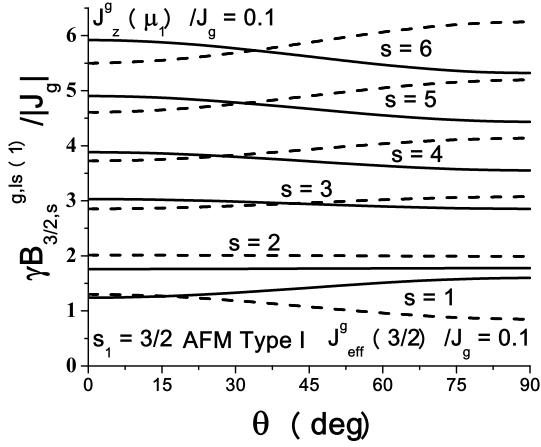}
\caption{Plots of the first-order level-crossing $\gamma
B_{3/2,s}^{g,\rm lc (1)}(\theta)/|\tilde{J}_g|$ for Type I,
$\tilde{J}_g-\tilde{J}_g'<0$ and $s_1=3/2$. Solid curves:
$J_z^g(\mu_1^g)/\tilde{J}_g=0.1$. Dashed curves: $J^g_{\rm
eff}(3/2)/\tilde{J}_g=0.1$.}\label{fig7}
\end{figure}

\begin{figure}
\includegraphics[width=0.45\textwidth]{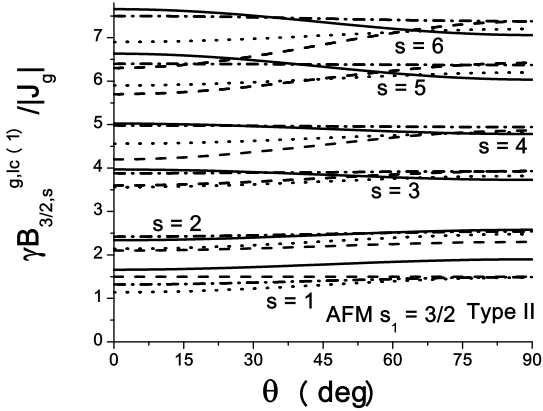}
\caption{Plots of the first-order level-crossing $\gamma
B_{3/2,s}^{g,\rm lc (1)}(\theta)/|\tilde{J}_g|$ for Type II with
$\tilde{J}_g-\tilde{J}_g'=0.5|\tilde{J}_g|$ and $s_1=3/2$. Solid
curves: $J_z^g(\mu_1^g)/\tilde{J}_g=0.2$. Dashed curves:
$J_{1,z}^g(\mu_{12}^g)/\tilde{J}_g=0.4$, $g=C_{4h}, D_{4h},
C_{4v}, S_4, D_{2d}$. Dotted curves: $J_{2,z}^g/\tilde{J}_g=0.4$
for $g=C_{4h}, D_{4h}, C_{4v}, S_4, D_{2d}$.
 Dash-dotted curves:  $J_{f,1}/\tilde{J}_g=0.4$ for $g=T_d$.}\label{fig8}
\end{figure}

\section{XII. The self-consistent Hartree approximation}

The self-consistent Hartree approximation, or strong exchange
limit,\cite{BenciniGatteschi} provides  accurate results for the
${\bm B}$ dependence of the specific heat and magnetization at low
$k_BT/|J|$ and $\gamma B/|J|$ not too small,\cite{ek2} where $k_B$
is Boltzmann's constant. In this approximation,
$E_{\nu}^g=E_{\nu,0}^g+E_{\nu,1}^g$ is given by Eqs. (\ref{E0}) and
(\ref{E1}), respectively. We shall present the self-consistent
Hartree approximation  of four measurable quantities in the
induction representation. We first define the trace in a way
sufficiently general as to encompass the four group symmetries under
consideration,
\begin{eqnarray}
{\rm
Tr}_{\nu}&\equiv&\sum_{\nu}=\sum_{s_{13},s_{24}=0}^{2s_1}\sum_{m=-s}^s\sum_{s=|s_{13}-s_{24}|}^{s_{13}+s_{24}},\label{trace}
\end{eqnarray}
  The partition
function in the self-consistent Hartree approximation may then be
written
 \begin{eqnarray} Z_{g}^{(1)}&=&{\rm Tr}_{\nu}e^{-\beta
E_{\nu}^{g}}, \end{eqnarray} where $\beta=1/(k_BT)$,
 and $E_{\nu}^{g}=E^g_{\nu,0}+E_{\nu,1}^{g,\tilde{\mu}^g}$ is given by Eqs. (\ref{E0}) and (\ref{E1}). In this compact notation, the self-consistent Hartree
magnetization $M_g^{(1)}({\bm B},T)$ and specific heat
$C_{g,V}^{(1)}({\bm B},T)$ are  given by
\begin{eqnarray}
M_g^{(1)}({\bm B},T)&=&\gamma{\rm
Tr}_{\nu}\Bigl(me^{-\beta E_{\nu}^g}\Bigr)/Z_g^{(1)},\\
\frac{C_{g,V}^{(1)}({\bm B},T)}{k_B\beta^2}&=&\frac{{\rm
Tr}_{\nu}\Bigl(\bigl(E_{\nu}^g\bigr)^2e^{-\beta
E_{\nu}^g}\Bigr)}{Z_g^{(1)}}\nonumber\\
& &-\biggl[\frac{{\rm Tr}_{\nu}\Bigl(E_{\nu}^{g} e^{-\beta
E_{\nu}^g}\Bigr)}{Z_g^{(1)}}\biggr]^2.
\end{eqnarray}

We note that there are strong differences between the low-$T$
behavior of FM and AFM tetramers.  We assume
$|\tilde{J}_g|>|\tilde{J}_g'-\tilde{J}_g|$.  For FM tetramers with
$\tilde{J}_g>0$, the low-$T$ thermodynamic behavior is dominated
by the $s=4s_1$, $m=-4s_1$ state, leading to
\begin{eqnarray}M_g^{(1)}({\bm
B},T)&{{\approx}\atop{T\rightarrow0}}&\gamma\hat{\bm B}
B_{4s_1}(\beta\gamma B),
\end{eqnarray} where $B_S(x)$ is the Brillouin function.  The universality
of this function renders thermodynamic studies useless for the
determination of the microscopic parameters.  For AFM tetramers with
$\tilde{J}_g<0$, however, there will be interesting level-crossing
effects, which can be employed to measure the microscopic
interaction parameters, as discussed in detail in Sec. V.  As for
dimers, $C_V(B,T)$ for AFM tetramers at sufficiently low $T$ exhibit
$2s_1$ central minima at the level-crossing inductions
$B_{s_1,s}^{g,{\rm lc}}(\theta)\approx B_{s_1,s}^{g,{\rm
lc}(1)}(\theta)$ that vanish as $T\rightarrow0$, equally surrounded
by peaks of equal height.\cite{ek2}  As for the magnetization,
$C_V(B,T)$ for FM tetramers at low $T$ reduces to that of a monomer
with spin $4s_1$, yielding a rather uninteresting Schottky anomaly.

However, the microscopic nature of FM tetramers can be better probed
either by EPR or INS techniques. The self-consistent Hartree EPR
absorption ${\Im}\chi_{-\sigma,\sigma}^{g,(1)}({\bm B},\omega)$ for
clockwise ($\sigma=1$) or counterclockwise ($\sigma=-1$) circularly
polarized oscillatory fields normal to ${\bm B}$  is
\begin{eqnarray}
{\Im}\chi^{g,(1)}_{-\sigma,\sigma}&=&\frac{\gamma^2}{Z_g^{(1)}}{\rm
Tr}_{\nu}{\rm Tr}_{\nu'} e^{-\beta E_{\nu}^g}\bigl|M_{\nu,\nu'}\bigr|^2\nonumber\\
&
&\times\bigl[\delta(E_{\nu}^g-E_{\nu'}^g+\omega)-\delta(E_{\nu'}^g-E_{\nu}^g+\omega)\bigr],\nonumber\\
\end{eqnarray}
where  $M_{\nu,\nu'}=A_s^{\sigma
m}\delta_{m',m+\sigma}\delta_{s',s}\delta_{s_{13}',s^{}_{13}}\delta_{s_{24}',s^{}_{24}}$
and ${\rm Tr}_{\nu'}=\sum_{\nu'}$. The strong resonant inductions
for  appear at
\begin{eqnarray}
\gamma B^{g,(1)}_{\rm
res}&=&\pm\omega+\frac{(2m+\sigma)}{2}(1-3\cos^2\theta)\tilde{J}_{z}^{g,\overline{\nu}}(\tilde{\mu}^g),\label{Bres}\nonumber\\
\end{eqnarray}
where $\tilde{J}_{z}^{g,\overline{\nu}}(\tilde{\mu}^g)$ is given by
Eq. (\ref{Jztildemu}).  We note that
$\tilde{J}_{z}^{g,\overline{\nu}}(\tilde{\mu}^g)$ contains the three
effective microscopic interactions, $J_z^g(\mu_1^g)$,
$J_{1,z}^g(\mu_{12}^g)$, and $J_{2,z}$, multiplied by the constants
$a_{\overline{\nu}}^{+}$,  $-c_{\overline{\nu}}^{-}$, and
$-a_{\overline{\nu}}^{-}/2$. Thus, for an experimental determination
of all three microscopic parameters, it is necessary to perform EPR
experiments in three different quantum state manifolds.  At low $T$,
the Boltzmann factor favors the ground state manifold, so to obtain
the desired measurements of three manifolds, one needs to access the
excited $s=4s_1-1$ states, as well.

As shown in the Appendix, for a FM tetramer with $\tilde{J}_g>0$ in
the lowest energy manifold of states,  nominally
$(s,s_{13},s_{24})=(4s_1,2s_1,2s_1)$, has the highest population at
very low $T$.  In this ground state manifold with $-4s_1\le m\le
4s_1$, the constants reduce to
\begin{eqnarray}
a_{\overline{\nu}}^{\pm}\rightarrow\frac{s-1\mp1}{2(2s-1)},\\
c_{\overline{\nu}}^{-}\rightarrow\frac{s-1}{2(2s-1)},
\end{eqnarray}
where $s=4s_1$. We note, of course, that when the single-ion and
anisotropic exchange interactions are sufficiently large, $s$ is not
really a good quantum number, as the  actual ground state, while
mostly the nominal  $s=4s_1$, contains a mixture of $s=4s_1-2$
states, to leading order in those interactions.\cite{ek2} Additional
mixtures are likely to be relevant for the excited states, as well.

However, assuming for simplicity that we can safely neglect such
mixing, and take $(s,s_{13},s_{24})$ to be a good quantum number
set, there is still some ambiguity in the determination of the order
of the lowest manifolds of excited states.  From Eq. (\ref{E0}), it
is evident that if $\tilde{J}_g>0$ and $\tilde{J}_g'=\tilde{J}_g>0$,
the states $(s,s_{13},s_{24}=(s,2s_1,2s_1)$ are energetically
favored for each $s$ value. We denote this a Type I FM tetramer.
However, for a Type II FM tetramer with $\tilde{J}_g>0$ and
$\tilde{J}g'-\tilde{J}_g<0$, the states with minimal allowed values
of $s_{13},s_{24}$ are favored.  Thus, for even $s$, these are
$(s,s/2,s/2)$, and for odd $s$ are $(s,(s+1)/2,(s-1)/2)$.

Thus, there are  two inequivalent lowest energy excited state
manifolds with $s=4s_1-1$ and $-4s_1\le m\le4s_1-1$ of excited
states with $s=4s_1-1$. For $|\tilde{J}_g-\tilde{J}_g'|\ll
\tilde{J}_g$, these are ordinarily both accessible.  One of these is
non-degenerate (at fixed $m$) with
$(s,s_{13},s_{24})=(4s_1-1,2s_1,2s_1)$ and the other is doubly
degenerate with $(s,s_{13},s_{24})=(4s_1-1,2s_1-1,2s_1),
(4s_1-1,2s_1,2s_1-1)$. In Type I FM tetramers,  the
$(4s_1-1,2s_1,2s_1)$ states have lower energy than the next excited
manifold with $(4s_1-1,2s_1-1,2s_1), (4s_1-1,2s_1,2s_1-1)$ states.
In Type II FM tetramers, the reverse order occurs.  From the results
presented in the Appendix, for the $(4s_1-1,2s_1-1,2s_1),
(4s_1-1,2s_1,2s_1-1)$  manifold of states with $-4s_1+1\le m\le
4s_1-1$, the constants are
\begin{eqnarray}
a_{\overline{\nu}}^{\pm}&\rightarrow&\frac{(s^2-s+1)\mp(2s-1)}{2s(2s-1)},\\
c_{\overline{\nu}}^{-}&\rightarrow&\frac{(s+1)(s-1)^2}{2s^2(2s-1)},
\end{eqnarray}
where $s=4s_1-1$. For the $(4s_1-1,2s_1,2s_1)$ manifold of states,
also with $-4s_1+1\le m\le 4s_1-1$, the constants are
\begin{eqnarray}
a_{\overline{\nu}}^{\pm}&\rightarrow&\frac{(s-1)(s\mp1)}{s(2s-1)}, \\
c_{\overline{\nu}}^{-}&\rightarrow&\frac{(s-2)}{2(2s-1)},
\end{eqnarray}
where again $s=4s_1-1$.  We note that these three sets of constants
are all distinct.  Thus, either by operating at higher temperature
or by some special pumping technique, it is possible to probe these
low-lying state manifolds, and thereby obtain a full measurement of
the first-order microscopic parameters.

The Hartree INS cross-section $S_g^{(1)}({\bm B},{\bm q},\omega)$
is
\begin{eqnarray}
S_g^{(1)}&=&{\rm Tr}_{\nu}{\rm Tr}_{\nu'}e^{-\beta E^g_{\nu}}\sum_{\tilde{\alpha},\tilde{\beta}}\bigl(\delta_{\tilde{\alpha},\tilde{\beta}}-\hat{q}_{\tilde{\alpha}}\hat{q}_{\tilde{\beta}}\bigr)\sum_{n,n'}\nonumber\\
& &\times e^{i{\bm q}\cdot({\bm r}_n-{\bm
r}_{n'})}\langle\nu|S_{n',\tilde{\alpha}}^{\dag}|\nu'\rangle\langle\nu'|S_{n,\tilde{\beta}}|\nu\rangle\nonumber\\
& &\qquad\times\delta(\omega+E^g_{\nu}-E^g_{\nu'}),
\end{eqnarray}
where $\tilde{\alpha},\tilde{\beta}=\tilde{x},\tilde{y},\tilde{z}$,
$\hat{q}_{\tilde{x}}=\sin\theta_{b,q}\cos\phi_{b,q}$,
$\hat{q}_{\tilde{y}}=\sin\theta_{b,q}\sin\phi_{b,q}$, and
$\hat{q}_{\tilde{z}}=\cos\theta_{b,q}$,  $\theta_{b,q}$ and
$\phi_{b,q}$ describe the  relative orientations of ${\bm B}$ and
${\bm q}$,\cite{ek2}  the ${\bm r}_n$ and
$\langle\nu'|S_{n,\tilde{\alpha}}|\nu\rangle$ are given by Eqs.
(\ref{rn}), (\ref{Mz}), and (\ref{Msigma}) respectively, and the
scalar ${\bm q}\cdot({\bm r}_n-{\bm r}_{n'})$ is invariant under the
rotation, Eq. (\ref{rotation}). After some algebra, we rewrite
$S_g^{(1)}({\bm q},\omega)$ as
\begin{eqnarray}
S_g^{(1)}&=&{\rm Tr}^g_{\nu}e^{-\beta
E^g_{\nu}}\sum_{\nu'}\delta(\omega+E_{\nu}^g-E_{\nu'}^g)\nonumber\\
& &\times\Bigl(\sin^2\theta_{b,q}L_{\nu,\nu'}({\bm
q})+\frac{(2-\sin^2\theta_{b,q})}{4}M_{\nu,\nu'}({\bm
q})\Bigr),\label{Sofqomega}\nonumber\\
\end{eqnarray}
where the Hartree functions $L_{\nu,\nu'}({\bm q})$ and
$M_{\nu,\nu'}({\bm q})$ are given in the Appendix. They are
independent of  ${\bm B}$.  Since $E_{\nu}^g$ is well-behaved as
${\bm B}\rightarrow0$, Eq. (\ref{Sofqomega}) is accurate for all
${\bm B}$.

 As for the dimer,\cite{ek2} additional EPR and INS transitions
with amplitudes higher order in the anisotropy parameters
$J_z^g(\mu_1^g)$, $J_{1,z}^g(\mu_{12}^g)$, and $J_{2,z}^g$
relative to $\tilde{J}_g$ are obtained in the extended Hartree
approximation, but will be presented elsewhere for
brevity.\cite{ekfuture}

\section{XIII. Discussion}

We note that the phenomenological total spin anisotropy model
widely used in fitting experimental data on SMM's is
\begin{eqnarray}
{\cal H}_p&=&-J_bS_z^2-J_d(S_x^2-S_y^2),\label{Hph}
\end{eqnarray}
containing axial and azimuthal contributions,
respectively,\cite{general} sometimes with additional quartic
terms.\cite{Hendrickson,Fe4} The terms are generally defined
relative to the total spin principal axes, which for equal spin,
high symmetry systems are the molecular axis vectors.  It is easy
to evaluate $E^p_{\nu}=\langle\nu|\tilde{\cal H}_p|\nu\rangle$ in
the induction representation. One obtains Eq. (\ref{E1}) with
\begin{eqnarray}
\tilde{J}_z^{g,\overline{\nu}}&\rightarrow& J_b,\\
J_d&=&0,\\
\delta\tilde{J}_z^{g,\overline{\nu}}&\rightarrow&0.
\end{eqnarray}
A non-vanishing $J_d$ would lead to a term in Eq. (\ref{E1})
proportional to $\sin^2\theta\cos(2\phi)$, as for the
dimer,\cite{ek2} which does not arise in the first-order
calculation for the high-spin tetramers under consideration, based
upon the microscopic parameters alone. Hence, $J_b$ alone
describes the $\theta$ and $m$ dependencies of $E_{\nu,1}^g$
correctly.

 In
addition, for the lowest energy states of FM tetramers, we have
$s=4s_1$, $s_{13}=s_{24}=2s_1$, so $\overline{\nu}$ is restricted to
a single set of values.  In this high-spin  case,  $J_b$ can provide
a correct phenomenology of the $s=4s_1$ ground state energy
manifold. However, if applied to the two low-lying excited state
manifolds with $s=4s_1-1$, for instance, one would infer two
different $J_b$ values from that obtained in the ground state
manifold. For AFM tetramers, the phenomenological model also
correctly provides a vanishing first-order correction to the $s=0$
manifold of states with
 $s_{13}=s_{24}=0,1,\ldots, 2s_1$. However, it also
has  problems describing the first excited  manifold of AFM states
with $s=1$, because it leads to the unphysical choice of $\delta
J_z^{g,\overline{\nu}}(\tilde{\mu})=0$ (or a constant, independent
of the quantum numbers $\overline{\nu}$). Hence, the
phenomenological model works best in describing only a single
manifold of states with fixed $(s,s_{13},s_{24})$.  This is more
restrictive than the usual assumption of its applicability to a
fixed set of $s$ states.\cite{Ni4,Fe4,CorniaFe4,Fe4spin5}

We note that for the FM Fe$_4$ SMM with $D_3$ symmetry,
Fe$_4$(thme)$_2$(dpm)$_6$, where H$_3$thme is
1,1,1-tris(hydroxymethyl)ethane and Hdpm is
dipivaloylmethane,\cite{Fe4,CorniaFe4} the $D_3$ symmetry also
precludes the $J_d$ term.  Nevertheless, in fits to INS data, it
was necessary to assume $J_d\ne0$,\cite{Fe4,Fe4spin5} to obtain
the appropriate anticrossing gaps, so that either the powdered
sample did not have pure $D_3$ symmetry, or the phenomenological
model they used, Eq. (\ref{Hph}) plus two quartic terms obeying
$D_3$ symmetry, was not appropriate.  Since single-ion
interactions appeared to be important,\cite{Fe4} the total spin
might not have been a well-defined quantum number, as in at least
one Fe$_2$ dimer and in Fe$_8$.\cite{ek2,Shapira,Dalal,Fe8} In
addition, second order effects might also provide a possible
explanation.\cite{ekfuture} To investigate this symmetry, it would
be necessary to evaluate the single-ion matrix elements for that
symmetry in compact form.\cite{ekfuture}

For simplicity, we have neglected  biquadratic exchange and higher
order anisotropic interactions.  Of these, those of the simplest
form are the isotropic biquadratic exchange interactions, which for
the symmetries under consideration take the form
\begin{eqnarray}
{\cal H}_{bq}^g&=&-J_{2,g}\sum_{n=1}^4({\bm S}_n\cdot{\bm
S}_{n+1})^2-J_{2,g}'\sum_{n=1}^2({\bm S}_n\cdot{\bm
S}_{n+2})^2,\nonumber\\
\end{eqnarray}
where $J_{2,g}'\ne J_{2,g}$ for $g=C_{4h},D_{4h},C_{4v},S_4$, and
$D_{2d}$, $J_{2,T_d}'=J_{2,T_d}$. The $J_{2,g}'$ term is diagonal
in the the $|\nu\rangle$ representation, but the remaining term
proportional to $J_{2,g}$ is not. However, it is important to note
that these isotropic interactions are rotationally invariant, so
they are independent of $\theta$ in the induction representation.
Hence, they  modify the positions but not  the
$\theta$-dependencies of the  AFM level-crossing inductions.
Although the diagonal $J_{2,g}'$ term does not affect the EPR
transitions, the non-diagonal $J_{2,g}$ term will modify the
positions but not the $\theta$-dependencies of the EPR resonant
inductions.

We note that our formulation of the single-ion matrix elements in
terms of a pair of dimers is applicable to low-symmetry systems
such as
Mo$_{12}$O$_{30}(\mu_2$-OH)$_{10}$H$_2$\{Ni(H$_2$O)$_3\}_4$,
abbreviated as $\{$Ni$_4$Mo$_{12}\}$,\cite{Schnack}
 systems such as  Ni$_4$
tetramers obtained from salts of
[Ni$_4$(H$_2$O)$_2$(PW$_9$O$_{34}$)$_2]^{10-}$ with $C_{2v}$
symmetry,\cite{Ni4C2v} and the unequal-spin systems
Mn$_2^{II}$Mn$_2^{III}$ and
Ni$_2^{II}$Mn$_2^{III}$.\cite{Lecren,Ni2Mn2} In the first system
with $C_{1v}$ symmetry, one would expect many more single-ion,
symmetric anisotropic exchange, and DM interactions, making
definitive fits to the existing powder magnetization data
problematic.\cite{Schnack} However, to improve the fits to the
$C_{3v}$ or $D_3$ symmetry systems, such as the Fe$_4$ compound
Fe$_4$(thme)$_2$(dpm)$_6$ and the Cr$^{III}$Ni$_3^{II}$ tetramer
with an $s=9/2$ ground state,\cite{Fe4,CrNi3} would require a
reformulation of the single-ion matrix elements as a trimer plus a
monomer.\cite{ekfuture}

\section{Summary}
We presented a microscopic theory of high-symmetry single molecule
magnets, including a compact form for the exact single-spin matrix
elements for four general spins.  We used the local axial and
azimuthal vector groups to construct the  invariant single-ion and
symmetric anisotropic exchange  Hamiltonians, and the molecular
representation to obtain the Dzyaloshinskii-Moriya interactions,
for equal-spin $s_1$ tetramers with site point group symmetries
$T_d$, $D_{4h}$, $D_{2d}$, $S_4$, $C_{4h}$,  or $C_{4v}$. Each
vector group introduces site-dependent molecular single-ion and
anisotropic exchange interactions. Assuming weak effective
site-independent single-ion and exchange anisotropy interactions,
we evaluated the first-order corrections to the eigenstate
energies, and provided analytic results and illustrations of the
antiferromagnetic level-crossing inductions. We also provided
Hartree expressions for the magnetization, specific heat, EPR
absorption, and INS cross-section, which are accurate at low
temperatures and arbitrary magnetic fields. For ferromagnetic
tetramers, we provided a procedure for a precise EPR determination
of the microscopic anisotropy parameters.  Our procedure is
extendable to more general systems.

We thank N. S. Dalal for helpful comments.  This work was
supported by the NSF under contract NER-0304665.

\section{Appendix}

\subsection{1. Site-dependent single-ion interactions}
For $g=D_{4h}$, the only non-vanishing site-dependent single-ion
interaction is
\begin{eqnarray}
K_{xy}^{D_{4h}}(n,\mu_1^{D_{4h}})&=&(-1)^{n+1}J_e.
\end{eqnarray}
For $g=C_{4h}$, the two non-vanishing site-dependent single-ion
interactions are
\begin{eqnarray}
J_{xy}^{C_{4h}}(\mu_1^{C_{4h}})&=&J_e\cos(2\chi_1^{C_{4h}}),\\
K_{xy}^{C_{4h}}(n,\mu_1^{C_{4h}})&=&(-1)^{n+1}J_e\sin(2\chi_1^{C_{4h}}).
\end{eqnarray}
For $g=C_{4v}$,  the non-vanishing site-dependent single-ion
interactions in Eq. (\ref{Hsimolecular}) are
\begin{eqnarray}
K_{xy}^{C_{4v}}(n,\mu_1^{C_{4v}})&=&\frac{(-1)^{n+1}}{2}\Bigl(J_a\sin^2\theta_1^{C_{4v}}\nonumber\\
& &-J_e(1+\cos^2\theta_1^{C_{4v}})\Bigr),\\
K_{xz}^{C_{4v}}(n,\mu_1^{C_{4v}})&=&-\frac{\gamma_n^{+}}{2\sqrt{2}}\sin(2\theta_1^{C_{4v}})(J_a-J_e),\\
K_{yz}^{C_{4v}}(n,\mu_1^{C_{4v}})&=&-\frac{\gamma_n^{-}}{2\sqrt{2}}\sin(2\theta_1^{C_{4v}})(J_a+J_e).
\end{eqnarray}

For $S_4$, the single-ion site-dependent interactions are
\begin{eqnarray}
J_{xy}^{S_4}(\mu_1^{S_4})&=&J_1(\mu_1^{S_4})\cos(2\phi_1^{S_4})\nonumber\\
& &\qquad+J_2(\mu_1^{S_4})\sin(2\phi_1^{S_4}) ,\label{JxyofmuS4}\\
K_{xy}^{S_4}(n,\mu_1^{S_4})&=&(-1)^n\Bigl(J_1(\mu_1^{S_4})\sin(2\phi_1^{S_4})\nonumber\\
& &\qquad-J_2(\mu_1^{S_4})\cos(2\phi_1^{S_4})\Bigr),\\
K_{xz}^{S_4}(n,\mu_1^{S_4})&=&(-1)^{n+1}\Bigl(J_3(\mu_1^{S_4})\delta_n^{-}(\phi_1^{S_4})\nonumber\\
& &\qquad-J_4(\mu_1^{S_4})\delta_n^{+}(\phi_1^{S_4})\Bigr),\\
K_{yz}^{S_4}(n,\mu_1^{S_4})&=&(-1)^{n+1}\Bigl(J_3(\mu_1^{S_4})\delta_n^{+}(\phi_1^{S_4})\nonumber\\
& &\qquad+J_4(\mu_1^{S_4})\delta_n^{-}(\phi_1^{S_4})\Bigr),\\
J_1(\mu_1^{S_4})&=&\frac{1}{2}\Bigl(J_a\sin^2\theta_1^{S_4}\nonumber\\
& &-J_e(1+\cos^2\theta_1^{S_4})\cos(2\psi_1^{S_4})\Bigr),\\
J_2(\mu_1^{S_4})&=&J_e\cos\theta_1^{S_4}\sin(2\psi_1^{S_4}),\\
J_3(\mu_1^{S_4})&=&\frac{1}{2}\sin(2\theta_1^{S_4})[J_a+J_e\cos(2\psi_1^{S_4})],\\
J_4(\mu_1^{S_4})&=&J_e\sin\theta_1^{S_4}\sin(2\psi_1^{S_4}).
\end{eqnarray}

 For $D_{2d}$, the non-vanishing site-dependent single-ion interactions are
\begin{eqnarray}
K_{xy}^{D_{2d}}(n,\mu_1^{D_{2d}})&=&\frac{(-1)^{n+1}}{2}\Bigl(J_a\sin^2\theta_1^{D_{2d}}\nonumber\\
& &+J_e(1+\cos^2\theta_1^{D_{2d}})\Bigr),\label{KxyD2d}\\
K_{xz}^{D_{2d}}(n,\mu_1^{D_{2d}})&=&\frac{\gamma_n^{-}}{2\sqrt{2}}(J_a-J_e)\sin(2\theta_1^{D_{2d}}),\label{KxzD2d}\\
K_{yz}^{D_{2d}}(n,\mu_1^{D_{2d}})&=&\frac{-\gamma_n^{+}}{2\sqrt{2}}(J_a-J_e)\sin(2\theta_1^{D_{2d}}).\label{KyzD2d}
\end{eqnarray}
The site-dependent interactions for $T_d$ symmetry are easily
obtained from those Eqs. (\ref{KxyD2d})-(\ref{KyzD2d}) by setting
$\theta_1^{D_{2d}}\rightarrow\tan^{-1}(\sqrt{2})$ and
$J_e\rightarrow0$.

\subsection{2. Site-dependent symmetric anisotropic exchange interactions}

For $g=C_{4h}$, the non-vanishing site-dependent symmetric
anisotropic exchange interactions in Eq. (\ref{Haemolecular}) are
\begin{eqnarray}
J_{m,xy}^{C_{4h}}(\mu_{1,m+1}^{C_{4h}})&=&J_{c,m}\cos(2\chi_{1,m+1}^{C_{4h}})
\end{eqnarray}
for $m=1,2$.  For $D_{4h}$, the non-vanishing site-dependent
symmetric anisotropic exchange interactions are
\begin{eqnarray}
J_{1,xy}^{D_{4h}}(\mu_{12}^{D_{4h}})&=&J_{c,1},\\
K_{2,xy}^{D_{4h}}(\mu_{13}^{D_{4h}})&=&-J_{c,2}.
\end{eqnarray}

 For $C_{4v}$, the non-vanishing site-dependent anisotropic exchange
interactions  are
\begin{eqnarray}
J_{1,xy}^{C_{4v}}(\mu_{12}^{C_{4v}})&=&\frac{1}{2}[J_{f,1}-J_{c,1}\cos(2\psi_{12}^{C_{4v}})],\label{J1xyC4v}\\
K_{2,xy}^{C_{4v}}(\mu_{13}^{C_{4v}})&=&J_{2,-},\label{J2xyC4v}\\
 K_{1,xz}^{C_{4v}}(n,\mu_{12}^{C_{4v}})&=&-\frac{1}{2}(\gamma_n^{+}-\gamma_n^{-})J_{c,1}\sin(2\psi_{12}^{C_{4v}}),\label{K1xzC4v}\\
 K_{1,yz}^{C_{4v}}(n,\mu_{12}^{C_{4v}})&=&\frac{1}{2}(\gamma_n^{+}+\gamma_n^{-})J_{c,1}\sin(2\psi_{12}^{C_{4v}}).\label{K1yzC4v}
\end{eqnarray}

Again, the more interesting group is $g=S_4$.  We find
\begin{eqnarray}
J_{1,xy}^{S_4}(\mu_{12}^{S_4})&=&-\tilde{J}_1(\mu_{12}^{S_4})\cos(2\phi_{12}^{S_4})\nonumber\\
& &+\tilde{J}_2(\mu_{12}^{S_4})\sin(2\phi_{12}^{S_4}),\label{J1xyS4}\\
J_{2,xy}^{S_4}(\mu_{13}^{S_4})&=&-J_{2,-}\sin(2\phi_{13}^{S_4}),\label{J2xyS4}\\
K_{2,xy}^{S_4}(\mu_{13}^{S_4})&=&J_{c,-}\cos(2\phi_{13}^{S_4}),\label{K2xyS4}\\
K_{1,xy}^{S_4}(n,\mu_{12}^{S_4})&=&(-1)^{n+1}\Bigl(\tilde{J}_1(\mu_{12}^{S_4})\sin(2\phi_{12}^{S_4})\nonumber\\
& &\qquad+\tilde{J}_2(\mu_{12}^{S_4})\cos(2\phi_{12}^{S_4})\Bigr),\label{K1xyS4}\\
K_{1,xz}^{S_4}(n,\mu_{12}^{S_4})&=&(-1)^{n}\Bigl(\tilde{J}_3(\mu_{12}^{S_4})\delta_n^{-}(\phi_{12}^{S_4})\nonumber\\
& &\qquad-\tilde{J}_4(\mu_{12}^{S_4})\delta_n^{+}(\phi_{12}^{S_4})\Bigr),\label{K1xzS4}\\
K_{1,yz}^{S_4}(n,\mu_{12}^{S_4})&=&(-1)^n\Bigl(\tilde{J}_3(\mu_{12}^{S_4})\delta_n^{+}(\phi_{12}^{S_4})\nonumber\\
&
&\qquad+\tilde{J}_4(\mu_{12}^{S_4})\delta_n^{-}(\phi_{12}^{S_4})\Bigr),\label{K1xyS4}
\end{eqnarray}
where
\begin{eqnarray}
\tilde{J}_1(\mu_{12}^{S_4})&=&\frac{1}{2}\Bigl(J_{f,1}\sin^2\theta_{12}^{S_4}\nonumber\\
& &\qquad-J_{c,1}(1+\cos^2\theta_{12}^{S_4})\cos(2\psi_{12}^{S_4})\Bigr),\label{tildeJ1}\\
 \tilde{J}_2(\mu_{12}^{S_4})&=&J_{c,1}\cos\theta_{12}^{S_4}\sin(2\psi_{12}^{S_4}),\label{tildeJ2}\\
\tilde{J}_3(\mu_{12}^{S_4})&=&\frac{1}{2}\sin(2\theta_{12}^{S_4})[J_{f,1}+J_{c,1}\cos(2\psi_{12}^{S_4})],\label{tildeJ3}\\
\tilde{J}_4(\mu_{12}^{S_4})&=&J_{c,1}\sin\theta_{12}^{S_4}\sin(2\psi_{12}^{S_4}),\label{tildeJ4}
\end{eqnarray}
and the  $\delta_n^{\pm}(\phi)$ are given by Eqs. (\ref{deltanp})
and (\ref{deltanm}), and listed in Table II.

For $g=D_{2d}$, the non-vanishing site-dependent anisotropic
exchange interactions are
\begin{eqnarray}
J_{1,xy}^{D_{2d}}(\mu_{12}^{D_{2d}})&=&J_{1,+}+J_{1,-}\cos^2\theta_{12}^{D_{2d}},\label{J1xyD2d}\\
K_{2,xy}^{D_{2d}}(\mu_{13}^{D_{2d}})&=&-J_{c,2},\label{J2xyD2d}\\
K_{1,xz}^{D_{2d}}(n,\mu_{12}^{D_{2d}})&=&-(\gamma_n^{+}+\gamma_n^{-})\frac{J_{1,-}}{4}\sin(2\theta_{12}^{D_{2d}}),\nonumber\\
& &\\
K_{1,yz}^{D_{2d}}(n,\mu_{12}^{D_{2d}})&=&(\gamma_n^{+}-\gamma_n^{-})\frac{J_{1,-}}{4}\sin(2\theta_{12}^{D_{2d}}).\nonumber\\
\end{eqnarray}
For $g=T_d$, the non-vanishing site-dependent anisotropic exchange
interactions reduce to
\begin{eqnarray}
J_{1,xy}^{T_d}&=&\frac{J_{f,1}}{4},\\
K_{2,xy}^{T_d}&=&-\frac{J_{f,1}}{2}.
\end{eqnarray}

\subsection{3. Compact single-ion matrix elements}

By using the Schwinger boson technique of representing a spin by two
non-interacting bosons, and checking our results using the standard
Clebsch-Gordan algebra with the assistance of symbolic manipulation
software,
 we find  the  single-spin
matrix elements  with  general $\{s_n\}=(s_1,s_2,s_3,s_4)$ to be
\begin{eqnarray}
\langle\nu'|S_{n,\tilde{z}}|\nu\rangle
&=&\delta_{m',m}\biggl(m\delta_{s',s}\Gamma_{s^{}_{13},s_{13}',s^{}_{24},s_{24}'}^{\{s_n\},s,n}\nonumber\\
&
&+\delta_{s',s+1}C_{-s-1}^m\Delta_{s^{}_{13},s_{13}',s^{}_{24},s_{24}'}^{\{s_n\},-s-1,n}\nonumber\\
& &+\delta_{s',s-1}C_s^m\Delta_{s^{}_{13},s_{13}',s^{}_{24},s_{24}'}^{\{s_n\},s,n}\biggr),\label{Mz}\\
\langle\nu'|S_{n,\tilde{\sigma}}|\nu\rangle&=&\delta_{m',m+\tilde{\sigma}}\biggl(A_s^{\tilde{\sigma}
m}\delta_{s',s}
\Gamma_{s^{}_{13},s_{13}',s^{}_{24},s_{24}'}^{\{s_n\},s,n}\nonumber\\
&
&-\delta_{s',s+1}D_{-s-1}^{\tilde{\sigma},m}\Delta_{s^{}_{13},s_{13}',s^{}_{24},s_{24}'}^{\{s_n\},-s-1,n}\nonumber\\
& &+\delta_{s',s-1}D_s^{\tilde{\sigma},m}\Delta_{s^{}_{13},s_{13}',s^{}_{24},s_{24}'}^{\{s_n\},s,n}\biggr),\label{Msigma}\\
C_s^m&=&\sqrt{s^2-m^2},\label{Csm}\\
D_s^{\tilde{\sigma},m}&=&\tilde{\sigma}\sqrt{(s-\tilde{\sigma}m)(s-\tilde{\sigma}m-1)},\label{Dsigmasm}\\
\Gamma_{s_{13},s_{13}',s_{24},s_{24}'}^{\{s_n\},s,n}&=&\delta_{s_{24}',s^{}_{24}}\epsilon_n^{-}\alpha_{s_1,s_3}^{s^{}_{24},s,n}(s^{}_{13},s_{13}')\nonumber\\
&
&+\delta_{s_{13}',s^{}_{13}}\epsilon_n^{+}\alpha_{s_2,s_4}^{s_{13},s,n}(s_{24},s_{24}'),\\
\Delta_{s^{}_{13},s_{13}',s^{}_{24},s_{24}'}^{\{s_n\},s,n}&=&\delta_{s_{24}',s^{}_{24}}\epsilon_n^{-}\beta_{s_1,s_3}^{s_{24},s,n}(s^{}_{13},s_{13}')\nonumber\\
&
&+\delta_{s_{13}',s^{}_{13}}\epsilon_n^{+}\beta_{s_2,s_4}^{s^{}_{13},s,n}(s^{}_{24},s_{24}'),\\
\alpha_{s_1,s_3}^{s^{}_{24},s,n}(s^{}_{13},s_{13}')&=&\frac{1}{4}(1+\xi_{s,s^{}_{13},s^{}_{24}})\delta_{s_{13}',s^{}_{13}}\nonumber\\
&
&+\gamma_n^{+}\Bigl(F^{s_{13},s_{24}}_{s_1,s_3,s}\delta_{s_{13}',s^{}_{13}-1}\nonumber\\
& &+F^{s_{13}+1,s_{24}}_{s_1,s_3,s}\delta_{s_{13}',s_{13}+1}\Bigr),\\
\beta_{s_1,s_3}^{s^{}_{24},s,n}(s_{13},s_{13}')&=&-\frac{(-1)^n}{4}\eta_{s,s_{13},s_{24}}\delta_{s_{13}',s^{}_{13}}\nonumber\\
&
&-\gamma_n^{+}\Bigl(G_{s_1,s_3,s}^{s_{13},s_{24}}\delta_{s_{13}',s^{}_{13}-1}\nonumber\\
&
&+G_{s_1,s_3,-s}^{s_{13}+1,s_{24}}\delta_{s_{13}',s^{}_{13}+1}\Bigr),\\
F_{s_1,s_3,s}^{s_{13},s_{24}}&=&\frac{\eta_{s_{13},s_1,s_3}A_{s+s_{13}}^{s_{24}}A_{s_{24}}^{s-s_{13}}}{4s(s+1)},\label{A}\nonumber\\
& &\\
G_{s_1,s_3,s}^{s_{13},s_{24}}&=&\frac{\eta_{s_{13},s_1,s_3}A_{s+s_{13}}^{s_{24}}A_{s+s_{13}-1}^{s_{24}}}{4s\sqrt{4s^2-1}},\label{B}\nonumber\\
& &\\
\eta_{z,x,y}&=&\frac{A_{x+z}^yA_y^{x-z}}{\sqrt{z^2(4z^2-1)}},\label{eta}\\
 \xi_{z,x,y}&=&\frac{x(x+1)-y(y+1)}{z(z+1)},\label{xi}
\end{eqnarray}
where $\gamma_n^{+}$ and $A_s^m$ are given by Eqs. (\ref{gamman})
and (\ref{Asm}).  The prefactors $m$, $A_s^{\tilde{\sigma} m}$,
$C_s^m$, $C_{-s-1}^m$, $D_s^{\tilde{\sigma},m}$, and
$D_{-s-1}^{\tilde{\sigma},m}$ are consequences of the
Wigner-Eckart theorem for a vector operator.\cite{Tinkham} The
challenge was to obtain the coefficients
$\Gamma_{s^{}_{13},s_{13}',s^{}_{24},s_{24}'}^{\{s_n\},s,n}$ and
$\Delta_{s^{}_{13},s_{13}',s^{}_{24},s_{24}'}^{\{s_n\},s,n}$.
Their hierarchical structure based upon  the unequal-spin dimer
suggests that analogous coefficients  with $n
> 4$ may be obtainable.\cite{ek2} Details will be presented
elsewhere.\cite{ekfuture}

\subsection{4. First-order eigenstate energy constants}

The constants appearing in the first-order eigenstate energies
(\ref{E1}) are
\begin{eqnarray}
c_{\overline{\nu}}^{\pm}&=&\frac{1}{4}\Bigl(1\pm\xi^2_{s,s_{13},s_{24}}-\eta_{s,s_{13},s_{24}}^2-\eta_{s+1,s_{13},s_{24}}^2\Bigr)\nonumber\\
a_{\overline{\nu}}^{\pm}&=&c_{\overline{\nu}}^{+} \pm2\biggl(\sum_{\sigma=\pm1}\Bigl[\Bigl(F_{s_1,s_1,s}^{s_{13}+(\sigma+1)/2,s_{24}}\Bigr)^2\nonumber\\
&
&-\sum_{\sigma'=\pm1}\Bigl(G_{s_1,s_1,\sigma\sigma's+\sigma(1+\sigma')/2}^{s_{13}+(1+\sigma)/2,s_{24}}\Bigr)^2\Bigr]\nonumber\\
&
&+(s_{13}\leftrightarrow s_{24})\biggr),\\
b_{\overline{\nu}}^{\pm}&=&\frac{1}{8}\sum_{\sigma'=\pm1}(2s+1+\sigma')^2\biggl(\eta^2_{s+(1+\sigma')/2,s_{13},s_{24}}\nonumber\\
&
&\pm8\sum_{\sigma=\pm1}\Bigl[\Bigl(G_{s_1,s_1,\sigma\sigma's+\sigma(1+\sigma')/2}^{s_{13}+(1+\sigma)/2,s_{24}}\Bigr)^2\label{b}\nonumber\\
& &+(s_{13}\leftrightarrow s_{24})\Bigr]\biggr),
\end{eqnarray}
where the $F_{s_1,s_3,s}^{s_{13},s_{24}}$,
$G_{s_1,s_3,s}^{s_{13},s_{24}}$, $\eta_{z,x,y}$, and $\xi_{z,x,y}$,
are given by Eqs.  (\ref{A})-(\ref{xi}), respectively.

\subsection{5. Type I First-order AFM level-crossing constants}

The Type I relevant both for $\tilde{J}_g'-\tilde{J}_g>0$
 AFM level-crossing inductions
 some low energy FM manifold states (with
$\tilde{J}_g'>\tilde{J}_g>0$) is $s_{13}=s_{24}=2s_1$ with $s$
arbitrary. Since all spins have $s_1$, we let
$a_{s,s_{13},s_{24}}^{\pm}\equiv a_{\overline{\nu}}^{\pm}$, etc. For
Type I, we have
\begin{eqnarray}
a_{s,2s_1,2s_1}^{s_1,\,\pm}&=&c_{s,2s_1,2s_1}^{s_1,\,-}\Bigl(1\mp\frac{1}{4s_1-1}\Bigr),\label{apm}\\
 b_{s,2s_1,2s_1}^{s_1,\,\pm}&=&\frac{1}{2(2s+3)(2s-1)}\biggl[s(s+1)
\nonumber\\
& &+[2s(s+1)-1][8s_1(2s_1+1)-s(s+1)]\nonumber\\
 & &\pm\frac{1}{4s_1-1}
\Bigl(2[16s_1^2+s(s+1)][s(s+1)-1]\nonumber\\
& &-8s_1[2s(s+1)-1]\Bigr)
\biggr],\label{bpm}\\
c_{s,2s_1,2s_1}^{s_1,\,\pm}&=&\frac{3[s(s+1)-1]-8s_1(2s_1+1)}{2(2s-1)(2s+3)}.\label{cpm}
\end{eqnarray}

From these expressions, we may evaluate the Type I first-order
level-crossing inductions for AFM tetramers. From the definitions of
the level-crossing constants in Eqs. (\ref{am})-(\ref{cm}), we
rewrite them to explicitly indicate the $s_1, s$ and type
dependencies, and for Type I, it is easy to show that
\begin{eqnarray}
a_{I,j}^{s_1,\,\pm}(s)&=&c_{I,j}^{s_1,\,-}(s)\Bigl(1\mp\frac{1}{4s_1-1}\Bigr),\\
b_{I}^{s_1,\,\pm}(s)&=&-\frac{2s[8s_1(2s_1+1)+4s^4-10s^2+3]}{(4s^2-1)(4s^2-9)}\nonumber\\
& &\times\Bigl(1\mp\frac{1}{4s_1-1}\Bigr),\\
c_{I,1}^{s_1,\,-}(s)&=&\frac{3[4s^3+5s^2-3s-3-8s_1(2s_1+1)]}{2(2s+1)(2s+3)},\nonumber\\
& &\label{c1mofs}\\
c_{I,2}^{s_1,\,-}(s)&=&\frac{c_{20}(s)+(4s^2-4s+3)8s_1(2s_1+1)}{2(4s^2-1)(4s^2-9)},\nonumber\\
& &\\ c_{20}(s)&=&3(4s^4-9s^2-s+3),\label{c20ofs}
\end{eqnarray}
for $j=1,2$.  For $s_1=1/2$, it is easy to see that
$a_{I,j}^{1/2,+}(s)=b_{I}^{1/2,+}(s)=0$ for $s=1,2$, as expected.

Moreover, it is easy to show that for Type I,
\begin{eqnarray}
c_2^{-}+\frac{1}{2}(b^{+}+b^{-})&=&-\frac{1}{3}c_1^{-},\\
a_2^{-}+b^{-}&=&-\frac{1}{3}a_1^{-}=-\frac{4s_1}{3(4s_1-1)}c_1^{-}.
\end{eqnarray}
This implies that for Type I, the axial near-neighbor and
next-nearest-neighbor axial anisotropic exchange interactions may be
combined to yield an effective axial anisotropic exchange
interaction given by Eq. (\ref{Jeff}). We emphasize that this is an
exact expression for the effective interactions affecting the
first-order level-crossing inductions, and is independent of the
level-crossing number $s$, provided that Type I applies.

For the single-ion contributions to the level-crossing inductions,
no such simple relation can be found.  We note that
\begin{eqnarray}
a_2^{+}+b^{+}&=&-\frac{1}{3}a_1^{+},
\end{eqnarray}
but the overall quantity $a_2^{+}+2b^{+}+a_1^{+}\cos^2\theta$ in Eq.
(\ref{levelcrossing}) contains the extra quantity $b^{+}$, which
depends upon $s,s_1$.

\subsection{6. Type II First-order AFM level-crossing constants}

For AFM tetramers with $\tilde{J}_g<0, \tilde{J}_g'-\tilde{J}_g<0$,
Type II, there are two classes of minimum energy configurations for
each $s$ manifold, depending upon whether $s$ is even or odd.  For
even $s$, the relevant states have $s_{13}=s_{24}=s/2$, and for $s$
odd, they are $s_{13},s_{24}=(s\pm1)/2, (s\mp1)/2$. This type is
also relevant for FM tetramers with $\tilde{J}_g>\tilde{J}_g'>0$,
especially for the first excited manifold of states with $s=4s_1-1$.
For even $s$ the relevant parameters are
\begin{eqnarray}
a_{s,s/2,s/2}^{s_1,\,\pm}&=&\frac{1}{2(2s-1)}\Bigl[s-1\mp\frac{f_1(s,s_1)}{2(s+3)}\Bigr],\\
f_1(s,s_1)&=&16s_1(s_1+1)-s^2-2s+6,\\
b_{s,s/2,s/2}^{s_1,\,\pm}&=&\frac{1}{2(2s-1)}\Bigl[s^2\pm\frac{f_2(s,s_1)}{s+3}\Bigr],\\
f_2(s,s_1)&=&16s_1(s_1+1)(s^2+2s-1)\nonumber\\
& &-s(s^3+4s^2+s-4),\\
c_{s,s/2,s/2}^{\pm}&=&\frac{s-1}{2(2s-1)}.
\end{eqnarray}

For odd $s$, the relevant parameters are
\begin{eqnarray}
a^{s_1,\,\pm}_{s,(s+1)/2,(s-1)/2}&=&\frac{1}{2s(2s-1)}\Bigl[s^2-s+1\nonumber\\
& &\qquad\mp\frac{f_3(s,s_1)}{2(s+2)(s+4)}\Bigr],\\
f_3(s,s_1)&=&16s_1(s_1+1)(s^2+3s-1)\nonumber\\
& &-s^4-5s^3+11s-11,\\
b^{s_1,\,\pm}_{s,(s+1)/2,(s-1)/2}&=&\frac{1}{2(2s-1)}\Bigl[s^2-1\nonumber\\
& &\qquad\pm\frac{f_4(s,s_1)}{(s+2)(s+4)}\Bigr],\\
f_4(s,s_1)&=&16s_1(s_1+1)(s^3+5s^2+4s-3)\nonumber\\
& &-(s+1)(s^4+6s^3+7s^2-3s+1),\nonumber\\
c^{+}_{s,(s+1)/2,(s-1)/2}&=&\frac{s^2-s+1}{2s(2s-1)},\\
c^{-}_{s,(s+1)/2,(s-1)/2}&=&\frac{(s+1)(s-1)^2}{2s^2(2s-1)}.
\end{eqnarray}
From these expressions, we obtain the level-crossing inductions for
the AFM type $\tilde{J}_g<0$ and $\tilde{J}_g'-\tilde{J}_g<0$.

For $s$ even, we have
\begin{eqnarray}
a_{II{\rm e},1}^{s_1,\pm}(s)&=&s(2s-1)a_{s,s/2,s/2}^{s_1,\pm}\nonumber\\
& &-(s-1)(2s-3)a_{s-1,s/2,(s-2)/2}^{s_1,\pm},\\
a_{||{\rm e},2}^{s_1,\pm}(s)&=&sa_{s,s/2,s/2}^{s_1,\pm}\nonumber\\
& &-(s-1)a_{s-1,s/2,(s-2)/2}^{s_1,\pm},\\
b_{II{\rm e}}^{s_1,\pm}(s)&=&b_{s,s/2,s/2}^{s_1,\pm}-b_{s-1,s/2,(s-2)/2}^{s_1,\pm},\\
c_{II{\rm e},1}^{s_1,\pm}(s)&=&s(2s-1)c_{s,s/2,s/2}^{\pm}\nonumber\\
& &-(s-1)(2s-3)c_{s-1,s/2,(s-2)/2}^{\pm},\\
c_{II{\rm e},2}^{\pm}(s)&=&sc_{s,s/2,s/2}^{\pm}\nonumber\\
& &-(s-1)c_{s-1,s/2,(s-2)/2}^{\pm}.
\end{eqnarray}

For $s$ odd, we have
\begin{eqnarray}
a_{II{\rm o},1}^{s_1,\pm}(s)&=&s(2s-1)a_{s,(s+1)/2,(s-1)/2}^{s_1,\pm}\nonumber\\
& &-(s-1)(2s-3)a_{s-1,(s-1)/2,(s-1)/2}^{s_1,\pm},\nonumber\\
& &\\
a_{II{\rm o},2}^{s_1,\pm}(s)&=&sa_{s,(s+1)/2,(s-1)/2}^{s_1,\pm}\nonumber\\
& &-(s-1)a_{s-1,(s-1)/2,(s-1)/2}^{s_1,\pm},\\
b_{II{\rm o}}^{s_1,\pm}(s)&=&b_{s,(s+1)/2,(s-1)/2}^{s_1,\pm}-b_{s-1,(s-1)/2,(s-1)/2}^{s_1,\pm},\nonumber\\
& &\\
c_{II{\rm o},1}^{s_1,\pm}(s)&=&s(2s-1)c_{s,(s+1)/2,(s-1)/2}^{\pm}\nonumber\\
& &-(s-1)(2s-3)c_{s-1,(s-1)/2,(s-1)/2}^{\pm},\nonumber\\
& &\\
c_{II{\rm o},2}^{s_1,\pm}(s)&=&sc_{s,(s+1)/2,(s-1)/2}^{\pm}\nonumber\\
& &-(s-1)c_{s-1,(s-1)/2,(s-1)/2}^{\pm}.
\end{eqnarray}

From these expressions, we may obtain the Type II AFM level-crossing
induction parameters.  For even $s$, we find
\begin{eqnarray}
a_{II{\rm e},1}^{s_1,\,\pm}(s)&=&\frac{2s-3}{2}\mp\frac{48s_1(s_1+1)+a_{10}^{\rm e}(s)}{4(s+1)(s+3)},\label{aII1pme}\\
a_{10}^{\rm e}(s)&=&-2s^3-5s^2+6s+18,\\
a_{II{\rm e},2}^{s_1,\,\pm}(s)&=&\frac{1}{2(2s-1)(2s-3)}\biggl[2s^2-6s+3\nonumber\\
&
&\pm\frac{16s_1(s_1+1)(2s^2-4s+3)+a_{20}^{\rm e}(s)}{2(s+1)(s+3)}\biggr],\label{aII2pme}\nonumber\\
& &\\
a_{20}^{\rm e}(s)&=&2s^4+2s^3-9s^2-18s+18,\\
b_{II{\rm e}}^{s_1,\,\pm}(s)&=&\frac{s}{(2s-1)(2s-3)}\biggl[s-1\nonumber\\
& &\pm\frac{16s_1(s_1+1)(s-2)+b_0^{\rm e}(s)}{2(s+1)(s+3)}\biggr],\label{bIIpme}\\
b_0^{\rm e}(s)&=&-4s^4-7s^3+20s^2+14s-18,\\
c_{II{\rm e},1}^{-}(s)&=&\frac{s(2s-3)}{2(s-1)},\label{cII1me}\\
c_{II{\rm
e},2}^{-}(s)&=&\frac{s(2s^2-4s+1)}{2(s-1)(2s-1)(2s-3)}.\label{cII2me}
\end{eqnarray}
Combining $a_{II{\rm e},2}^{s_1,\,-}(s)$ and $b_{II{\rm
e}}^{s_1,\,-}(s)$, we find
\begin{eqnarray}
a_{II{\rm e},2}^{s_1,\,-}+b_{II{\rm
e}}^{s_1,\,-}&=&\frac{1}{2}-\frac{16s_1(s_1+1)-d_0^{\rm
e}(s)}{4(s+1)(s+3)},\\
d_0^{\rm e}(s)&=&2s^3+7s^2+2s-6.
\end{eqnarray}
We note that  this expression differs substantially from that for
$a_{II{\rm e},1}^{s_1,\,-}(s)$, except for $s_1=1/2$ and $s=2$.
Similarly, it is elementary to combine $c_{II{\rm e},2}^{-}(s)$ and
$[b_{II{\rm e}}^{s_1,\,+}(s)+b_{II{\rm e}}^{s_1,\,-}(s)]/2$.  We
find
\begin{eqnarray}
c_{II{\rm e},2}^{-}(s)+\frac{1}{2}\Bigl(b_{II{\rm
e}}^{s_1,\,+}(s)+b_{II{\rm
e}}^{s_1,\,-}(s)\Bigr)&=&\frac{s}{2(s-1)}.
\end{eqnarray}
This simple expression differs from that for $c_{II{\rm
e},1}^{-}(s)$ by the factor $2s-3$.  However, at $s=2$, the only
even $s$ value for $s_1=1/2$,  they are equivalent. In addition, as
for Type I, there is no simple relation between the single-ion
parameters $a_{II{\rm e},2}^{s_1,\,+}(s)+2b_{II{\rm
e}}^{s_1,\,+}(s)$ and $a_{II{\rm e},1}^{s_1,\,+}(s)$.

For odd $s$, the Type II level-crossing induction parameters are
\begin{eqnarray}
a_{II{\rm o},1}^{s_1,\,\pm}(s)&=&\frac{2s-1}{2}\mp\frac{48s_1(s_1+1)+a_{10}^{\rm o}(s)}{4(s+2)(s+4)},\label{aIIpmo}\\
a_{10}^{\rm o}(s)&=&-2s^3-11s^2-10s+17,\\
a_{II{\rm o},2}^{s_1,\,\pm}(s)&=&\frac{1}{2(2s-1)(2s-3)}\biggl[2s^2-2s-1\nonumber\\
& &\pm\frac{16s_1(s_1+1)(2s^2+1)+a_{20}^{\rm
o}(s)}{2(s+2)(s+4)}\biggr],\label{aII2pmo}\\
a_{20}^{\rm o}(s)&=&2s^4+10s^3+9s^2-22s-5,\\
b_{II{\rm
o}}^{s_1,\,\pm}(s)&=&\frac{1}{(2s-1)(2s-3)}\biggl[(s-1)(s-2)\nonumber\\
& &\pm\frac{16s_1(s_1+1)(s^2-4s+1)+b_0^{\rm
o}(s)}{2(s+2)(s+4)}\biggr],\label{bIIpmo}\nonumber\\
& &\\
b_0^{\rm o}(s)&=&-4s^5-19s^4+54s^2-2s-5,\\
c_{II{\rm o},1}^{-}(s)&=&\frac{(s-1)(2s-1)}{2s},\label{cII1o}\\
c_{II{\rm
o},2}^{-}(s)&=&\frac{(s-1)(2s^2-4s+3)}{2s(2s-1)(2s-3)}.\label{cII2o}
\end{eqnarray}
We note that $a_{II{\rm e},j}^{1/2,\,+}(2)=a_{II{\rm
o},j}^{1/2,\,+}(1)=b_{II{\rm e}}^{1/2,\,+}(2)=b_{II{\rm
o}}^{1/2,\,+}(1)=0$ for $j=1,2$, as expected.  However, by combining
$a_{II{\rm o},2}^{s_1,\,-}(s)$ and $b_{II{\rm o}}^{s_1,\,-}(s)$, we
have
\begin{eqnarray}
a_{II{\rm o},2}^{s_1,\,-}+b_{II{\rm
o}}^{s_1,\,-}&=&\frac{1}{2}-\frac{16s_1(s_1+1)-d_0^{\rm
o}(s)}{4(s+2)(s+4)},\\
d_o^{\rm o}(s)&=&2s^3+13s^2+22s+5,
\end{eqnarray}
which differs substantially from the version with $s_1=1/2$.  In
addition, it is elementary to combine $c_{II{\rm
o},2}^{-}(s)+[b_{II{\rm o}}^{s_1,\,+}(s)+b_{II{\rm
o}}^{s_1,\,-}(s)]/2$.  We find
\begin{eqnarray}
c_{II{\rm o},2}^{-}(s)+\frac{1}{2}\Bigl(b_{II{\rm
o}}^{s_1,\,+}(s)+b_{II{\rm
o}}^{s_1,\,-}(s)\Bigr)&=&\frac{(s-1)}{2s},\end{eqnarray} which
differs from $c_{II{\rm o},1}^{-}(s)$ by the factor $2s-1$.  At
$s=1$, the only relevant odd $s$ value for $s_1=1/2$, these are
equivalent.  In addition, as for Type I and the even crossings of
Type II, there is no simple relation between the single-ion
parameters $a_{II{\rm o},2}^{s_1,\,+}(s)+2b_{II{\rm
o}}^{s_1,\,+}(s)$ and $a_{II{|rm o},1}^{s_1,\,+}(s)$.

\subsection{7. Hartree INS functions}

The functions $L_{\nu,\nu'}({\bm q})$ and $M_{\nu,\nu'}({\bm q})$ in
the self-consistent Hartree  INS $S_g^{(1)}({\bm q},\omega)$ in the
induction representation are given by

\begin{eqnarray}
L_{\nu,\nu'}({\bm
q})&=&\delta_{m',m}\delta_{s_{24}',s_{24}^{}}\biggl[m^2\delta_{s',s}\nonumber\\
& &\times\Bigl(\delta_{s_{13}',s_{13}^{}}f_{\overline{\nu},0}({\bm
q})+\sum_{\sigma''=\pm1}\delta_{s_{13}',s_{13}^{}+\sigma''}f^{\sigma''}_{\overline{\nu},1}({\bm
q})\Bigr)\nonumber\\
&
&+\sum_{\sigma'=\pm1}\delta_{s',s+\sigma'}\Bigl(C^m_{-\sigma's-(\sigma'+1)/2}\Bigr)^2\nonumber\\
& &\times\Bigl(\delta_{s_{13}',s_{13}^{}}
f^{\sigma'}_{\overline{\nu},2}({\bm
q})\nonumber\\
&
&\qquad+\sum_{\sigma''=\pm1}\delta_{s_{13}',s_{13}^{}+\sigma''}f^{\sigma',\sigma''}_{\overline{\nu},3}({\bm
q})\Bigr)\biggr]\nonumber\\
& &+\left(\begin{array}{c} s_{13}\leftrightarrow s_{24}\\
s_{13}'\leftrightarrow s_{24}'\\
q_y\rightarrow -q_y\end{array}\right),\\
M_{\nu,\nu'}({\bm
q})&=&\sum_{\sigma=\pm1}\delta_{m',m+\sigma}\delta_{s_{24}',s_{24}^{}}\biggl[\Bigl(A_s^{\sigma m}\Bigr)^2\delta_{s',s}\nonumber\\
& &\times\Bigl(\delta_{s_{13}',s_{13}^{}}f_{\overline{\nu},0}({\bm
q})+\sum_{\sigma''=\pm1}\delta_{s_{13}',s_{13}^{}+\sigma''}f^{\sigma''}_{\overline{\nu},1}({\bm
q})\Bigr)\nonumber\\
&
&+\sum_{\sigma'=\pm1}\delta_{s',s+\sigma'}\Bigl(D^{\sigma,m}_{-\sigma's-(\sigma'+1)/2}\Bigr)^2\nonumber\\
& &\times\Bigl(\delta_{s_{13}',s_{13}^{}}
f^{\sigma'}_{\overline{\nu},2}({\bm
q})\nonumber\\
&
&\qquad+\sum_{\sigma''=\pm1}\delta_{s_{13}',s_{13}^{}+\sigma''}f^{\sigma',\sigma''}_{\overline{\nu},3}({\bm
q})\Bigr)\biggr]\nonumber\\
& &+\left(\begin{array}{c} s_{13}\leftrightarrow s_{24}\\
s_{13}'\leftrightarrow s_{24}'\\
q_y\rightarrow -q_y\end{array}\right),\\
f_{\overline{\nu},0}({\bm q})&=&\frac{1}{8}\Bigl(f_{+}({\bm
q})+\xi^2_{s,s_{13},s_{24}}f_{-}({\bm q})\nonumber\\
& &-2\xi_{s,s_{13},s_{24}}\sin(q_xa)\sin(q_ya)\Bigr),\\
f_{\overline{\nu},1}^{\sigma''}({\bm
q})&=&2\Bigl(1-\cos[a(q_x+q_y)]\Bigr)\nonumber\\
&
&\times\Bigl(F_{s_1,s_1,s}^{s_{13}+(\sigma''+1)/2,s_{24}}\Bigr)^2\\
f_{\overline{\nu},2}^{\sigma'}({\bm q})&=&\frac{1}{8}f_{-}({\bm
q})\eta^2_{s+(\sigma'+1)/2,s_{13},s_{24}},\\
f_{\overline{\nu},3}^{\sigma',\sigma''}({\bm
q})&=&2\Bigl(1-\cos[a(q_x+q_y)]\Bigr)\nonumber\\
&
&\times\Bigl(G_{s_1,s_1,\sigma'\sigma''s+\sigma''(\sigma'+1)/2}^{s_{13}+(\sigma''+1)/2,s_{24}}\Bigr)^2,\\
f_{\pm}({\bm
q})&=&1+\cos(q_xa)\cos(q_ya)\nonumber\\
& &\pm\cos(q_zc)[\cos(q_xa)+\cos(q_ya)],
\end{eqnarray}
where the $A_s^m$, $C_s^m$, $D_s^{\tilde{\sigma},m}$,
$F_{s_1,s_3,s}^{s_{13},s_{24}}$,  $G_{s_1,s_3,s}^{s_{13},s_{24}}$,
$\eta_{z,x,y}$, $\xi_{z,x,y}$,
 are given by Eqs. (\ref{Asm}),
(\ref{Csm}), (\ref{Dsigmasm}),  and (\ref{A})-(\ref{xi}),
respectively.

\end{document}